\documentclass[12pt,preprint]{aastex}
\usepackage{rotating}

\newcommand{\dg}{$^{\circ}$}
\newcommand{\ncrit}{n$_{crit}$}
\newcommand{\um}{$\mu$m}

\newcommand{\Msun}{M$_{\odot}$}

\newcommand{\Kkms}{K\,km~s$^{-1}$}
\newcommand{\kms}{km~s$^{-1}$}
\newcommand{\Spitzer}{{\it Spitzer}}
\newcommand{\Herschel}{{\it{Herschel}}}
\newcommand{\cms}{${\rm cm}^{-2}$}
\newcommand{\cmc}{${\rm cm}^{-3}$}
\newcommand{\hii}{\mbox{$\mathrm{H\,{\scriptstyle {II}}}$}}

\newcommand{\cloud}{G0.253+0.016}
\newcommand{\vlsr}{V$_{LSR}$}

\newcommand{\nthpnt}{N$_{2}$H$^{+}$}
\newcommand{\nthp}{N$_{2}$H$^{+}$\,(1--0)}
\newcommand{\hcopnt}{HCO$^{+}$}
\newcommand{\hcop}{HCO$^{+}$\,(1--0)}
\newcommand{\siont}{\mbox{SiO}}
\newcommand{\sio}{\mbox{SiO~(2--1)}}
\newcommand{\htcopnt}{H$^{13}$CO$^{+}$}
\newcommand{\htcop}{H$^{13}$CO$^{+}$\,(1--0)}

\newcommand{\tctfs}{$^{13}$C$^{34}$S\,(2--1)}
\newcommand{\tcsnt}{$^{13}$CS}
\newcommand{\tcs}{$^{13}$CS~(2--1)}
\newcommand{\chtcnnt}{CH$_{3}$CN}
\newcommand{\chtcn}{CH$_{3}$CN\,5(1)--4(1)}
\newcommand{\hctn}{HC$_{3}$N\,(10--9)}
\newcommand{\hctnnt}{HC$_{3}$N}
\newcommand{\hncnt}{HNC}
	
\newcommand{\hcnnt}{HCN}
	
\newcommand{\hncofznt}{HNCO}
\newcommand{\cchnt}{C$_{2}$H} 	   
\newcommand{\hntcnt}{HN$^{13}$C}
\newcommand{\hnc}{HNC~(1--0)}
\newcommand{\hctccn}{HC$^{13}$CCN\,(10--9)}	
\newcommand{\hcn}{HCN (1--0)}
\newcommand{\hncofo}{HNCO\,4(1,3)--3(1,2)}	
\newcommand{\hncofz}{HNCO\,4(0,4)--3(0,3)}
\newcommand{\cch}{C$_{2}$H\,(1--0)~3/2--1/2} 	   
\newcommand{\hntc}{HN$^{13}$C\,(1--0)}
\newcommand{\halpha}{H\,41$\alpha$}

\newcommand{\tastar}{T$^{*}_{A}$}
\newcommand{\tmb}{T$_{mb}$}
\newcommand{\hh}{H$_{2}$}
\slugcomment{}

\shorttitle{G0.253+0.016}
\shortauthors{Rathborne et al.}

\begin{document}
%%%%%%%%%%%%%%%%%%%%%%%%%%%%%%%%%%%%%%%%%%%%%%%%%%%%%%%%%%%%%%%%%%%%%%%%%%%%%%%%%%%%%%%%%%%%%%%%%%%%%%%%%%%%%%%%%%%%%%%%%%%%%%%%%%%%%%%%%%%%%%%%%%%%%%%%%%
\title{G0.253+0.016: A centrally condensed, high-mass protocluster}
%\title{Global chemistry and kinematics within the high-mass protocluster G0.25+0.02}
%Hot, shocked gas within G0.25+0.02: a molecular cloud progenitor of an Arches-like cluster}
%%%%%%%%%%%%%%%%%%%%%%%%%%%%%%%%%%%%%%%%%%%%%%%%%%%%%%%%%%%%%%%%%%%%%%%%%%%%%%%%%%%%%%%%%%%%%%%%%%%%%%%%%%%%%%%%%%%%%%%%%%%%%%%%%%%%%%%%%%%%%%%%%%%%%%%%%%
\author{J. M. Rathborne}
\affil{CSIRO Astronomy and Space Science,  P.O. Box 76, Epping NSW, 1710, Australia; Jill.Rathborne@csiro.au}
\and
\author{S. N. Longmore\footnote{Current Address: Astrophysics Research Institute, Liverpool John Moores University, Egerton Wharf, Birkenhead CH41 1LD, UK}}
\affil{European Southern Observatory, Karl-Schwarzschild-Str. 2,  85748 Garching bei Munchen, Germany}
\and
\author{J. M. Jackson}
\affil{Institute for Astrophysical Research, Boston University, Boston, MA 02215, USA}
\and
\author{J. B. Foster\footnote{Current Address: Department of Astronomy, Yale University, P.O. Box 208101 New Haven, CT 06520-8101, USA}}
\affil{Institute for Astrophysical Research, Boston University, Boston, MA 02215, USA}
\and
\author{Y. Contreras}
\affil{CSIRO Astronomy and Space Science,  P.O. Box 76, Epping NSW, 1710, Australia}
\and
\author{G. Garay}
\affil{Universidad de Chile, Camino El Observatorio1515, Las Condes, Santiago, Chile}
\and
\author{L. Testi}
\affil{European Southern Observatory, Karl-Schwarzschild-Str. 2,  85748 Garching bei Munchen, Germany; INAF-Arcetri, Largo E. Fermi 5, I-50125 Firenze, Italy; Excellence Cluster Universe, Boltzmannstr. 2, D-85748, Garching, Germany}
\and
\author{J. F. Alves}
\affil{University of Vienna, T\"urkenschanzstrasse 17, 1180 Vienna, Austria}
\and
\author{J. Bally}
\affil{Center for Astrophysics and Space Astronomy, University of Colorado, UCB 389, Boulder, CO 8030}
\and
\author{N. Bastian}
\affil{Astrophysics Research Institute, Liverpool John Moores University, Egerton Wharf, Birkenhead CH41 1LD, UK}
\and
\author{J. M. D. Kruijssen}
\affil{Max-Planck Institut fur Astrophysik, Karl-Schwarzschild-Strasse 1, 85748, Garching, Germany}
\and
\author{E. Bressert}
\affil{CSIRO Astronomy and Space Science,  P.O. Box 76, Epping NSW, 1710, Australia}
%%%%%%%%%%%%%%%%%%%%%%%%%%%%%%%%%%%%%%%%%%%%%%%%%%%%%%%%%%%%%%%%%%%%%%%%%%%%%%%%%%%%%%%%%%%%%%%%%%%%%%%%%%%%%%%%%%%%%%%%%%%%%%%%%%%%%%%%%%%%%%%%%%%%%%%%%%
\begin{abstract}
Despite their importance as stellar nurseries and the building blocks of galaxies, very little is known about 
the formation of the highest mass clusters. The dense clump \cloud\,  represents an example of 
a clump that may form an Arches-like, high-mass cluster. Here we present molecular line maps toward \cloud\, taken as part of the 
MALT90 molecular line survey, complemented with APEX observations. Combined, these data 
reveal the global physical properties and kinematics of \cloud. Recent {\it{Herschel}} data show that while the
dust temperature is low ($\sim$19\,K) toward its centre,  the dust temperature on the exterior is 
higher ($\sim$27\,K) due to external heating. 
Our new molecular line data reveal that, overall, the morphology of dense gas detected toward \cloud\,
matches very well its IR extinction and dust continuum emission.  
An anti-correlation between the dust and gas  column densities  toward its centre indicates that the clump is centrally condensed 
with a cold, dense interior in which the molecular gas is  chemically  depleted. 
The velocity field shows a strong gradient along the clump's major axis, with
the blue-shifted side at higher Galactic longitude.  The optically thick gas tracers are systematically red-shifted with respect to 
the optically thin and hot gas tracers, indicating radial motions. 
 The gas kinematics and line ratios support the recently proposed scenario in which G0.253+0.016 results from a tidal 
compression during a recent pericentre passage near SgrA$^{*}$. 
Because \cloud\, represents an excellent example of a clump that may form a high-mass cluster,  its detailed study 
should reveal a wealth of knowledge about the early stages of cluster formation.
\end{abstract}
%%%%%%%%%%%%%%%%%%%%%%%%%%%%%%%%%%%%%%%%%%%%%%%%%%%%%%%%%%%%%%%%%%%%%%%%%%%%%%%%%%%%%%%%%%%%%%%%%%%%%%%%%%%%%%%%%%%%%%%%%%%%%%%%%%%%%%%%%%%%%
%%%%%%%%%%%%%%%%%%%%%%%%%%%%%%%%%%%%%%%%%%%%%%%%%%%%%%%%%%%%%%%%%%%%%%%%%%%%%%%%%%%%%%%%%%%%%%%%%%%%%%%%%%%%%%%%%%%%%%%%%%%%%%%%%%%%%%%%%%%%%

%%%%%%%%%%%%%%%%%%%%%%%%%%%%%%%%%%%%%%%%%%%%%%%%%%%%%%%%%%%%%%%%%%%%%%%%%%%%%%%%%%%%%%%%%%%%%%%%%%%%%%%%%%%%%%%%%%%%%%%%%%%%%%%%%%%%%%%%%%%%%
\section{Introduction}
%%%%%%%%%%%%%%%%%%%%%%%%%%%%%%%%%%%%%%%%%%%%%%%%%%%%%%%%%%%%%%%%%%%%%%%%%%%%%%%%%%%%%%%%%%%%%%%%%%%%%%%%%%%%%%%%%%%%%%%%%%%%%%%%%%%%%%%%%%%%%

\cloud, an infrared dark cloud (IRDC) located in close proximity to the Galactic Centre 
\citep{Lis94,Lis98,Lis01,Immer12}, has recently gained attention as a potential candidate of 
a high-mass cluster in a very early stage of formation \citep{Longmore12}. \cloud\, is
clearly seen as an extinction feature from mid- to far-IR wavelengths.  
Recent modelling of \Herschel\, data reveals a central dust temperature of 
$\sim$20\,K, peak \hh\, column density of $\sim$ 3.3 $\times$ 10$^{23}$\,\cms, and mass of 
$\sim$ 1.3$\times$ 10$^{5}$\,\Msun\, \citep{Molinari11,Longmore12}. Despite
its high mass and density, it shows little evidence for wide-spread star formation,
consistent with the low dust temperature. This combination of properties, little star 
formation and a low dust temperature, yet with a high mass and column density, 
make \cloud\, extreme when compared to other known Galactic molecular clumps. 
Indeed, \cite{Ginsburg12} find no other examples of cold, starless clumps with masses 
$>$ 10$^{4}$\,\Msun\, in the recent Bolocam Galactic Plane Survey. Because it 
contains $\sim$ 10$^{5}$\,\Msun\,  of material,  \cloud\, has sufficient mass to form 
a young massive cluster (YMC) through direct collapse. As such, its detailed study 
may reveal the initial conditions within a protocluster and the processes by which 
a high-mass cluster is formed.
 
Young massive clusters are gravitationally bound stellar systems with masses $>$10$^{4}$\,\Msun\,and 
ages $<$ 100\,Myr \citep{Portegies-Zwart10}. Only a few YMCs have  been 
identified within our Galaxy (e.g., Arches, Quintuplet, Westerlund 1, RSCG, 
GLIMPSE-CO1: \citealp{Figer99,Clark05,Figer06,Davies11}). Recent work suggests 
that YMCs may be the `missing link'  between open clusters and globular clusters and, 
as such, may be the local-universe analogs of the progenitors of globular clusters 
(e.g.\, \citealp{Elmegreen97,Bastian08,Kruijssen12}). Characterising and understanding 
how these YMCs form is critical to connect  Galactic and extra-galactic cluster formation. 
Identifying a sample of Galactic molecular clouds that may form YMCs is important because 
their detailed study can shed light not only on how these high-mass clusters form, but potentially 
on how all clusters form.  To  date, \cloud\, is one of a handful of known molecular clumps with enough 
mass to form a cluster of similar mass to these Galactic YMCs through direct collapse. Because 
\cloud\, shows evidence for structure on small spatial scales (7.5\arcsec), \cite{Longmore12} 
speculate that it may be destined to form a YMC through hierarchical fragmentation, in a scaled 
up version of open cluster formation. Indeed, recent models predict that \cloud\, should form a cluster 
through hierarchical fragmentation in which $>$80\% of the stars should remain bound after the 
expulsion of the residual gas by feedback \citep{Kruijssen12b}.

Located at a distance of $\sim$8.5 kpc, \cloud\, lies $\sim$100\,pc from the Galactic Centre 
(GC; \citealp{Molinari11}). Its location within the harsh GC environment may provide clues to 
the formation of such a high-mass protocluster and whether star formation can progress within it. 
What remains unknown is if the influence of strong tidal forces, high turbulence, and extreme 
magnetic fields, radiation fields, and cosmic ray fluxes  within the GC region helps or hinders 
the formation of a high-mass cluster. While some YMCs within the Galaxy are located close to the 
GC (e.g. Arches, Quintuplet), recent work finds that given the large reservoir of dense gas available, 
the broader GC region appears to be under-producing stars compared to commonly assumed 
relations between gas mass and star formation (e.g., \citealp{Longmore13}).

Molecular line studies show that the bulk of the gas in the Central Molecular Zone (CMZ), the region
within $\sim$ 200\,pc of the Galactic Centre, has temperatures of $\sim$ 80\,K and 
densities $> 10^{4}$\,\cmc\, \citep{Walmsley86, Ao13}. Somewhat more 
surprising is that many complex organic molecules show bright, widespread emission across the 
CMZ, leading to the speculation that shock chemistry might dominate in the region \citep{Wilson82,Martin-Pintado01}.
Molecular clouds located in the CMZ have exceptional physical properties; they are denser, warmer, more 
turbulent, and more massive compared to molecular clouds in the Galactic disk.
Such high densities are required for the molecular clouds to survive in the 
steep Galactic Center gravitational potential \citep{Bania86}. 

Previous molecular line observations toward \cloud\, (also known as GCM0.25+0.011 and M0.25+0.01; 
\citealp{Gusten81,Lis94}) reveal large line-widths, indicating a high degree of turbulence, similar to other 
molecular clouds near the GC. Given its location in the CMZ where the interstellar radiation field (ISRF) and 
cosmic ray ionisation rate (CRIR) are high \citep{Clark13,Yusef-zadeh13}, one might expect a high-density clump like \cloud\,  to be externally 
heated. Indeed, observations have shown that it has a low internal luminosity and that its derived dust temperature 
increases smoothly from $\sim$ 19\,K in its centre to $\sim$ 27\,K toward its edges \citep{Lis01,Longmore12}. 

These observations are consistent with recent SPH modelling of the dust and gas temperature distribution 
in \cloud\, \citep{Clark13}. In these models, the gas and dust temperatures are derived for \cloud\, as the 
ISRF and CRIR vary. This modelling suggest that a very high ISRF, combined with a high CRIR, reproduces 
well the observed discrepant dust and gas temperatures: both the ISRF and CRIR are predicted to be 1000 times higher than 
the values measured in the solar neighbourhood. Indeed, the models of \cite{Clark13} find that the gas and dust are not coupled, 
have different temperatures throughout the clump, and that \cloud\, is externally heated with a relatively 
cooler interior that is highly sub-structured. 

Despite its high mass, there appears to be very little active high-mass star formation within \cloud. The detection 
of a weak water maser toward it \citep{Lis94,Breen11} supports the idea that star formation may be `turning on'  
within this clump \citep{Lis98}.  JVLA radio continuum observations reveal three compact \hii\, 
regions located toward the periphery of the clump; their fluxes suggest that if they lie at the distance of \cloud,
then they are powered by B0.5 ZAMS stars \citep{Rodriguez13}.
These \hii\, regions, however, have not yet been definitively shown to be associated with \cloud. 
Recent CARMA and SMA observations of N$_{2}$H$^{+}$ line and dust continuum 
emission show little evidence for dense material on small scales, leading \cite{Kauffmann13} to speculate that \cloud\, 
may lack the potential to form a cluster.
 
A measured gas temperature of $\sim$80\,K, significantly higher than its derived dust 
temperature ($\sim$20\,K),  led \cite{Lis01} to speculate that shocks associated with clump-clump collisions 
might be the dominant heating source for the gas, rather than reprocessed UV radiation. 
Indeed, recent H$_{2}$CO mapping of the
dense gas across the CMZ measure gas kinetic temperatures ranging from 50--100\,K \citep{Ao13}, leading to the conclusion
that the gas may be heated by the dissipation of turbulent energy and/or cosmic rays rather than by photon heating.
This idea is also supported by other recent work that suggests that the high cosmic ray ionisation rate in the GC is responsible 
for the high gas temperatures (e.g. \citealp{Yusef-zadeh13}).
Limited  observations of SiO emission toward \cloud\, show an overall correlation between the SiO emission and 
high column density gas \citep{Lis01}, however, fully-sampled maps are needed to understand this 
apparent correspondence and to determine the details of the gas kinematics.

 To provide a more detailed picture of the dense and shocked gas within \cloud\, we utilise molecular line data obtained 
recently as part of the Millimetre Astronomy Legacy Team 90\,GHz (MALT90) survey. 
This survey covers many important spectral lines within the 3\,mm (90\,GHz) regime. We complement 
these data with additional, higher frequency APEX observations (at 230 and 345 GHz).
When combined, these data can be used to measure the global physical properties and kinematics of the gas within
 \cloud\,  and to determine whether it may be in the early stages of forming a high-mass cluster triggered by a recent close 
 passage to Sgr\,A$^{*}$, as suggested recently by \cite{Longmore13b}.
 
% suggest a possible mechanism for the formation of \cloud\, which has been triggered by the passage of a gas stream close to Sgr\,A$^{*}$. 

 %The broad line-widths measured 
 %toward \cloud\, suggests that turbulence may in fact be preventing its rapid fragmentation. 

%%%%%%%%%%%%%%%%%%%%%%%%%%%%%%%%%%%%%%%%%%%%%%%%%%%%%%%%%%%%%%%%%%%%%%%%%%%%%%%%%%%%%%%%%%%%%%%%%%%%%%%%%%%%%%%%%%%%%%%%%%%%%%%%%%%%%%%%%%%%%
\section{The Millimetre Astronomy Legacy Team 90\,GHz (MALT90) survey}
%%%%%%%%%%%%%%%%%%%%%%%%%%%%%%%%%%%%%%%%%%%%%%%%%%%%%%%%%%%%%%%%%%%%%%%%%%%%%%%%%%%%%%%%%%%%%%%%%%%%%%%%%%%%%%%%%%%%%%%%%%%%%%%%%%%%%%%%%%%%%

\cloud\, was observed in July 2010 during the first season of the MALT90\footnote{http://malt90.bu.edu} survey 
(\citealp{Foster-malt90,Jackson-malt90}, Rathborne et al. in prep). By observing thousands of high-mass 
star-forming clumps, MALT90 aims to characterise their global chemical and physical evolution.

\subsection{Survey strategy}
MALT90 utilizes the Mopra 22m telescope located near Coonabarabran in New South Wales, Australia to obtain
 small maps around clumps identified in the recent 870\,\um\, continuum APEX (Atacama Pathfinder Experiment) 
 Telescope Large Area Survey of the Galaxy (ATLASGAL; \citealp{Schuller09,Contreras13}).  For all maps, the `zoom mode' of the 
 MOPS spectrometer was configured to a central frequency of  89.690 GHz.  The zoom mode allows 16 individual 
 IFs to be tuned to any molecular transition within a range of 8 GHz, allowing us to simultaneously map their distribution. 
 Each zoom band was 137.5 MHz wide, with 4096 spectral channels, corresponding to a velocity resolution of 0.11\,\kms. At 
 90\,GHz, Mopra's beam is $\sim$ 38\arcsec\, and the typical pointing accuracy is better than 8 \arcsec\, \citep{Foster-malt90}.

MALT90 has obtained 3\arcmin$\times$3\arcmin\, maps around $\sim$ 2,000 ATLASGAL clumps. 
The maps were obtained using the On-The-Fly (OTF) mode,  scanning in 
Galactic coordinates. To facilitate the removal of any 
striping artefacts or baseline variations in the maps that result from weather fluctuations, two maps were obtained by scanning in 
orthogonal directions: one scanning in Galactic Latitude, the other in Galactic Longitude. At the end of each scanning row 
a reference position was observed. An individual absolute reference position was used, located at a position 
a degree away from the source position in Galactic Latitude. The pointing was checked using a nearby, bright SiO maser 
immediately before and after the maps were obtained. The system temperature was measured via a paddle every 15 mins 
(two per map). Each map took $\sim$ 31 mins to complete. 
The data reduction was performed using an automated python-based script (ASAP, Livedata and 
Gridzilla; see \citealp{Jackson-malt90} for more details). The processed data cubes were 
downloaded from the MALT90 on-line archive\footnote{http://atoa.atnf.csiro.au/MALT90}. The  typical  rms 
noise of \tastar\, is $\sim$ 0.25 K  per channel. To convert to main-beam brightness temperature (\tmb), 
we divide the antenna temperature (\tastar) by the main-beam efficiency ($\eta_{mb}$) of 0.49 
\citep{Ladd05}. To cover the full extent of \cloud\, we combined two adjacent MALT90 maps. In all images the (0,0) offset in Galactic 
Longitude and Latitude corresponds to the peak in the 870\,\um\, dust continuum emission at $l$=0.257\dg, $b$=0.016\dg.

\subsection{Observed molecular transitions}
The 16 available IFs for the MALT90 survey were tuned to the specific molecular transitions listed 
in Table~\ref{malt90-lines} (\citealp{Jackson-malt90}). These molecules and transitions 
were selected because they span a range in excitation energies and critical densities, and, thus, can be 
used to reveal the physical and chemical state of the gas.  The four brightest transitions 
that are typically detected towards clumps within the survey are \nthpnt, \hncnt, \hcopnt, and \hcn. These 
are ground-state rotational transitions of molecules with similar critical densities and are typically detected 
in dense star-forming regions independent of the star formation activity and, thus, internal 
temperatures.  Because \nthpnt\, better resists depletion at low temperatures and 
high-densities compared to carbon-bearing species, it traces both hot and cold 
gas.  As such, it is a useful probe of the gas associated with all phases in the evolutionary 
sequence of a high-mass star-forming clump. \hcopnt\, is a good tracer of kinematics, as it often shows 
broad line wings indicative of outflows or line asymmetries indicative of infall. However, because these 
ground-state transitions are often optically thick, also included within the survey are four optically thin isotopologues: \tcs, 
\tctfs, \hntc, and \htcop. 
  
Several molecular transitions that trace more complex chemistry are also included within the survey: 
\chtcn, \hctn, \hctccn, \hncofz, and \hncofo. These high excitation energy, complex carbon-bearing species trace heated gas 
and are often associated with `hot molecular cores' (HMCs): regions of molecular material that 
are compact ($<$~0.1\,pc), dense (10$^{5}$--10$^{8}$\,\cmc), with high temperatures 
(T$_{gas}$ $>$ 100\,K; \citealp{Garay99,Kurtz00,Churchwell02}). Because
their environment is heated by the UV radiation from a high-mass star, HMCs
trace the immediate vicinity around a recently formed high-mass star and are an excellent 
signpost to the earliest phases in the formation of a high-mass star.

Also included is the \halpha\, recombination line which traces ionized gas; \sio\, which traces 
shocked gas and outflows; and \cch\, which, in addition to tracing cold dense gas \citep{Vasyunina11}, 
can also be produced in photodissociation regions (PDRs; \citealp{Lo09,Gerin11}) when the UV radiation 
from a high-mass star heats and dissociates molecules on the edge of dense molecular material. 

Of the molecules that have been observed,  HNC, \nthpnt, \cchnt, HCN, and \chtcnnt\, all show hyperfine 
splitting (listed in approximate order of increasing separation between the hyperfine components). Because 
the measured line-widths toward \cloud\, are broad it is often difficult to separate the hyperfine components 
as their offset in their hyperfine splitting is significantly less than the line-width. 

\subsection{Complementary APEX observations: higher J transitions}

While the molecular transitions observed as part of the MALT90 survey can reveal the location, kinematics, 
and opacity of the dense and shocked/heated gas, by themselves, they are insufficient to determine the 
physical conditions reliably. Only with complementary observations of the higher J transitions of 
the same molecular species can we perform a multi-level excitation analysis which provides information on
the physical properties. To obtain these complementary data we utilised the APEX telescope, located in 
Llano de Chajnantor, Chile, to map emission from the \hncnt, \hcnnt, \hcopnt, and \nthpnt\, (3--2) and (4--3), 
and SiO\,(5--4) transitions.

%Moreover, the given the expected complex velocity structure of the gas within the clump, additional the kinematical information
%provided by these observations will also be beneficial.

Using the OTF mapping mode, we obtained half-beam sampled 3\arcmin $\times$3\arcmin\, maps across \cloud\, in 
each of the transitions listed above. The observations were conducted in September 2012. In all cases the maps 
were centred on the peak in the dust continuum emission and, thus, only cover the brightest part of the clump (i.e. 
the coverage is similar to that of the individual map from the MALT90 survey in the lower right of the moment maps 
in Fig.~\ref{momentmaps-densegas}). We used the APEX-1 ($\sim$230\,GHz, beam size of $\sim$27\arcsec) and 
APEX-2 ($\sim$350\,GHz, beam size of $\sim$18\arcsec) heterodyne receivers tuned to the transitions listed above 
which resulted in a velocity resolution of $\sim$ 1.2\,\kms.
The maps took between 1--2.5 hours to complete, depending on the observing frequency. A  fixed off-source position 
was used for all maps.  A nearby SiO maser was used to check the pointing accuracy approximately every hour and 
was typically better than 4\arcsec.  During our observations the water vapour column density ranged from 0.6--1.2\,mm 
and the system temperatures were typically 200--300K. The data were baseline subtracted before being combined and 
gridded into data cubes within the GILDAS package. While each cube was obtained so that it was Nyquist sampled at 
the corresponding frequency, all cubes were smoothed and re-gridded in ($l$,$b$) to match the beam and pixel size of 
the MALT90 maps. The final maps have a \tastar\, 1$\sigma$ rms noise sensitivity of $\sim$ 0.2 K channel$^{-1}$ in 
the 38\arcsec\, beam. To convert from \tastar\, to \tmb\, we used a beam efficiency, $\eta_{mb}$, of 0.73 and 0.75 
for the 230 and 350\,GHz data respectively \citep{Gusten06}.

\subsection{Moment analysis, channel maps, position-velocity diagrams}
\label{moment-analysis}

The observed emission was characterised via a moment analysis \citep{Sault95}. This analysis is
independent of Gaussian deconvolution and assumptions about individual velocity components. 
However, care needs to be taken in the interpretation when more than one velocity component is 
apparent and with molecules that show hyperfine splitting. 

For each molecular transition we compute the first three moments: the zeroth moment (M$_{0}$, or 
integrated intensity), the first moment (M$_{1}$, the intensity weighted velocity field), and 
the second moment (M$_{2}$, the intensity weighted velocity dispersion, $\sigma$). These are defined as

\[M_{0} = \int T_{A}^{*}(v)\,dv, \hspace{8mm}
M_{1} = \frac{\int T_{A}^{*}(v)v\, dv}{\int T_{A}^{*}(v)\,dv}, \hspace{8mm}
M_{2} = \sqrt{\frac{\int T_{A}^{*}(v) (v-M_{1})^{2}\,dv}{\int T_{A}^{*}(v)\,dv}},\]

\noindent where \tastar$(v)$ is the measured brightness at a given velocity $v$. 
The channel maps show the emission integrated over $\sim$ 6\,\kms\, around the listed velocity.
In all cases, the major axis of the clump is shown as the diagonal solid line on the three moment maps.
The position-velocity diagrams were made by averaging the emission over the clump's minor axis 
and shows the emission along the major axis (position) as a function of velocity. 

%mention the maps are smaller - include an outline on one of the image.

%%%%%%%%%%%%%%%%%%%%%%%%%%%%%%%%%%%%%%%%%%%%%%%%%%%%%%%%
\section{Results}
%%%%%%%%%%%%%%%%%%%%%%%%%%%%%%%%%%%%%%%%%%%%%%%%%%%%%%%%

\subsection{Dense, hot gas: morphological match to the IR extinction}\label{moments_section}

\begin{figure}
\centering
\includegraphics[width=0.45\textwidth,clip=true]{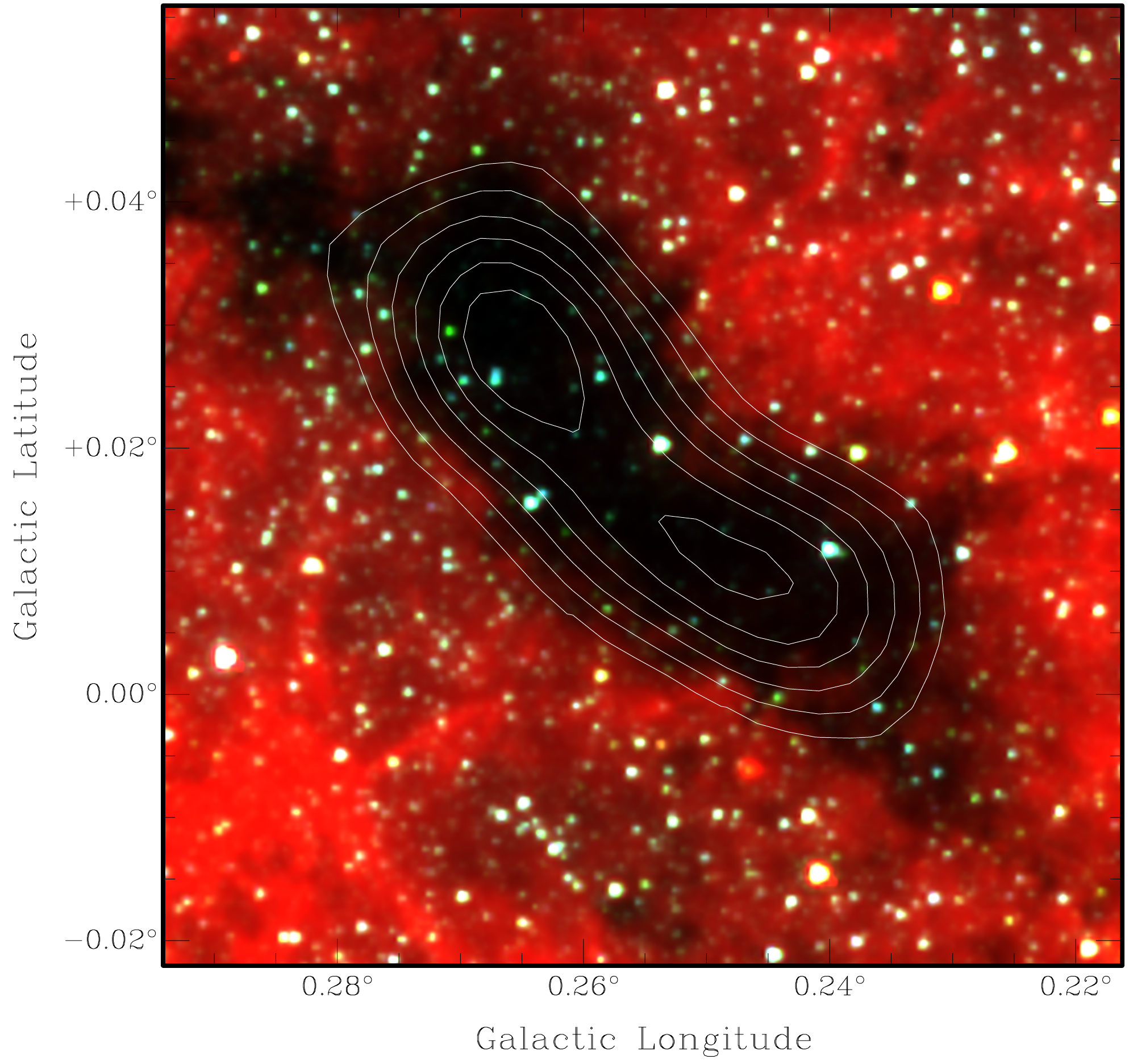}
\includegraphics[width=0.45\textwidth,clip=true]{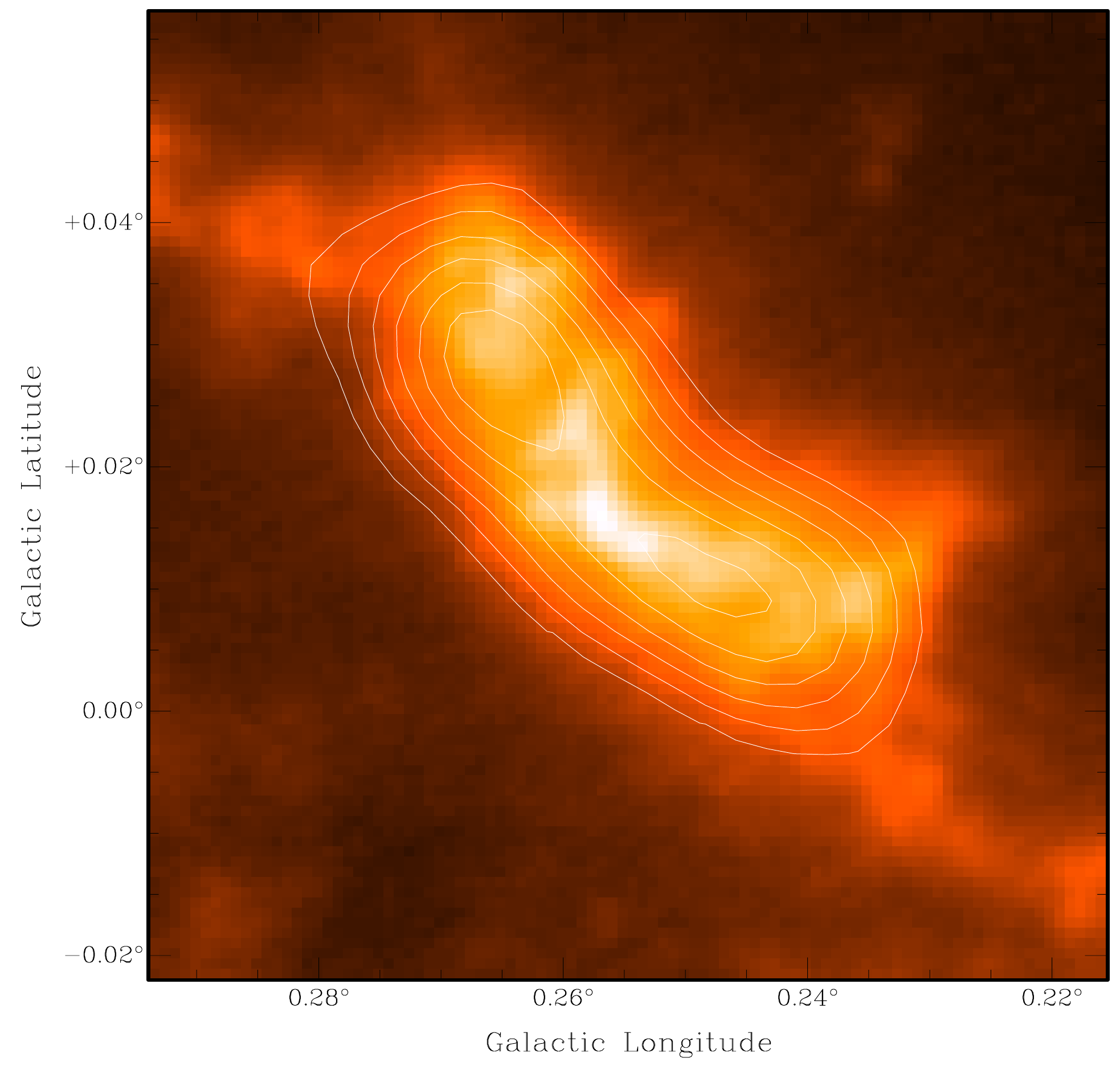}
\caption{\label{images} Continuum images toward \cloud\, ({\it{left}}: \Spitzer/IRAC three colour 
  and {\it{right}}: JCMT 450\,\um; \citealp{Longmore12}) overlaid with the \hncofz\, integrated intensity 
  image (this image was generated by integrating the emission in the range $-$25\,\kms $<$ \vlsr $<$ 
  60\,\kms; see section~\ref{spectra_section}; contours levels are from 10 to 90\% of the peak in steps of 
  10\%.  The 450\,\um\, emission ranges from 10 Jy\, beam$^{-1}$ at the clump's edge to 50 Jy\, beam$^{-1}$ at the central peak).  
  The clump is seen as an extinction feature in the IR but is a strong emitter at sub-mm/mm wavelengths. Overall the morphology 
  of the molecular line emission matches well the mid-IR extinction and sub-mm dust continuum emission.}
\end{figure}

The MALT90 data reveal extended emission toward \cloud\, from 12 of the 16 transitions observed as 
part of the survey. In all cases, emission was also detected toward \cloud\, in the higher J transitions observed 
using APEX. The four main density tracers (\nthpnt, \hcnnt, \hcopnt, and \hncnt; (1--0), (3--2) and (4--3)) show the 
brightest emission. Also detected were three isotopologues (\htcop, \hntc, and \tcs) and 
extended emission from a number of tracers of complex 
chemistry and/or shocks (\chtcn, \cch, \hncofz, \hctn, \siont\, (2--1) and (5--4)). Table~\ref{malt90-lines} lists, for each of 
the detected molecular line transitions, the position of the peak (in Galactic Longitude, Galactic Latitude and LSR velocity) 
and the corresponding measured peak brightness temperature (\tmb).  Figure~\ref{images} shows the integrated intensity 
image for \hncofz\, overlaid on the mid-IR extinction and sub-mm dust continuum emission toward \cloud. 

\begin{sidewaysfigure}
\centering
\hspace{-0.3cm}
\includegraphics[angle=0,width=0.245\textwidth,clip=true,trim=13mm 70mm 20mm 65mm]{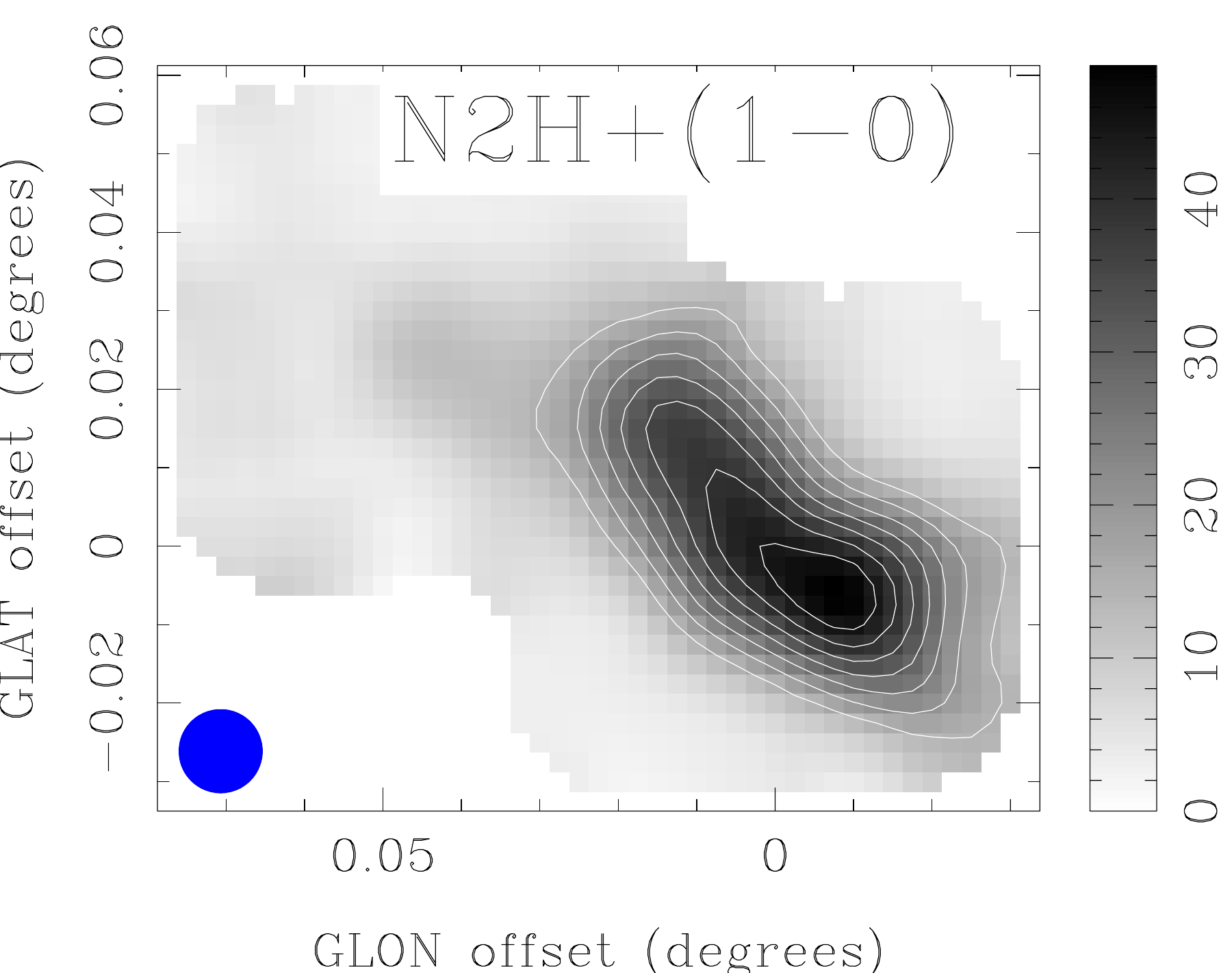}
\includegraphics[angle=0,width=0.245\textwidth,clip=true,trim=13mm 70mm 20mm 65mm]{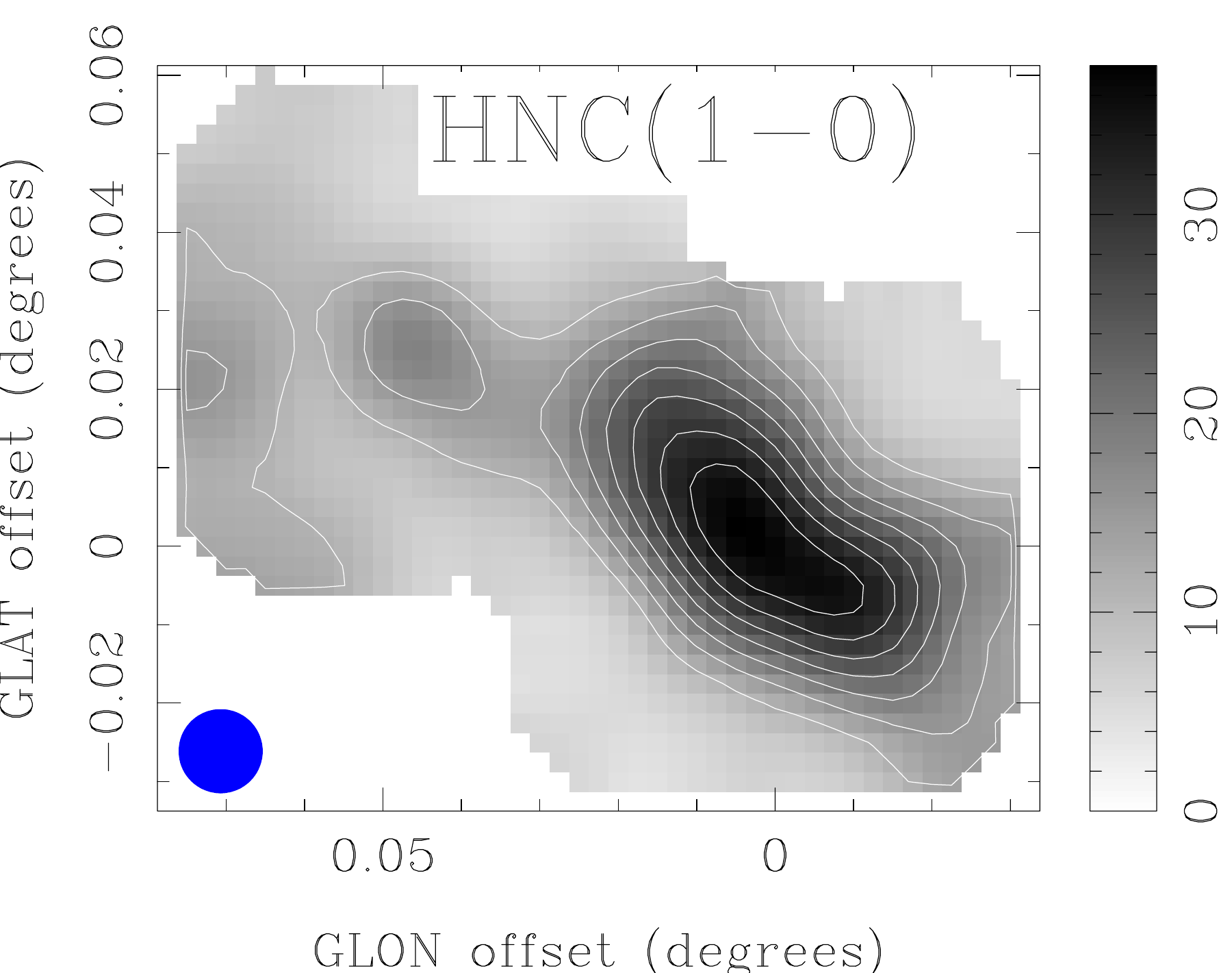}
\includegraphics[angle=0,width=0.245\textwidth,clip=true,trim=13mm 70mm 20mm 65mm]{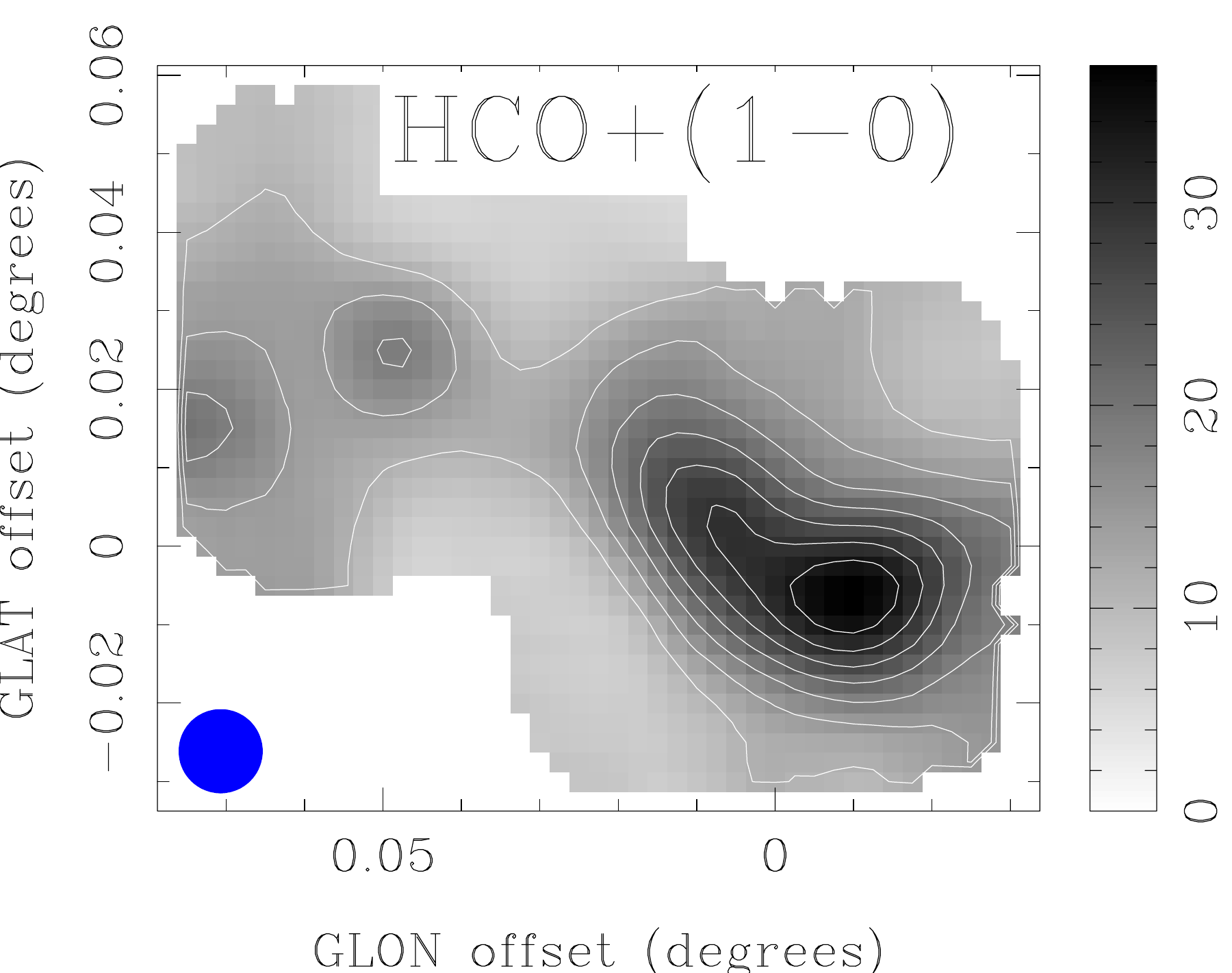}
\includegraphics[angle=0,width=0.245\textwidth,clip=true,trim=13mm 70mm 20mm 65mm]{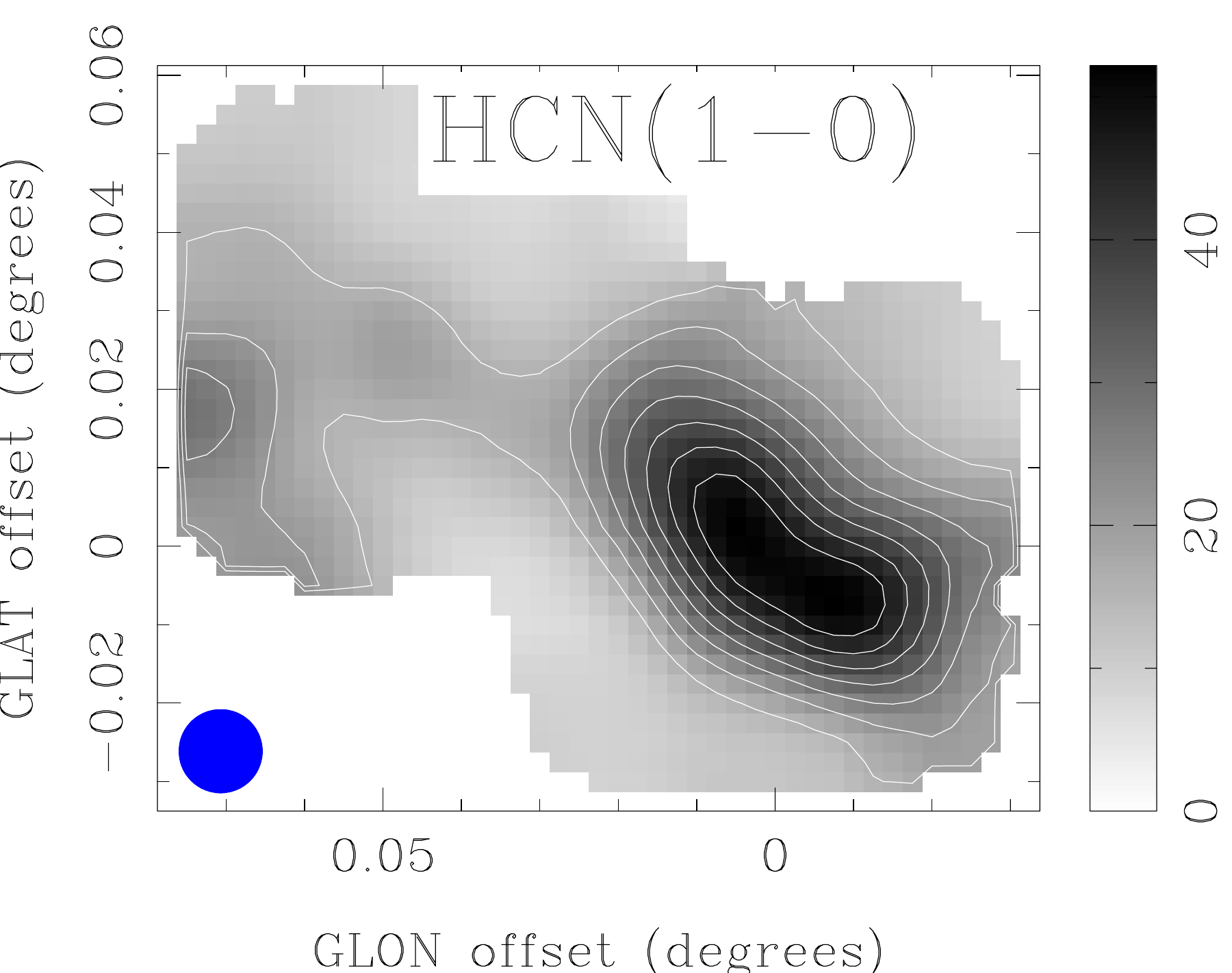}\\
\hspace{-0.3cm}
\includegraphics[angle=0,width=0.245\textwidth,clip=true,trim=13mm 70mm 20mm 65mm]{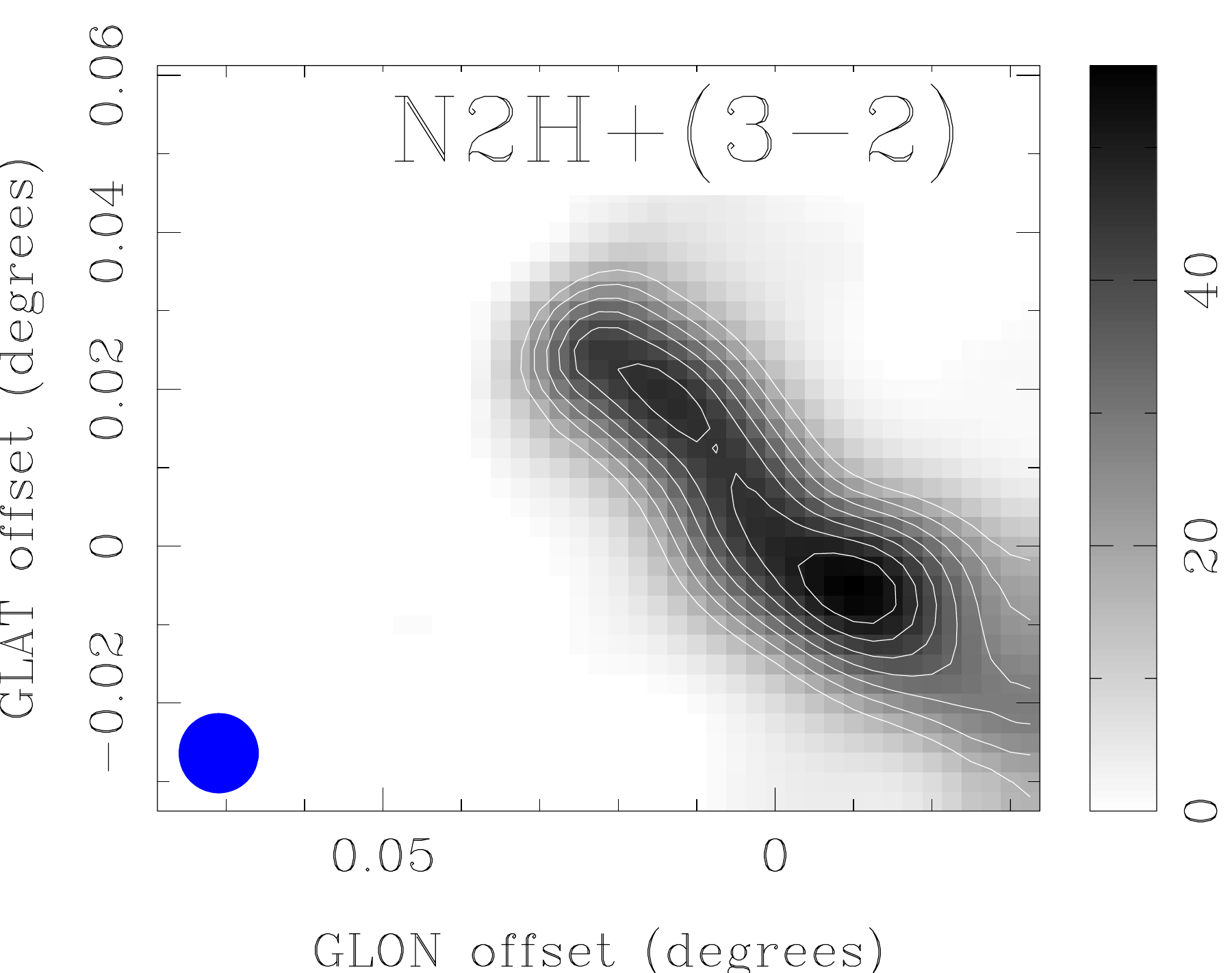}
\includegraphics[angle=0,width=0.245\textwidth,clip=true,trim=13mm 70mm 20mm 65mm]{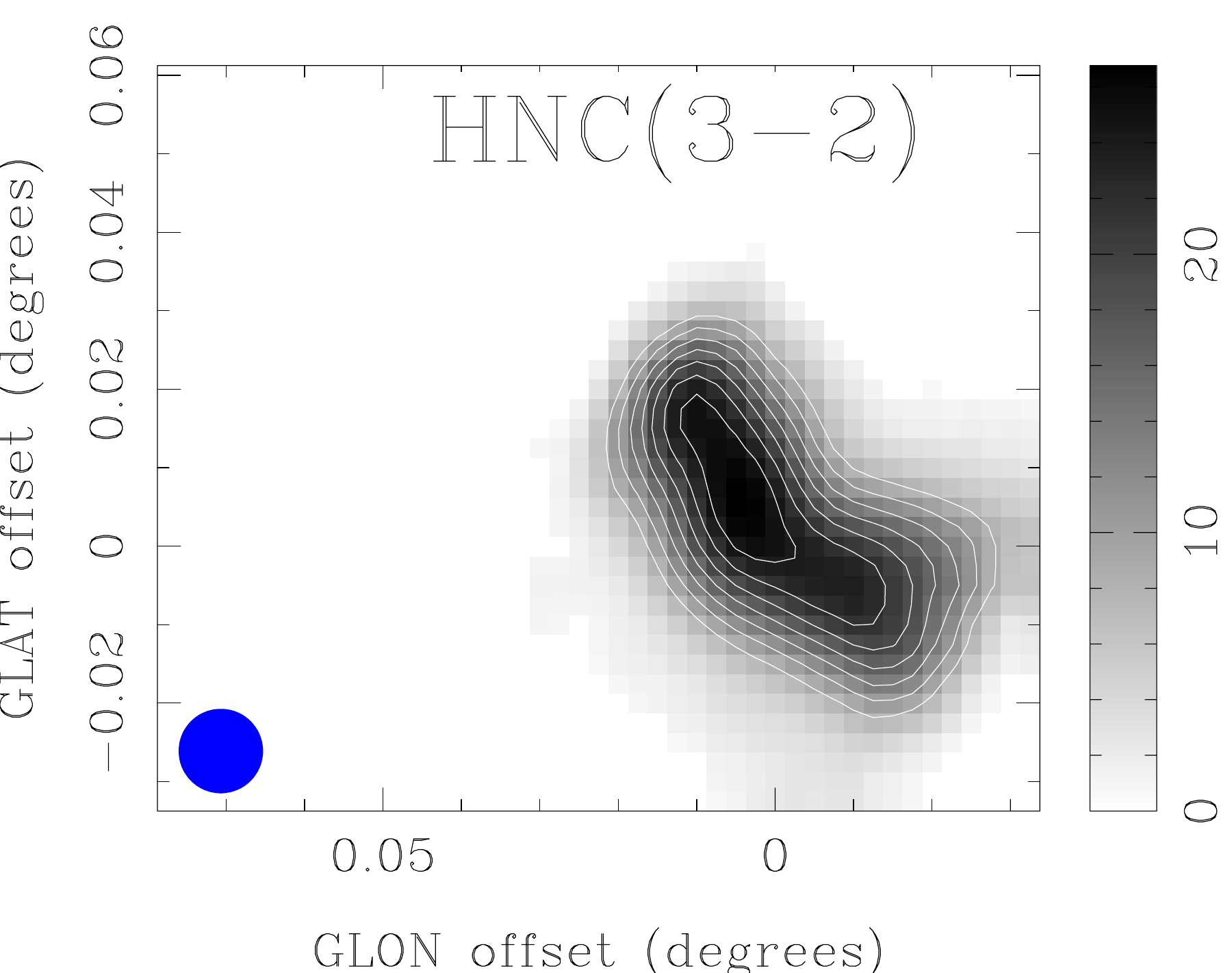}
\includegraphics[angle=0,width=0.245\textwidth,clip=true,trim=13mm 70mm 20mm 65mm]{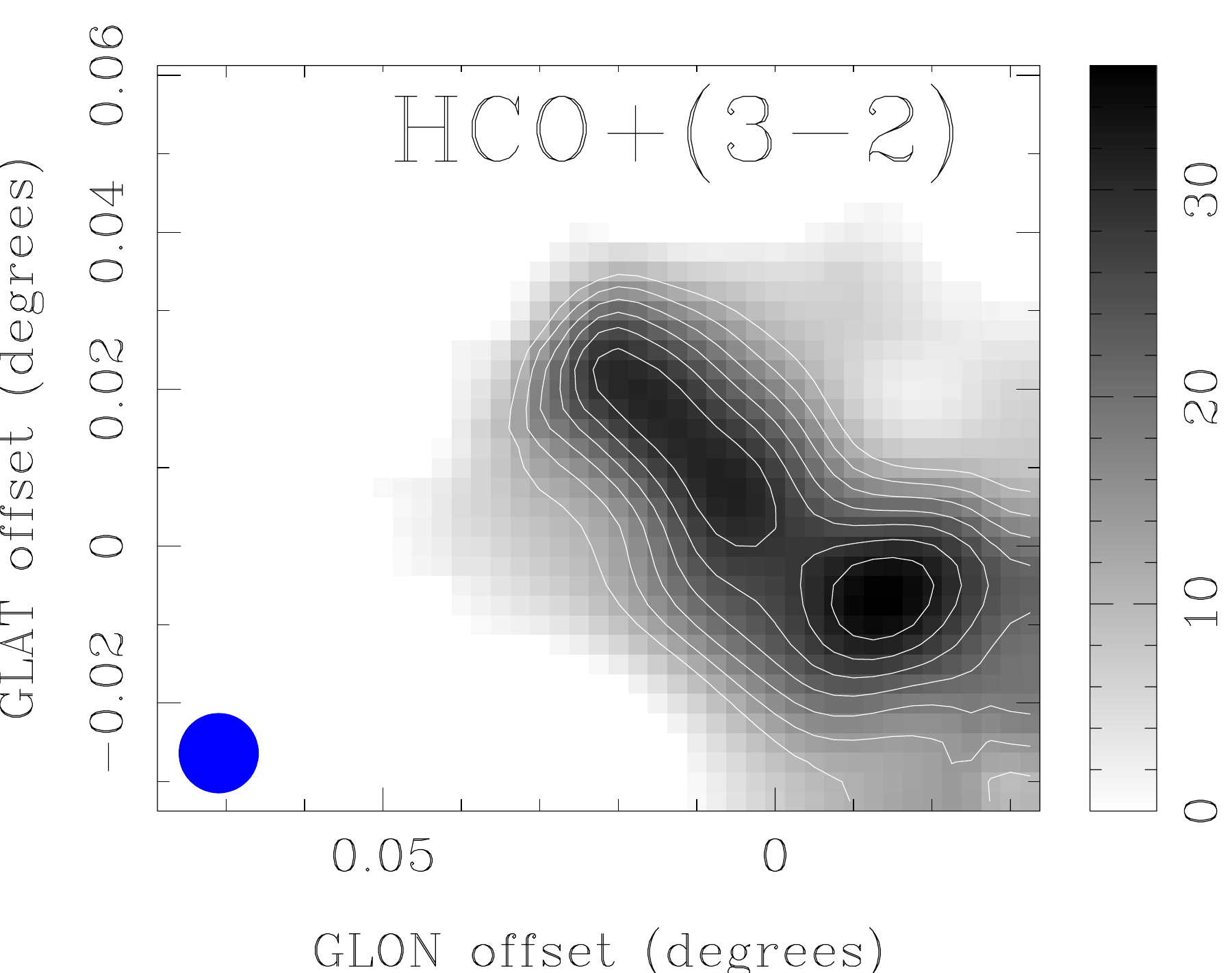}
\includegraphics[angle=0,width=0.245\textwidth,clip=true,trim=13mm 70mm 20mm 65mm]{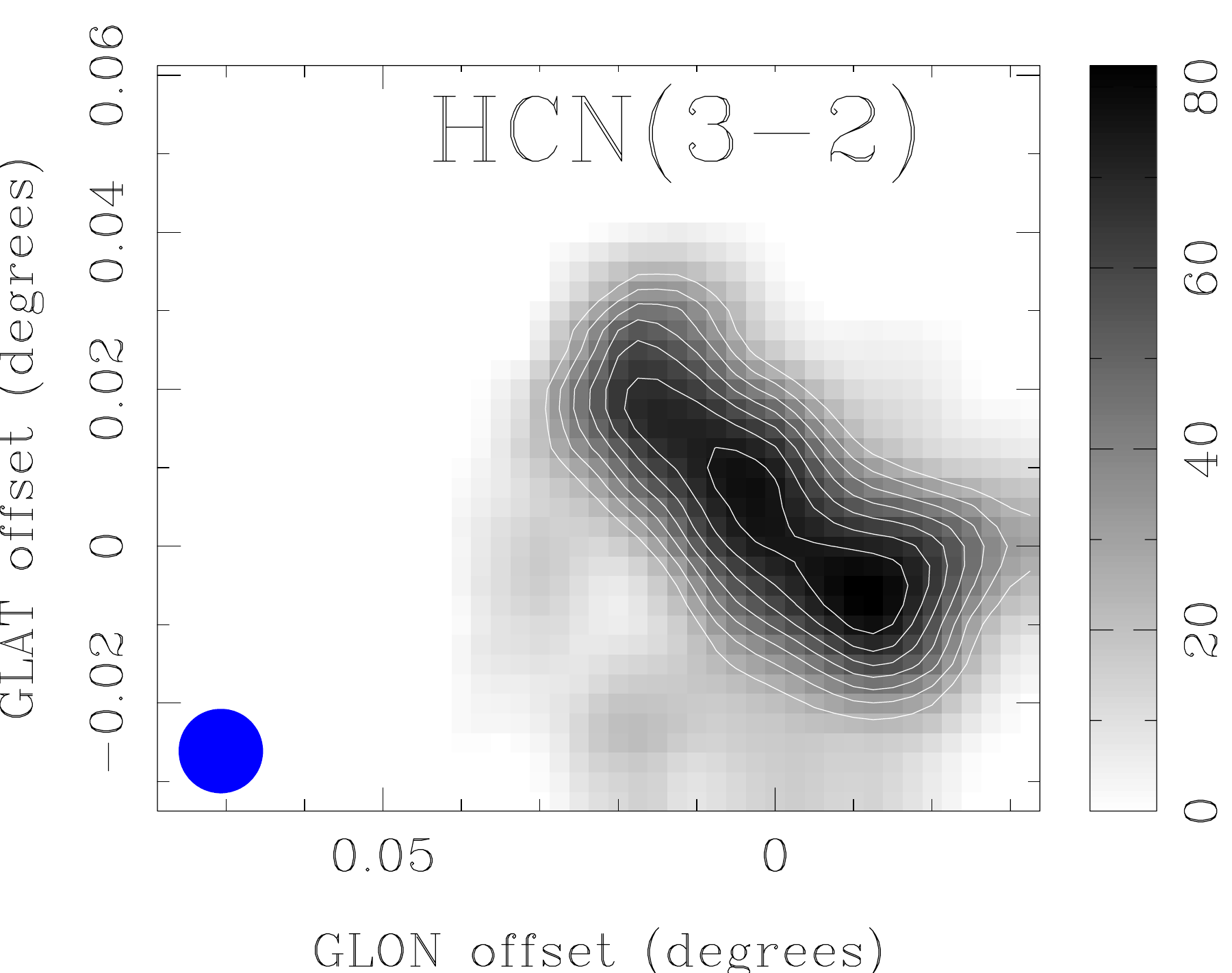}\\
\hspace{-0.3cm}
\includegraphics[angle=0,width=0.245\textwidth,clip=true,trim=13mm 70mm 20mm 65mm]{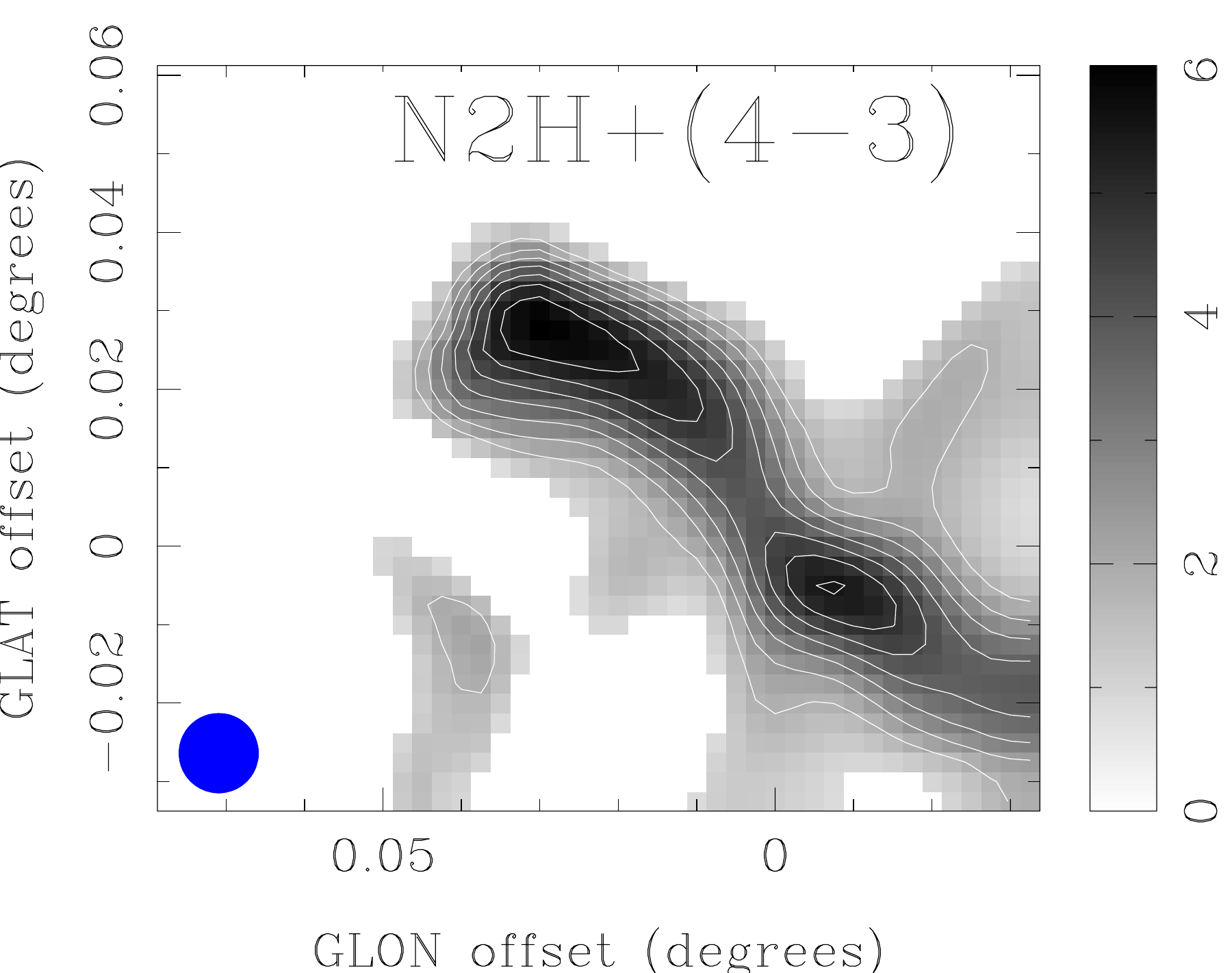}
\includegraphics[angle=0,width=0.245\textwidth,clip=true,trim=13mm 70mm 20mm 65mm]{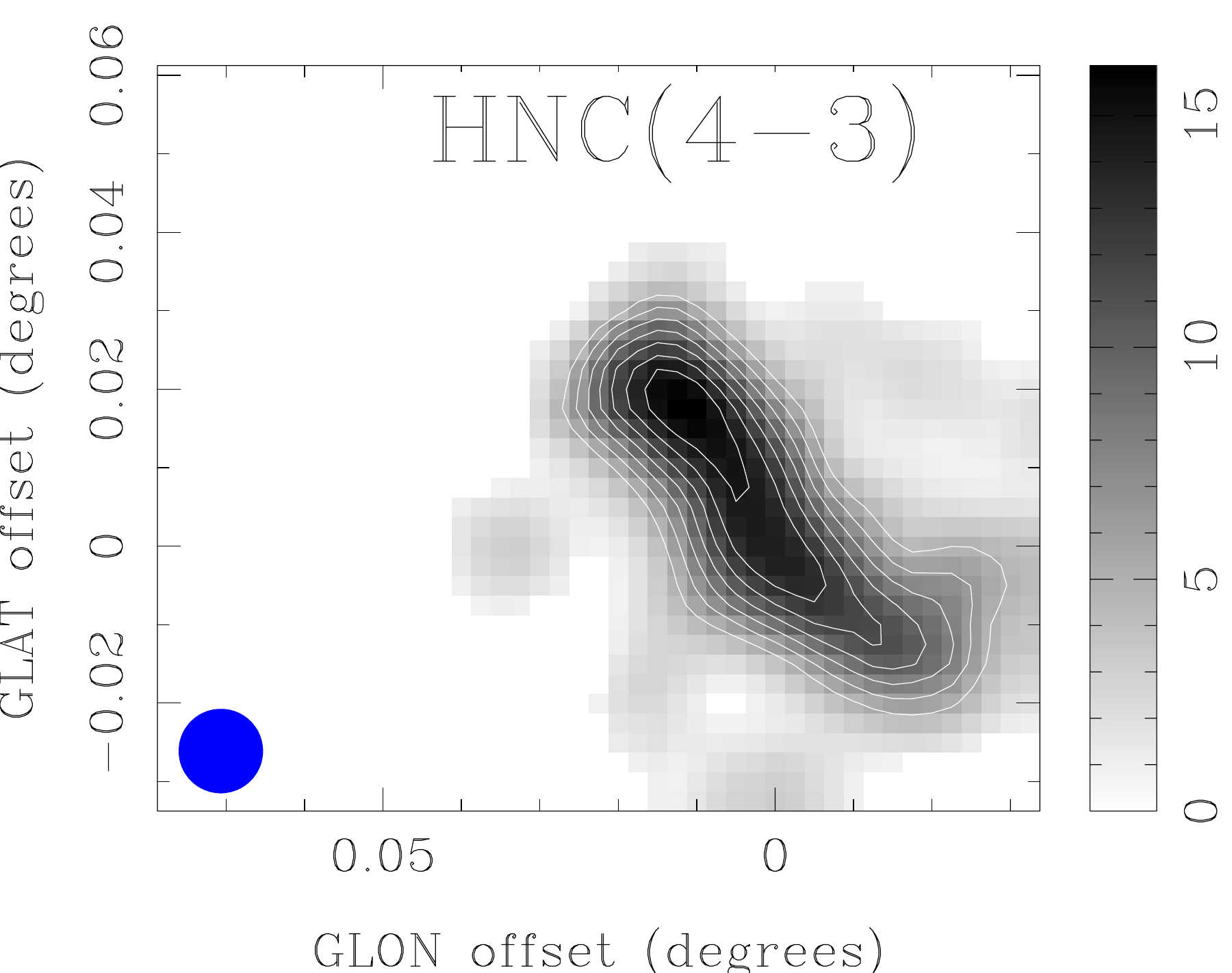}
\includegraphics[angle=0,width=0.245\textwidth,clip=true,trim=13mm 70mm 20mm 65mm]{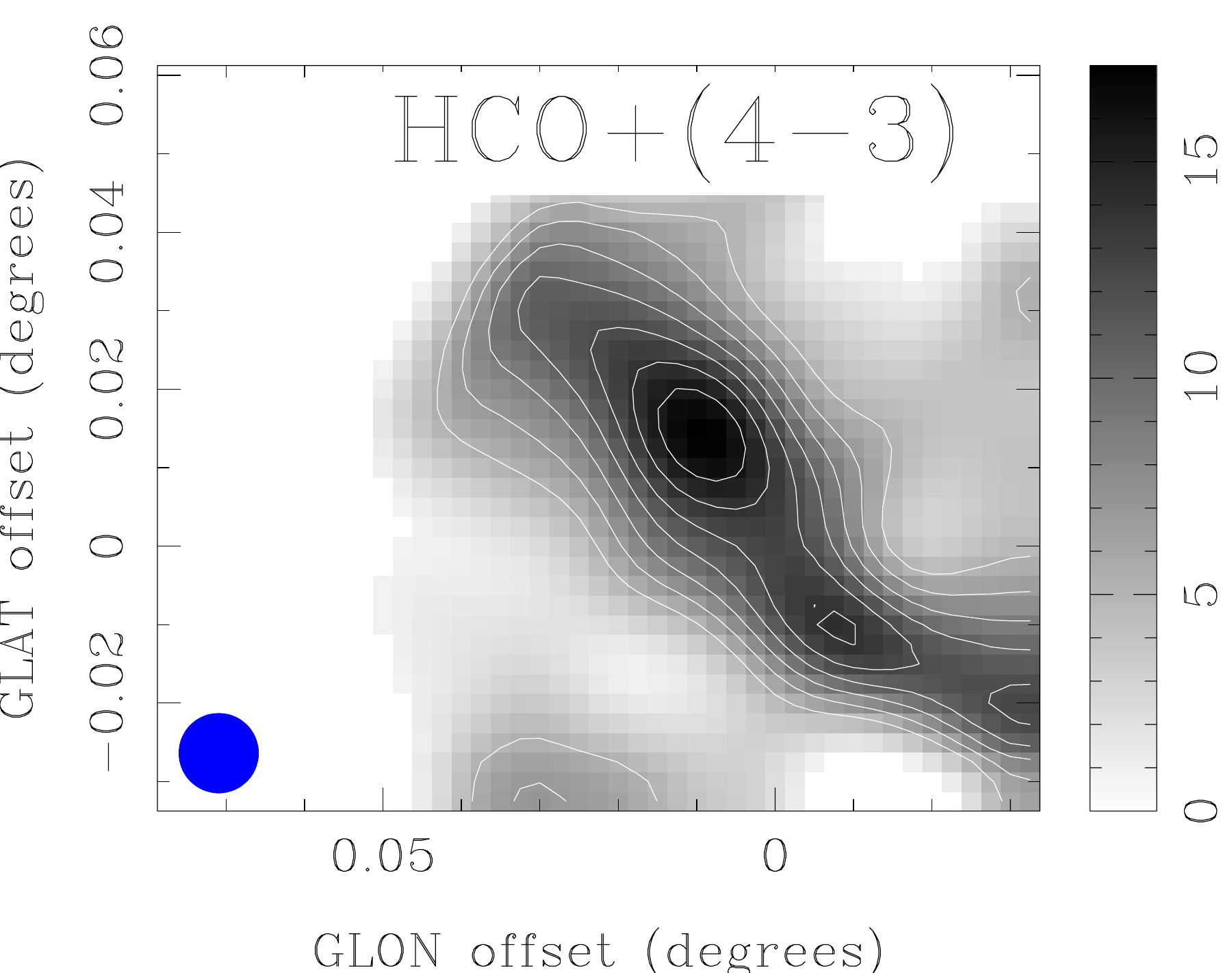}
\includegraphics[angle=0,width=0.245\textwidth,clip=true,trim=13mm 70mm 20mm 65mm]{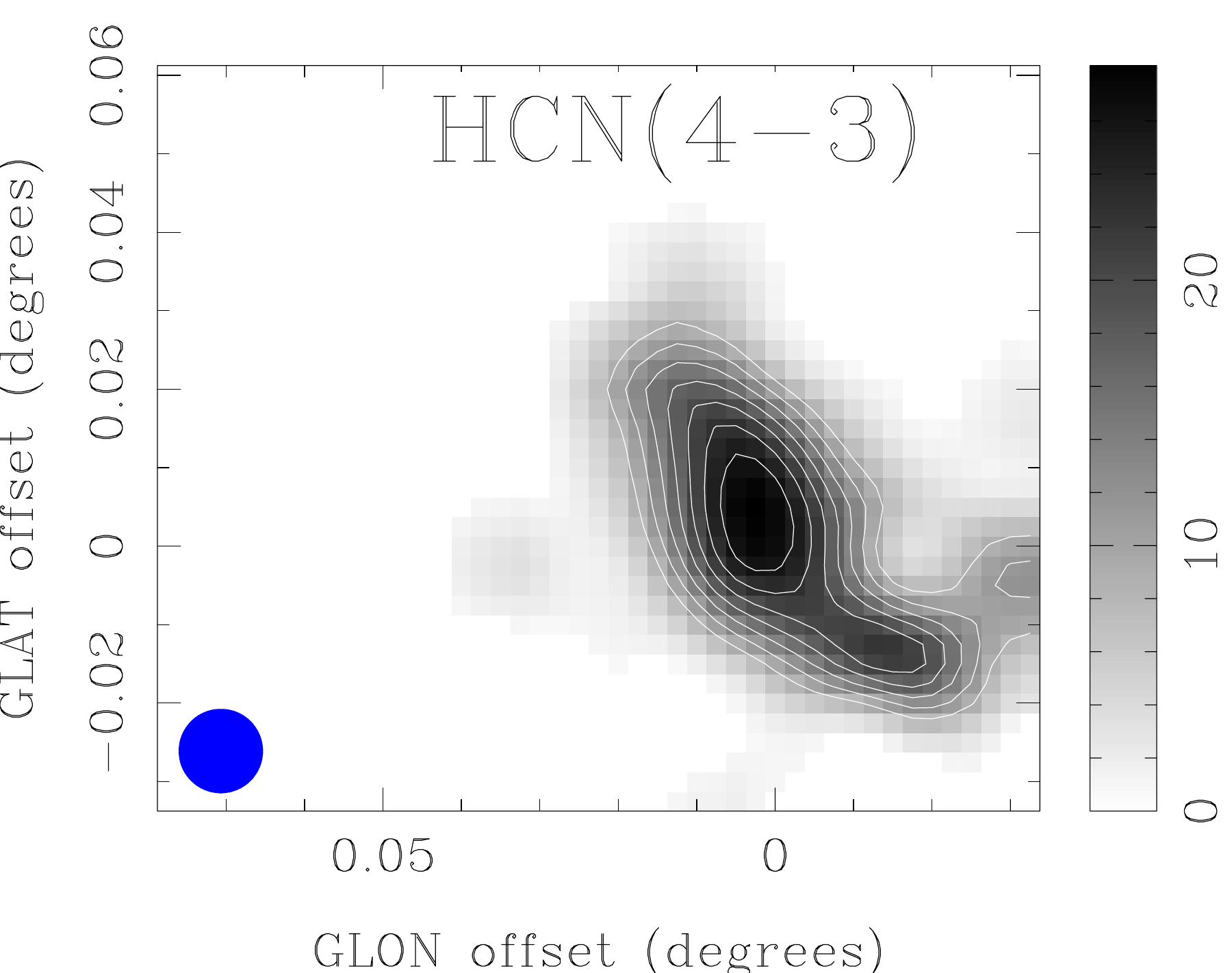}\\
\caption{\label{momentmaps-densegas} Integrated intensity images of the detected emission toward \cloud\, from the four main dense 
	gas tracers (emission is integrated over the velocity range $-$25\,\kms\, to 60\,\kms; contour levels start at 30\% of the peak 
	and increase in steps of 10\%; the adjacent colour bars show the value of the peak; units are \Kkms). The (0,0) offset position corresponds to 
	the peak in dust continuum emission. The beam size is shown in the lower left. These  tracers (\nthpnt, \hcnnt, \hcopnt, and 
	\hncnt) show the brightest emission, with their overall morphology matching well the mid-IR extinction (note that the maps of 
	the emission from the (3--2) and (4--3) transitions cover a smaller area).}
\end{sidewaysfigure}

\begin{sidewaysfigure}
\centering
\hspace{-0.3cm}
\includegraphics[angle=0,width=0.245\textwidth,clip=true,trim=13mm 70mm 20mm 65mm]{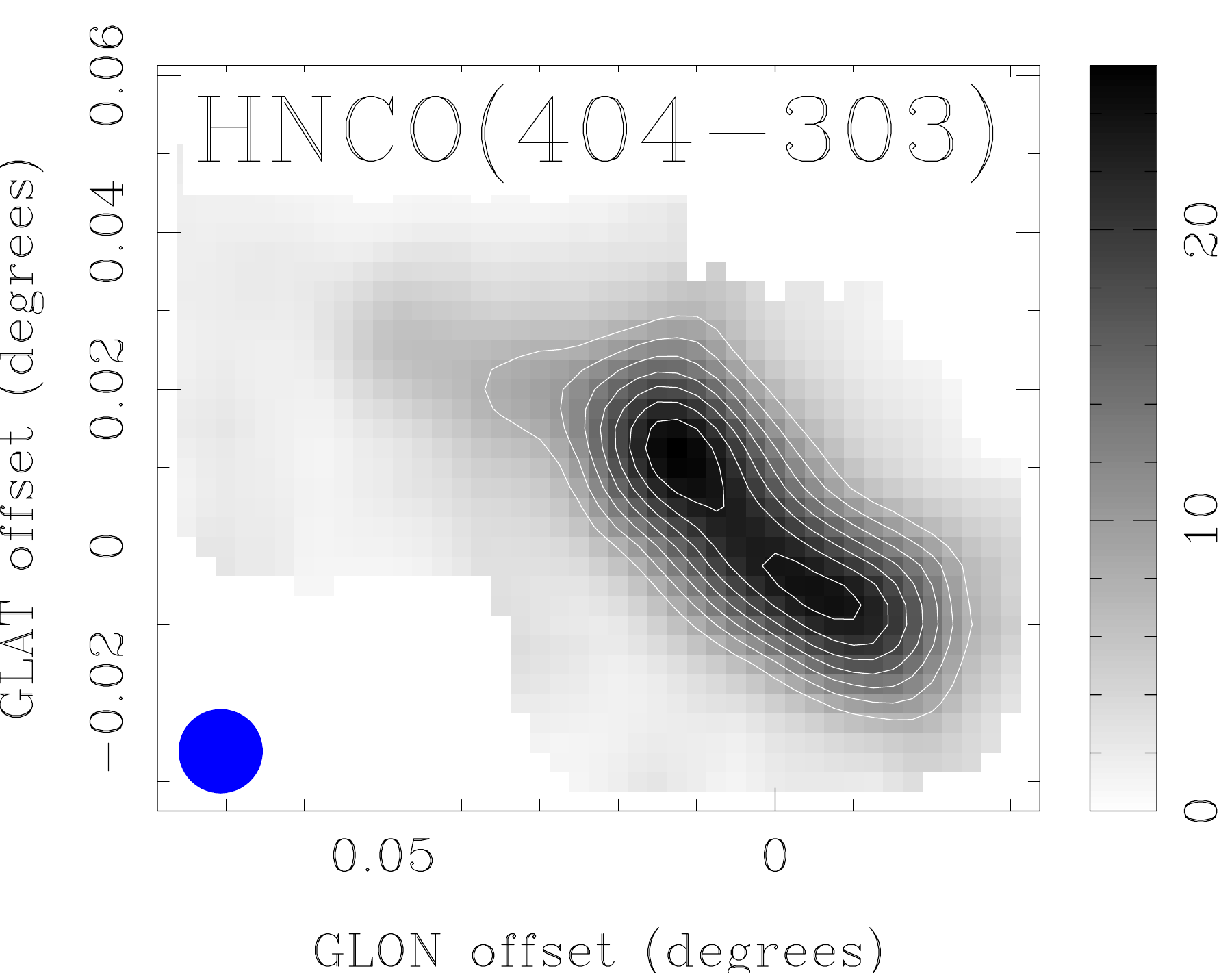}
\includegraphics[angle=0,width=0.245\textwidth,clip=true,trim=13mm 70mm 20mm 65mm]{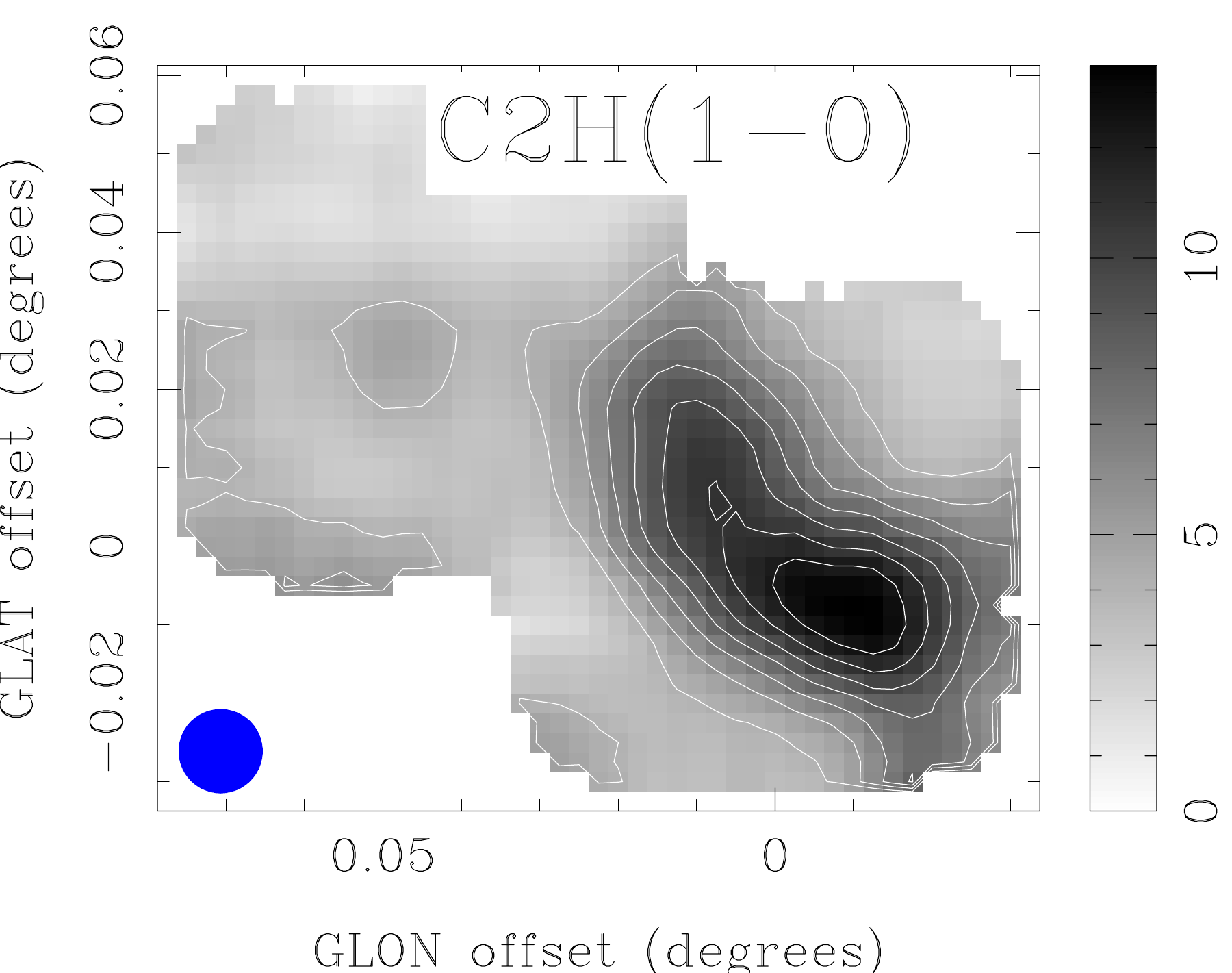}
\includegraphics[angle=0,width=0.245\textwidth,clip=true,trim=13mm 70mm 20mm 65mm]{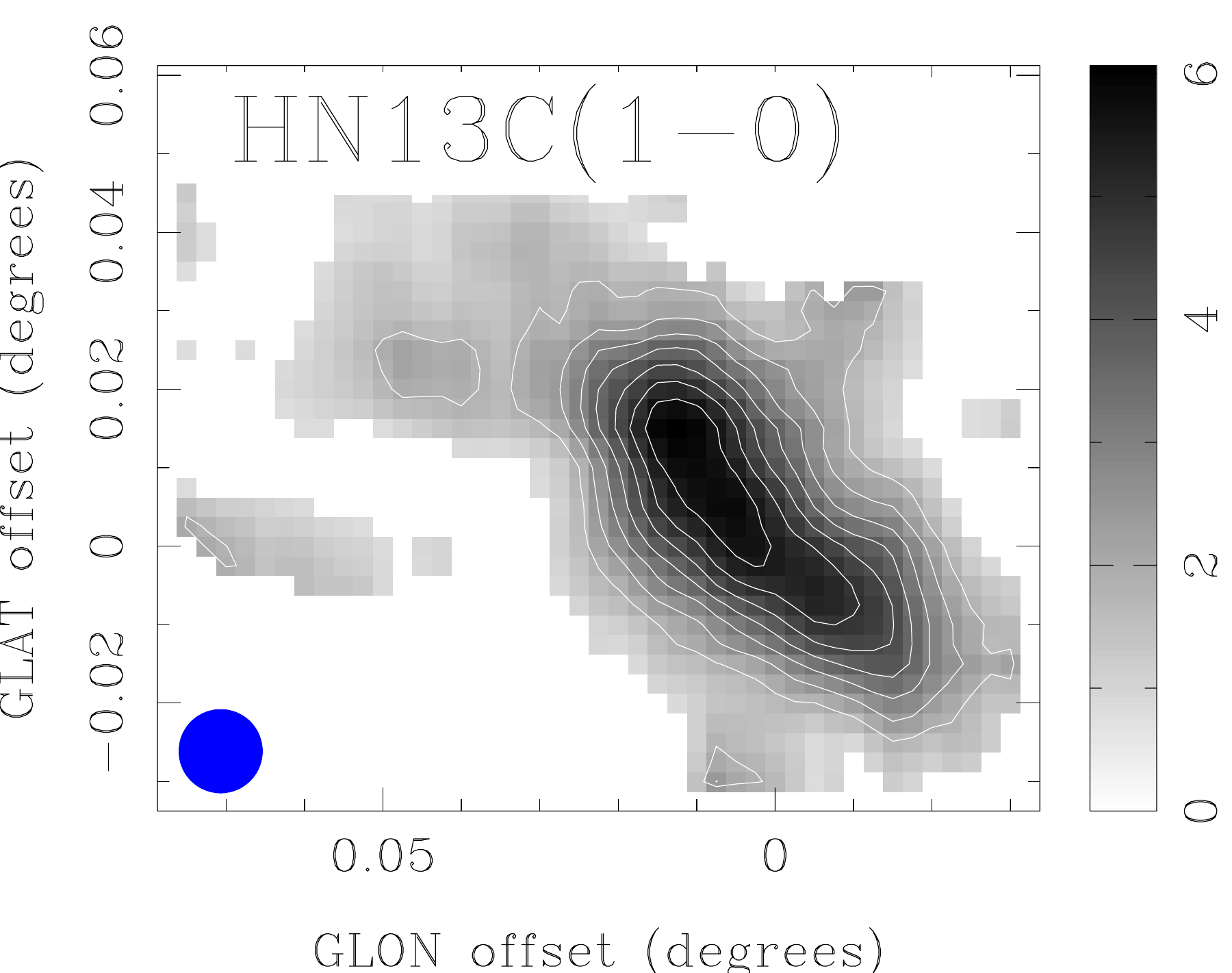}\\
\hspace{-0.3cm}
\includegraphics[angle=0,width=0.245\textwidth,clip=true,trim=13mm 70mm 20mm 65mm]{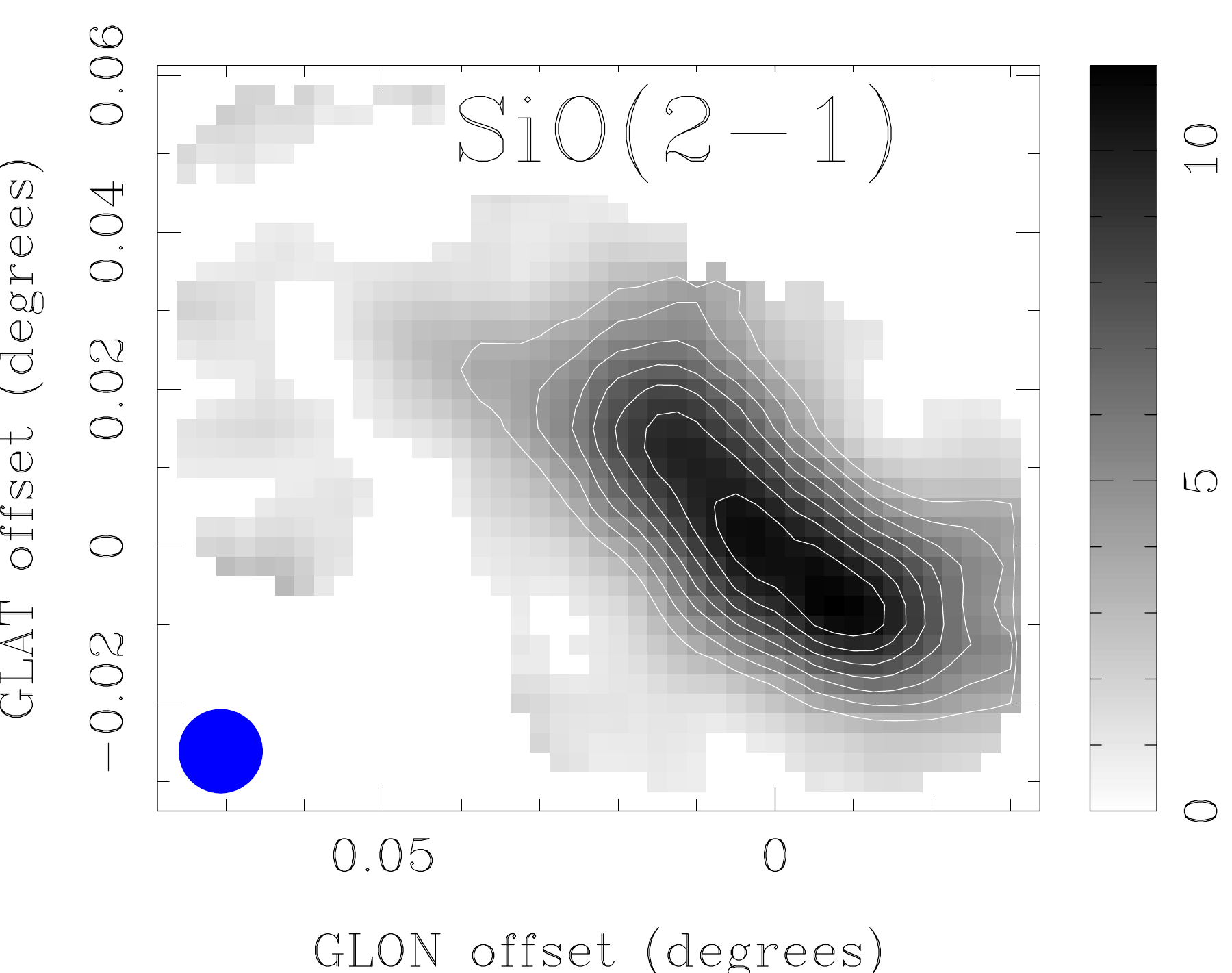}
\includegraphics[angle=0,width=0.245\textwidth,clip=true,trim=13mm 70mm 20mm 65mm]{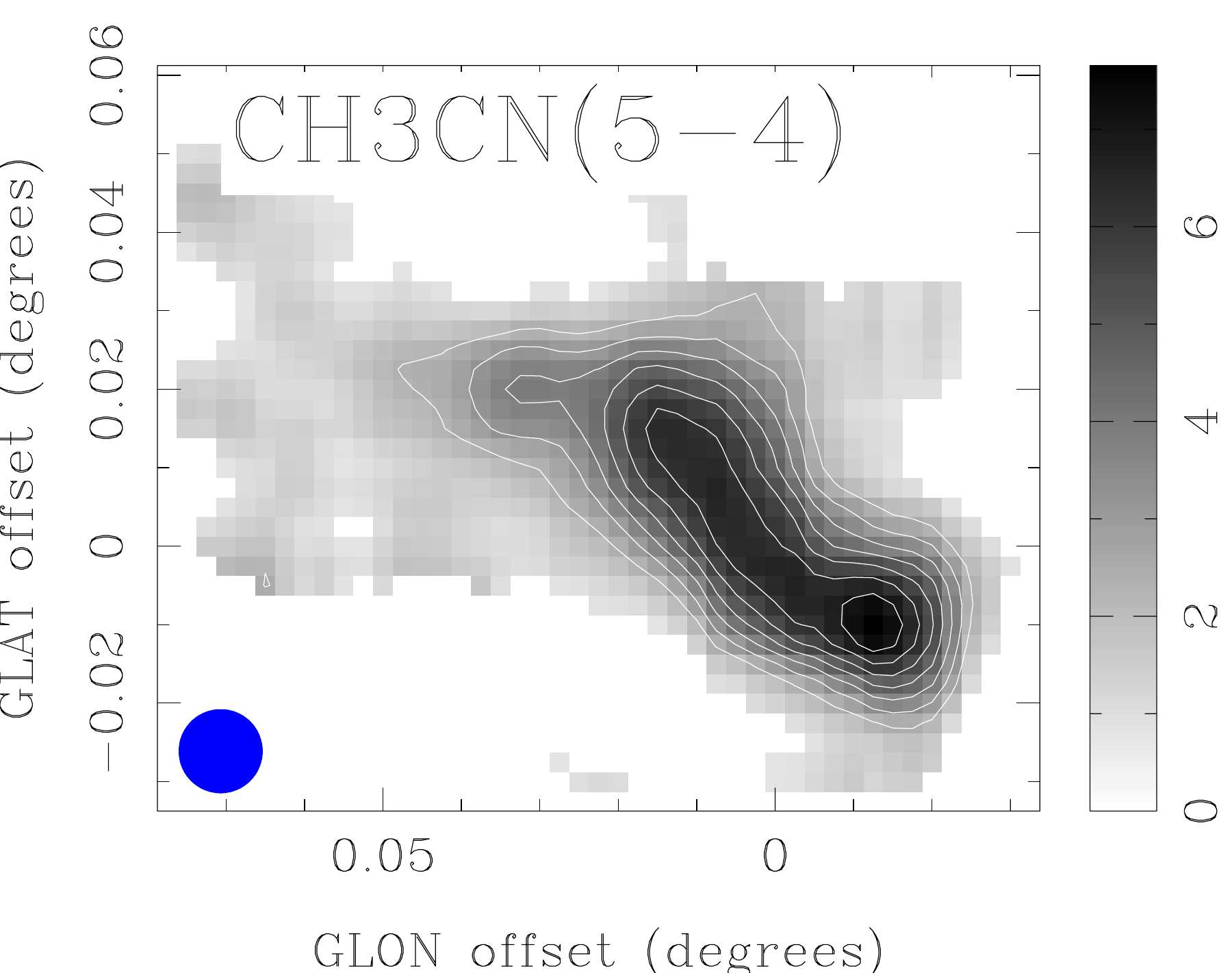}
\includegraphics[angle=0,width=0.245\textwidth,clip=true,trim=13mm 70mm 20mm 65mm]{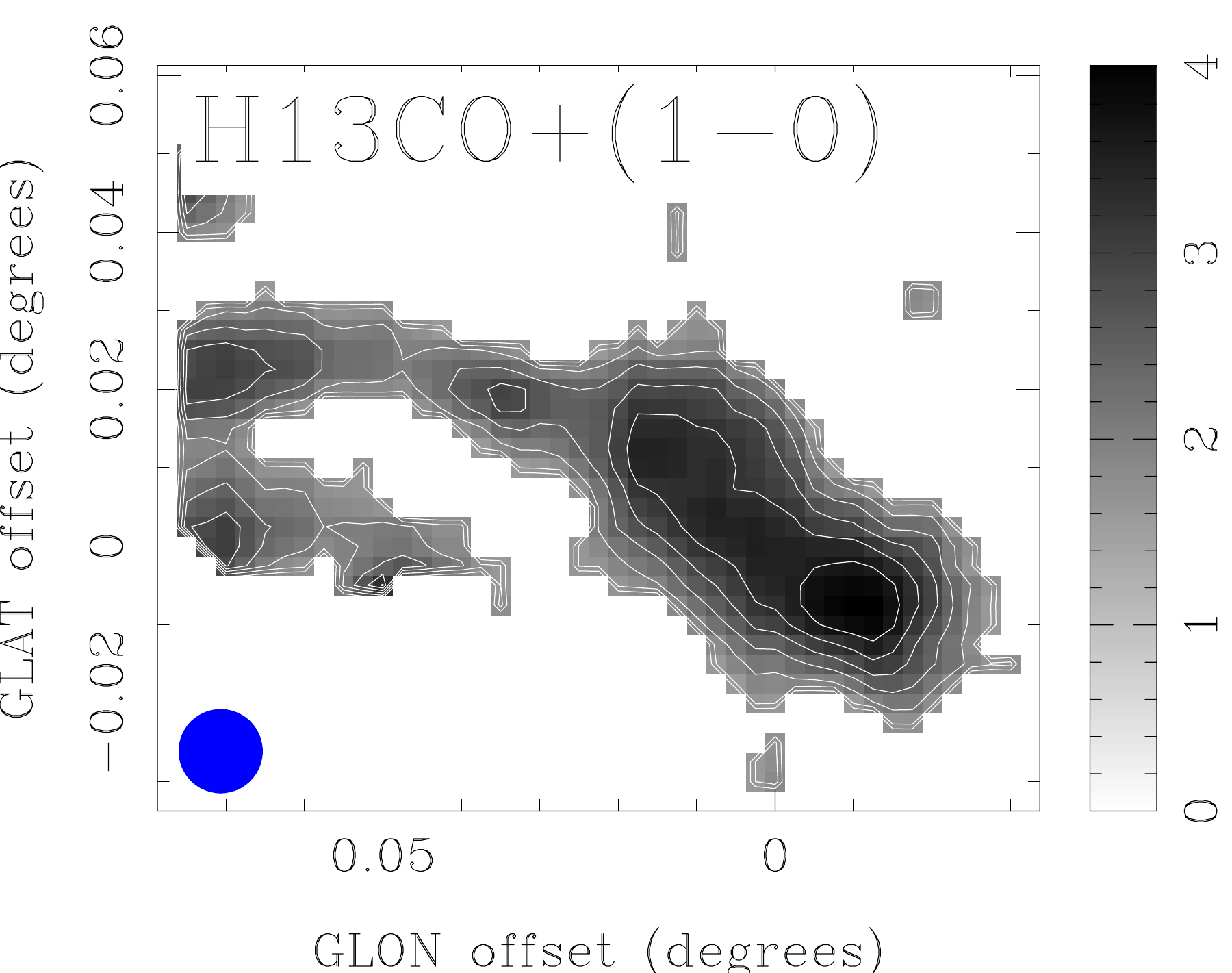}\\
\hspace{-0.3cm}
\includegraphics[angle=0,width=0.245\textwidth,clip=true,trim=13mm 70mm 20mm 65mm]{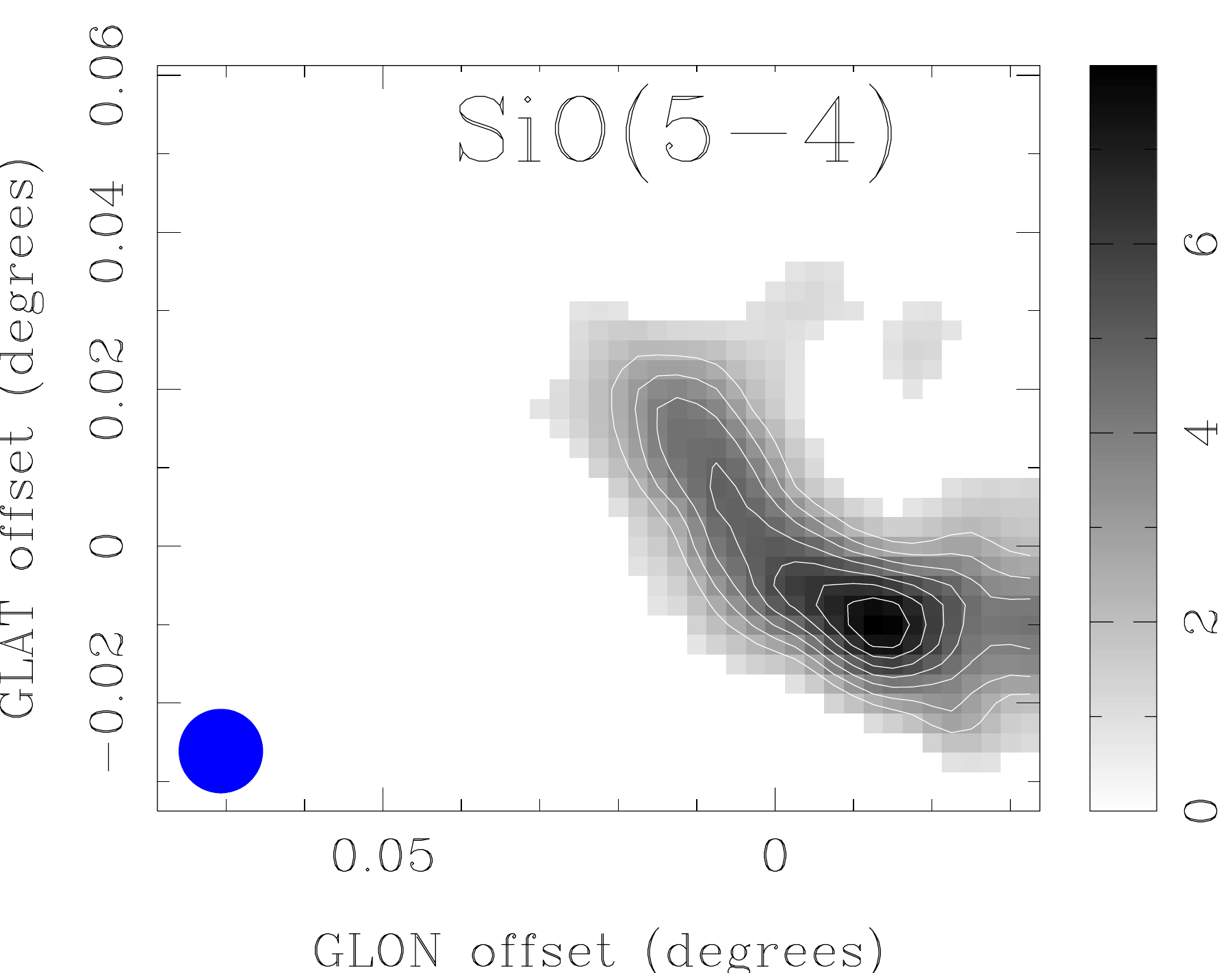}
\includegraphics[angle=0,width=0.245\textwidth,clip=true,trim=13mm 70mm 20mm 65mm]{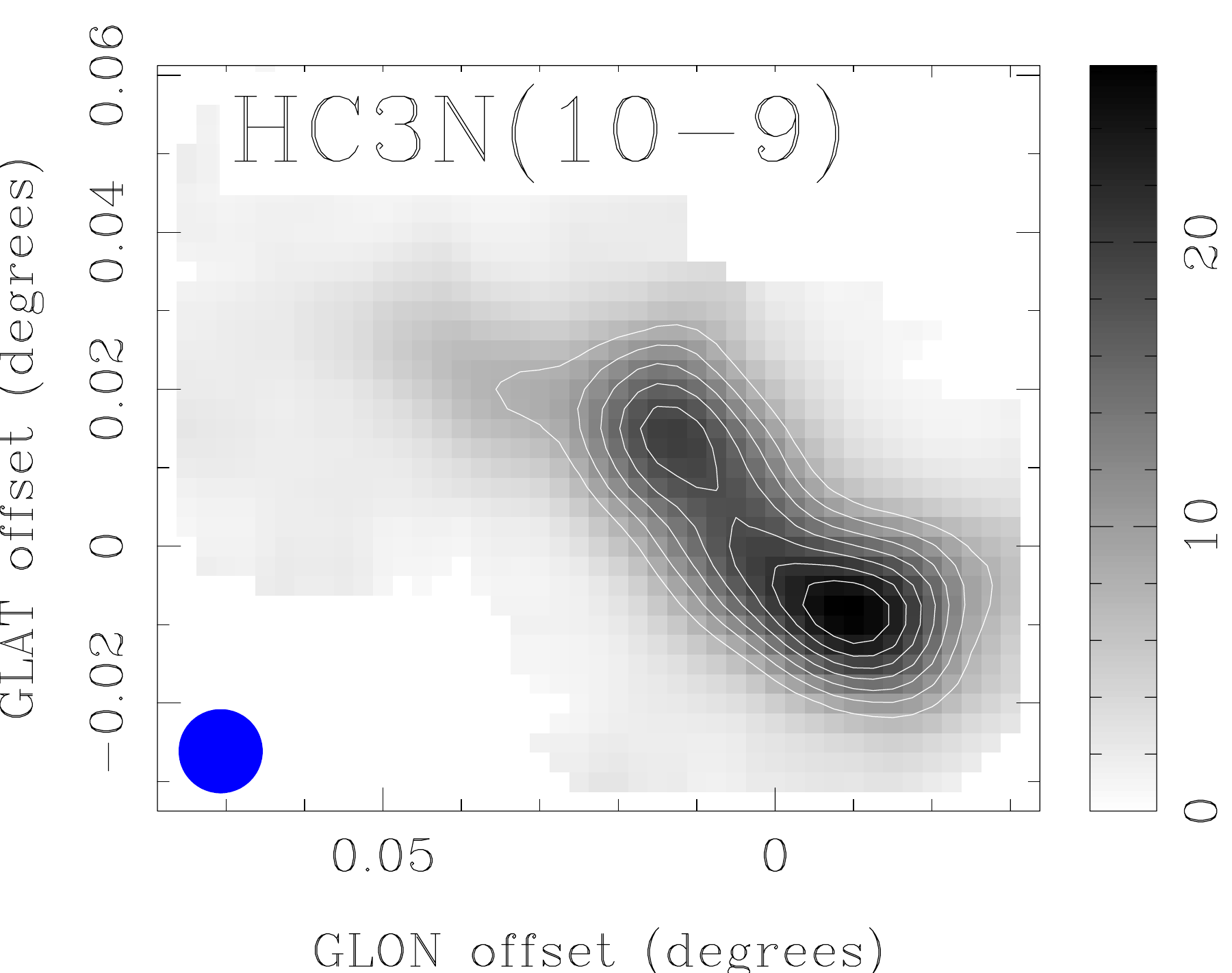}
\includegraphics[angle=0,width=0.245\textwidth,clip=true,trim=13mm 70mm 20mm 65mm]{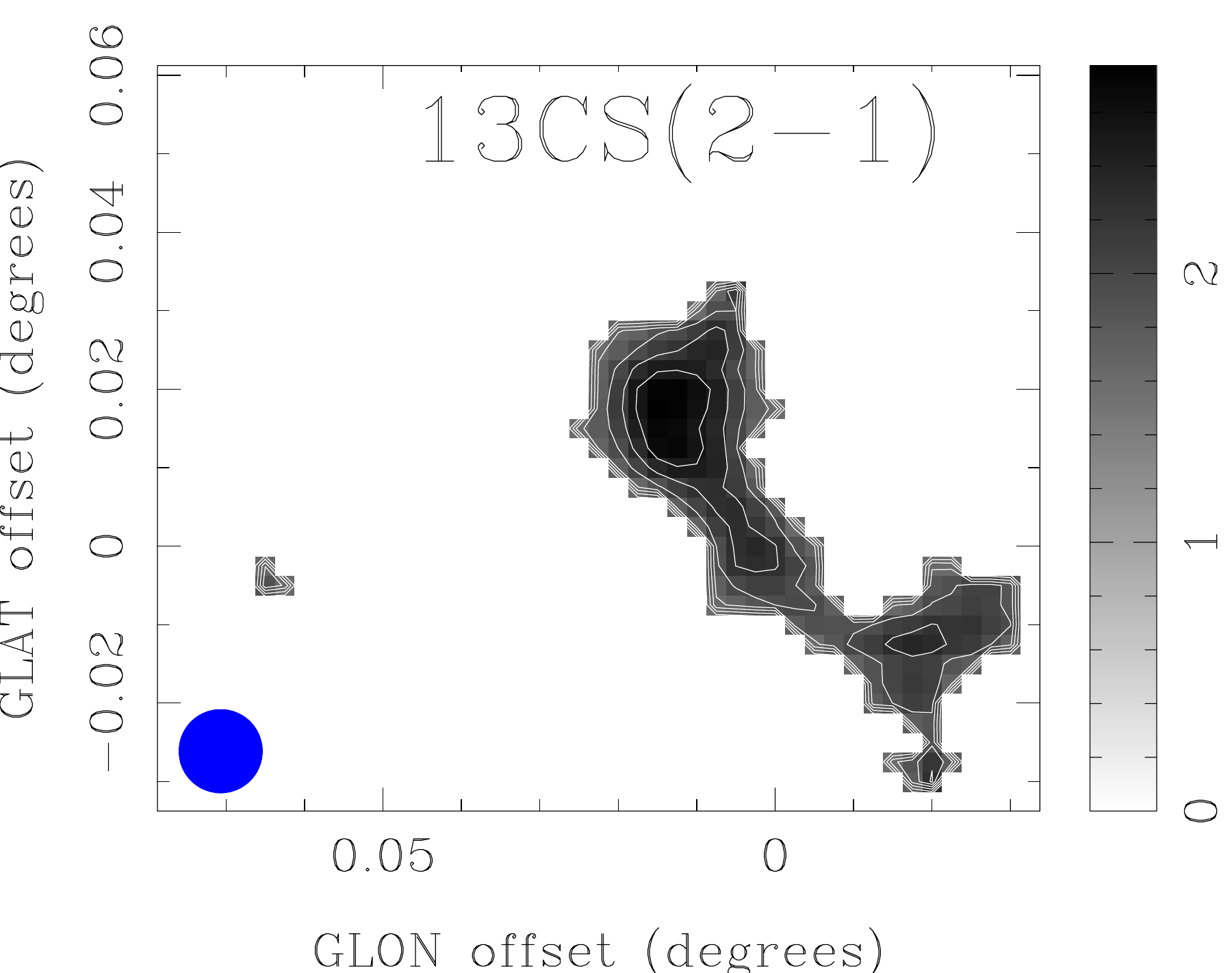}\\
\caption{\label{momentmaps-hot-shockedgas} Integrated intensity images of the detected emission toward \cloud\, from the isotopomers and hot/shocked gas 
	tracers (emission is integrated over the velocity range $-$25\,\kms\, to 60\,\kms; contour levels start at 30\% of the peak and increase in steps of 10\%; 
	the adjacent colour bars show the value of the peak; units are \Kkms). The (0,0) offset position corresponds to the peak in dust continuum emission. The beam size is 
	shown in the lower left. Similar to the main dense gas tracers, the morphology of the emission from the isotopomers (\htcopnt, \hntcnt, and \tcsnt) and 
	hot/shocked gas tracers (\cchnt, \siont, \hncofznt, \hctnnt, and \chtcnnt) matches well the the mid-IR extinction  (note that the maps of the emission from 
	SiO (5--4) covers a smaller area).}
\end{sidewaysfigure}

Figures~\ref{momentmaps-densegas} and \ref{momentmaps-hot-shockedgas} show the integrated intensity 
images for each of the detected molecular transitions. Similar to the dense gas tracers, the overall morphology 
of the emission from the isotopologues also match well the emission from more 
complex molecules that typically trace hot/shocked gas. 
The fact that the morphology of the emission from the dense gas tracers and complex molecules 
matches well the morphology of the dust continuum emission and IR extinction demonstrates that the molecular 
line emission originates from the clump rather than from the surrounding large-scale GC environment.

Despite the overall correspondence, there are clear differences in the location of the peaks and relative 
extents of the integrated emission from the various molecules and transitions. Transitions with higher critical densities 
and excitation temperatures are less extended compared to those with lower critical 
densities and excitation temperatures (i.e. compare HCN (4--3) and (1--0) in Fig.~\ref{momentmaps-densegas}).
 Moreover, while the integrated intensities for most transitions show a single peak toward the 
lower part of the clump, the integrated intensities for \tcs, \hntc, and \hncnt\,(3--2) peak toward the clump's centre. Moreover,
the integrated intensities
from \hncofz, \hctn, \hcopnt\,(3--2), and \hcnnt\,(3--2) and (4--3) show two peaks. These 
differences  arise from variations in the excitation, optical depth, and chemistry within the clump.

\subsection{Gas density and temperature}

While the critical densities of the observed transitions are all high (10$^{5}$--10$^{7}$\,\cmc), the excitation energies 
of the `hot core' molecules E$_{u}$/$k$ of 10--20\,K are much higher compared to 
E$_{u}$/$k$ of $\sim$ 4\,K for the dense gas tracers (see Table~\ref{malt90-lines}). The presence of the 
hot core molecules usually indicates the location of gas that has either a high temperature (i.e. $>$ 100\,K) 
and/or high density (i.e. the gas density is well above the critical density, $n_{crit}$): these conditions are 
needed for both their formation and excitation. For the observed molecules, $n_{crit}$ is $\sim$ 10$^{5}$--10$^{7}$\,\cmc\, 
(Table~\ref{malt90-lines}).  Toward \cloud\, the average volume density is $\sim$ 2 $\times$10$^{4}$\,\cmc\, 
(calculated using a radius equal to the geometric mean of the major and minor axes and that assuming the same extent along the line of sight).
If the gas is not highly clumped, this volume density
falls well below the critical densities for these molecules, and thus, the excitation would be sub-thermal  (i.e. n$<$n$_{crit}$) .   In other
words, if the gas is approximately uniformly distributed, the excitation temperatures (T$_{ex}$) will be significantly lower
than the gas kinetic temperatures (T$_{K}$).  However, 
because this volume density is averaged over parsec size-scales, the local volume density may be significantly 
higher if the gas is clumpy and highly structured on small scales.  Thus, small regions with higher volume densities may
lie at or above the critical density   (i.e. n$\sim$n$_{crit}$) , and in these regions T$_{ex}$ will approach T$_{K}$.

\begin{figure}
\centering
\includegraphics[width=0.55\textwidth,clip=true,trim=1mm 1mm 1mm 1mm]{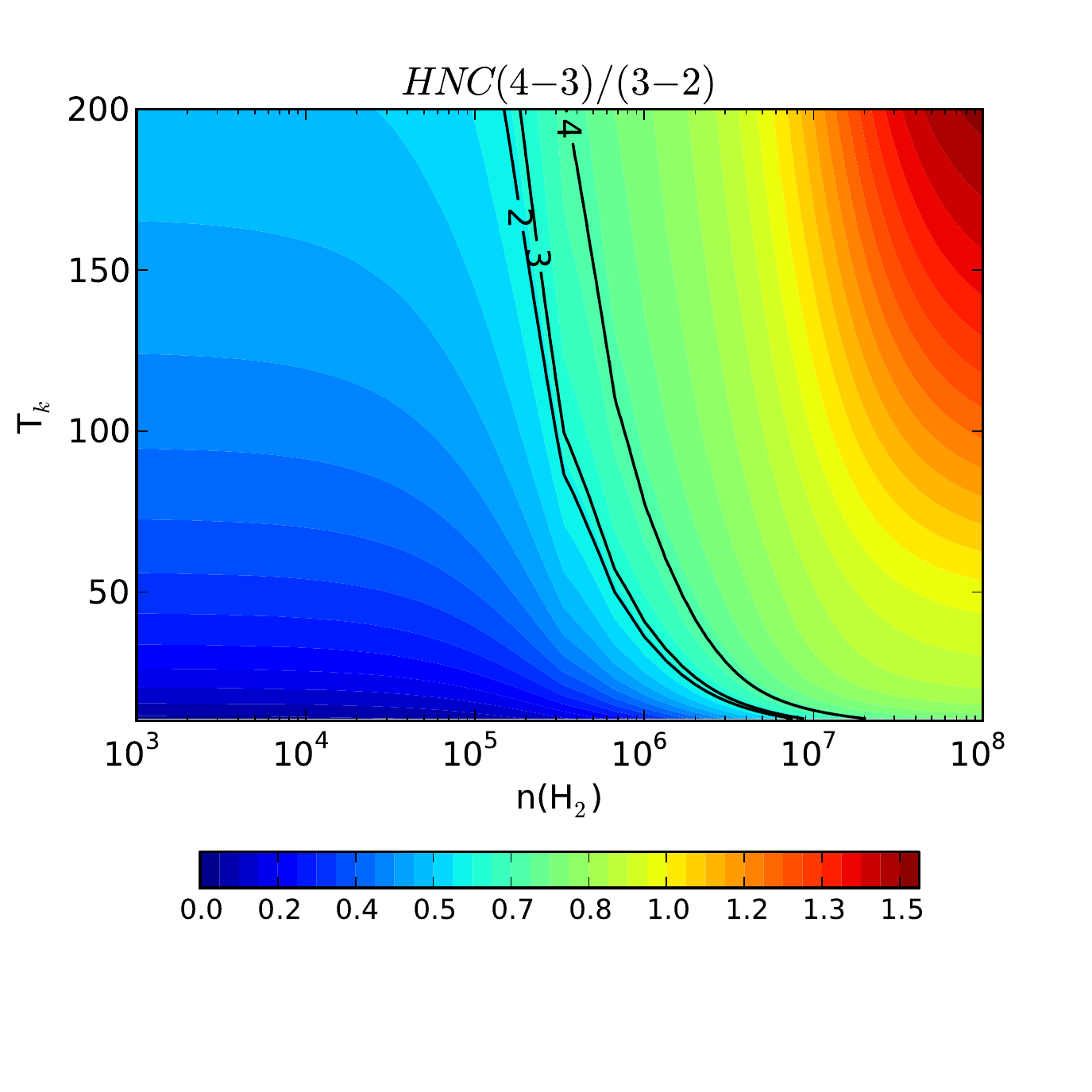}\\
\caption{\label{radex}  The predicted HNC line ratios as a function of volume density and kinetic temperature.  
 	This plot was generated using RADEX, a radiative transfer model under the assumption of non-LTE conditions.  
	The solid black lines mark the values of the ratio calculated from the data toward three of the 5 positions within \cloud\, (P2, P3, and P4). While we
	are unable to constrain the kinetic temperature, we measure gas densities of $\sim$10$^{6}$\,\cmc.}
\end{figure}

Using the combination of the MALT90 and APEX datasets, we can investigate the gas density and temperature 
within \cloud\, by means of excitation and radiative transfer modelling.  Figure~\ref{radex} plots the expected line ratios for HNC 
as a function of volume density and kinetic temperature. This plot was generated using RADEX,  a non-LTE radiative 
transfer model \citep{vanderTak07} assuming a spherical cloud and a uniform density and temperature. The input 
column density of HNC (10$^{15}$\,\cms) was estimated from the dust-derived column density measured toward 
\cloud \, \citep{Longmore12} and assuming an abundance of HNC relative to \hh\, of 3$\times 10^{-9}$ \citep{Schilke92}.

For transitions that are both optically thin  and with the same excitation conditions , the line intensity ratio equals 
the abundance ratio of the two species 
(abundance ratio of $^{12}$C/$^{13}$C is $\sim$ 20 for the GC region; \cite{Wilson99,Savage02}). If both 
transitions are optically thick and their emitting regions have the same excitation conditions, then the intensity 
ratio is unity. Using the detected isotopologues we find line intensity ratios of $\sim$ 4--6 from HNC and 
HN$^{13}$C (1--0) and slightly higher ratios of $\sim$ 6--8 from  \hcopnt\, and \htcop.  These ratios give an 
estimate of the opacity of the $^{12}$C molecules of 2--3, which implies that the emission is moderately optically thick.

%Indeed, if the opacity was very high then the derived filling factors of the (3--2) and (4--3) transitions 
%would be significantly different from each other, indicating that the assumed input column density was 
%not a true measure of the clump's column density.

Overlaid on Figure~\ref{radex} are the observed line ratios toward three positions within \cloud\, (solid lines, P2, P3 and P4, see 
Fig.~\ref{hnco_mom_chan}). Assuming that the beam filling factor is similar for both transitions, the 
HNC (4--3) to (3--2) line ratios indicate a gas density of $\sim$ 10$^{6}$\,\cmc.  This derived density is an order of 
magnitude higher than that derived toward \cloud\, from LVG modelling of the emission from CS and C$^{34}$S \citep{Lis01}.
The large disparity between the average volume density and the implied local volume density indicated by the line ratios
indicates that the gas is highly clumped, at least in the gas traced by HNC.  The models, however, should be treated
with caution if multiple gas components with different temperatures and densities are present along the same line of sight,
especially for optically thick lines, where radiative transfer effects can significantly alter the observed line ratios.
Indeed, no single component solution exists for HCO$^+$ emission.

While the kinetic temperature cannot be well constrained by our data, recent H$_{2}$CO observations that 
are sensitive to the gas temperature indicate a kinetic temperature of 65 $^{+20}_{-10}$ or 
70$^{+25}_{-15}$ (depending on the assumed abundance; \citealp{Ao13}). Within the errors, this kinetic 
temperature is indistinguishable from the global kinetic temperature of the molecular gas within the broader GC region 
(65 $\pm$ 10 K; \citealp{Ao13}).  

\subsection{Comparison between the dust and gas column density distribution}

\begin{figure}
\centering
%\hspace{-1.15cm}
%\includegraphics[width=0.480\textwidth,clip=true,trim=28mm 1mm 57mm 1mm]{fig8a.pdf}
%\includegraphics[width=0.39\textwidth,clip=true,trim=35mm 1mm 57mm 1mm]{fig4a.pdf}
%\includegraphics[width=0.58\textwidth,clip=true,trim=1mm 1mm 1mm 1mm]{fig4d.pdf}\\
%\includegraphics[width=0.39\textwidth,clip=true,trim=35mm 1mm 57mm 1mm]{fig4a.pdf}
%\includegraphics[width=0.58\textwidth,clip=true,trim=1mm 1mm 1mm 1mm]{fig4f.pdf}\\
%\includegraphics[width=0.39\textwidth,clip=true,trim=35mm 1mm 57mm 1mm]{fig4a.pdf}
%\includegraphics[width=0.58\textwidth,clip=true,trim=1mm 1mm 1mm 1mm]{fig4b.pdf}\\
\includegraphics[width=0.49\textwidth,clip=true,trim=1mm 1mm 1mm 1mm]{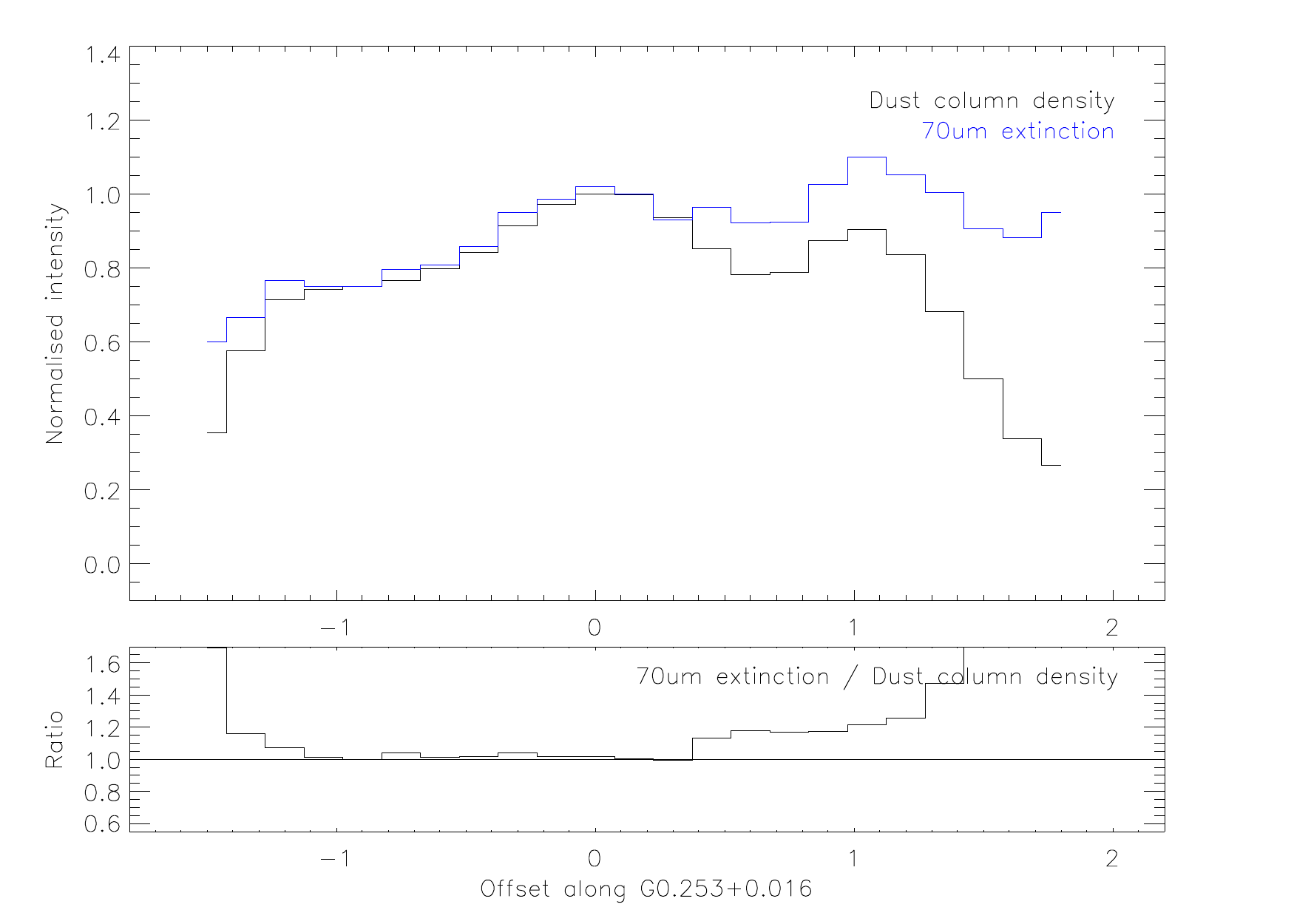}
\includegraphics[width=0.49\textwidth,clip=true,trim=1mm 1mm 1mm 1mm]{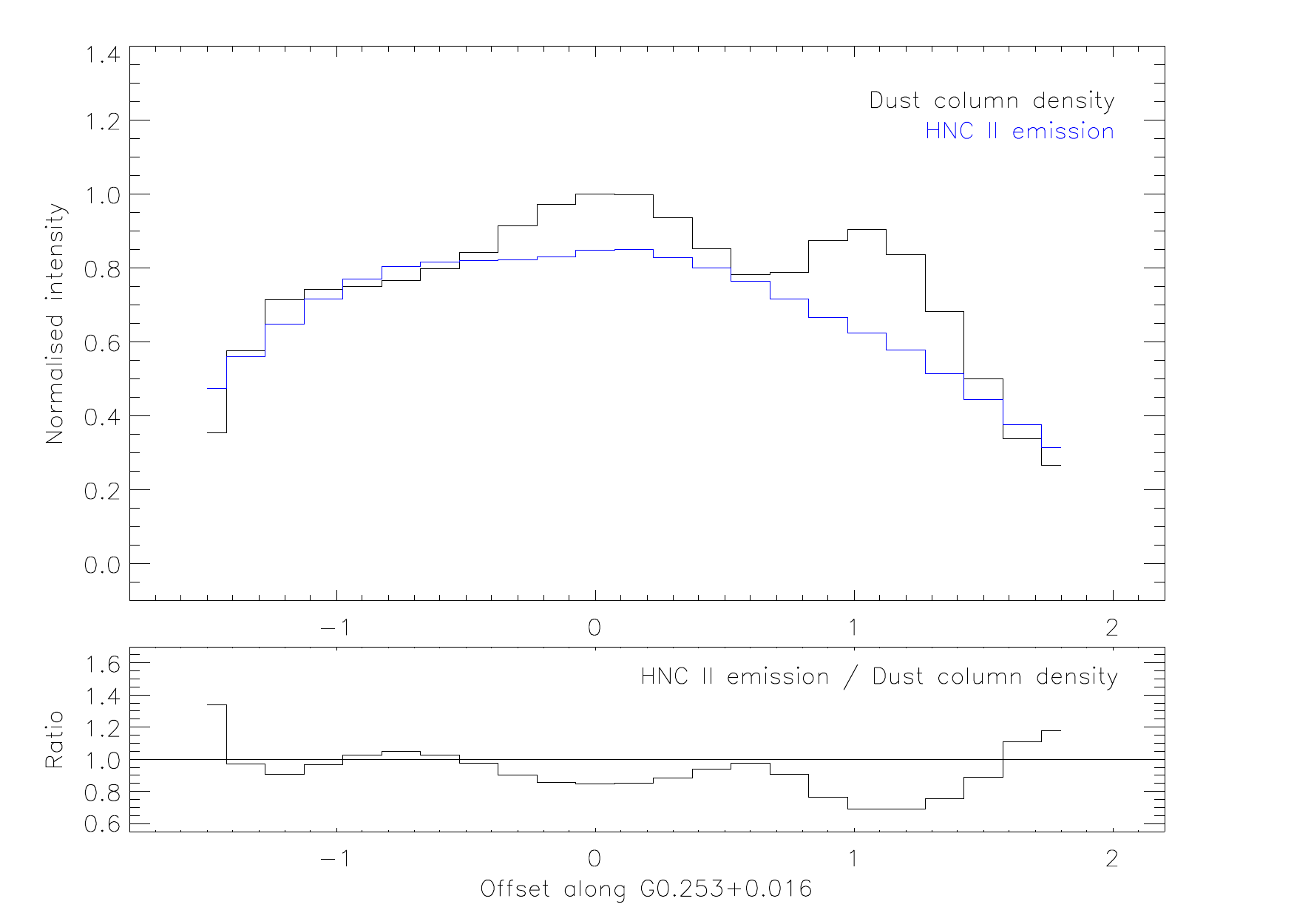}\\
\includegraphics[width=0.49\textwidth,clip=true,trim=1mm 1mm 1mm 1mm]{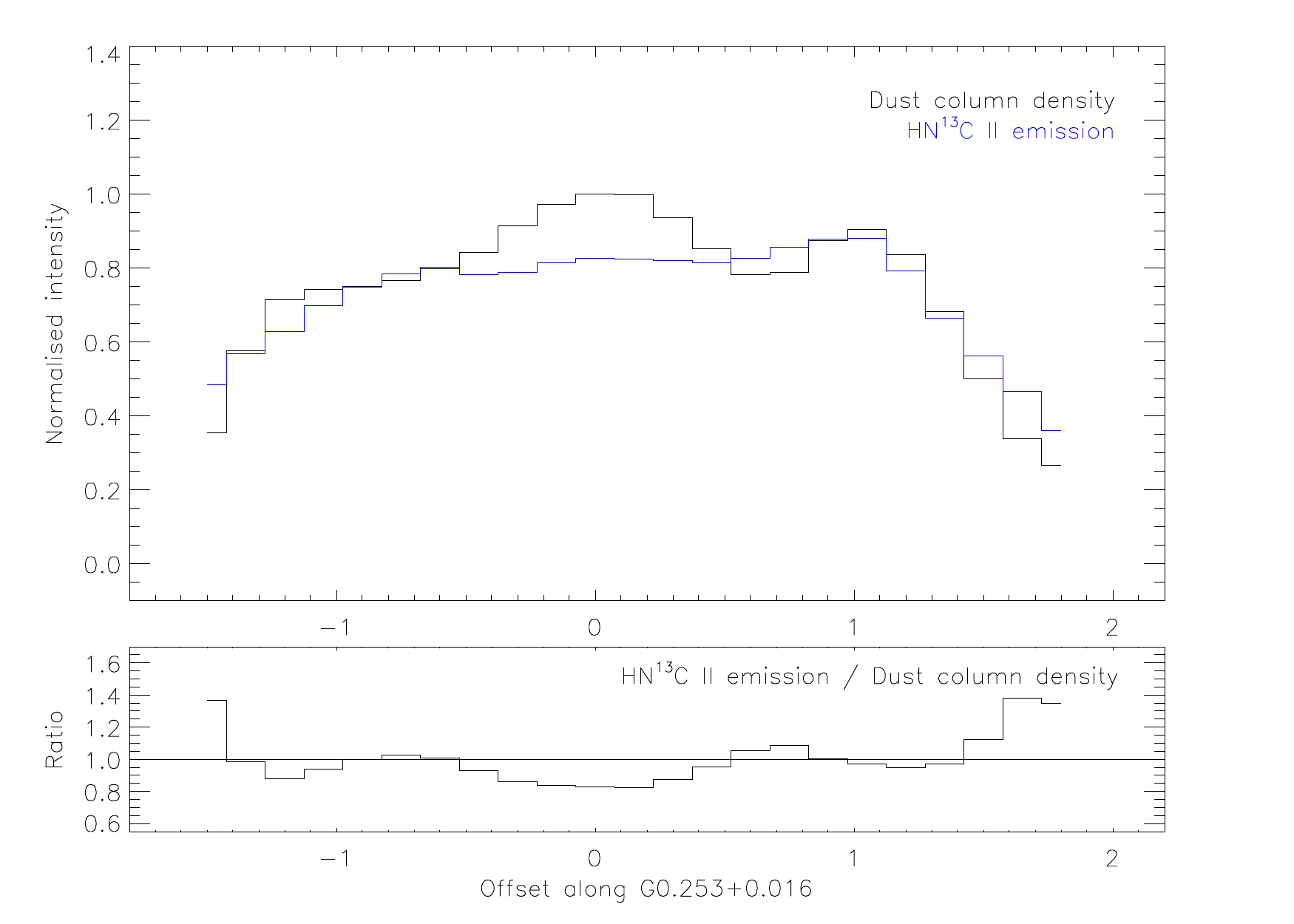}
\includegraphics[width=0.49\textwidth,clip=true,trim=1mm 1mm 1mm 1mm]{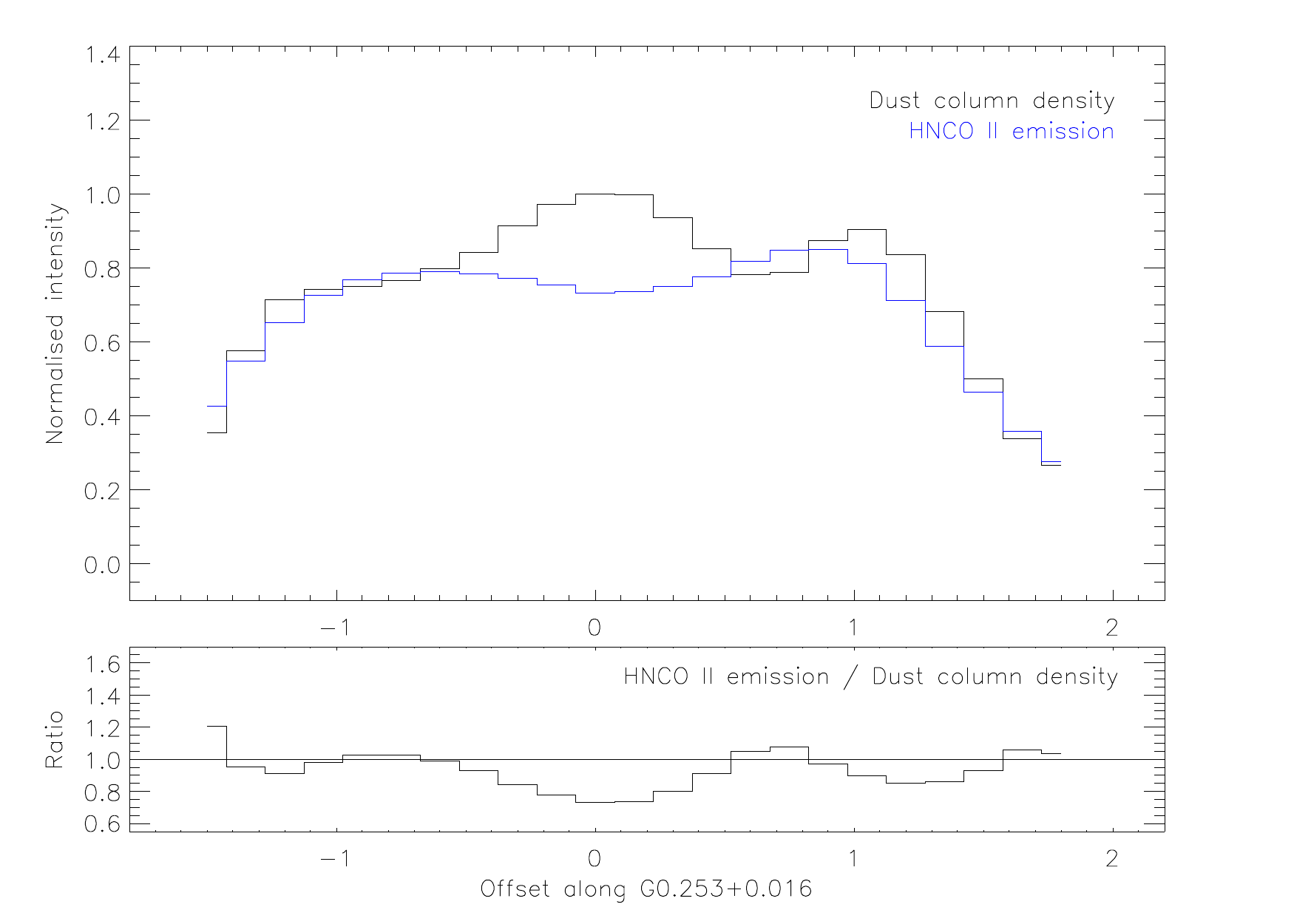}
\caption{\label{columns} Normalised dust column density profile along the major axis of \cloud\, overlaid with 70\,\um\, extinction profile (left upper panel), 
\hncnt\, (right upper panel), \hntcnt\, (left lower panel) and \hncofznt\, (right lower panel) integrated intensity emission profiles (in all cases the emission 
has been normalised to match the normalised dust column density toward the clump's edges). Under each panel is the ratio of the normalised profiles.
Overall, the dust column density profile is matched well by the 
70\,\um\, extinction profile (there is relatively more extinction toward the upper portion of the clump, positive offsets, compared to the centre and lower portion). 
In contrast, the line emission profiles appear anti-correlated with the dust column density toward the 
clump's centre, regardless of the gas tracer (i.e. HNC, \hntcnt, or \hncofznt; these three molecules were selected to represent the emission from the 
dense gas tracers, isotopolgues, and hot gas tracers respectively).  If the line emission is optically thin, then its profile ought to match the dust column 
density profile. Given the density of the clump, the emission from HNC is likely optically thick. However, the isotopologues and hot gas tracers (which 
are less likely to be optically thick) also show an absence of emission toward the clump's centre. Thus, if they are indeed optically thin, then we interpret 
this anti-correlation to arise due to gas depletion in the clump's cold interior.}
\end{figure}

If both the dust and gas were optically thin and traced the same material throughout the clump, then the 
integrated intensity of an optically thin gas tracer ought to follow the column density profile derived from 
the dust. Instead, we find that  regardless of the gas tracer, their integrated intensities peak at different 
locations when compared to the dust column density. Figure~\ref{columns} shows the normalised dust 
column density profile along the major axis of \cloud\, overlaid with 70\,\um\, extinction profile and the 
integrated intensity emission profiles from three different gas tracers (\hncnt, \hntcnt, and \hncofznt). 

 Overall, the dust column density profile is matched well by the 70\,\um\, extinction profile.
In contrast, the line emission profiles appear anti-correlated with the dust column density toward the 
clump's centre, regardless of the gas tracer (i.e. HNC, \hntcnt, or \hncofznt; these three molecules 
were selected to represent the emission from the dense gas tracers, isotopolgues, and hot gas tracers 
respectively). 
 
% 
%While overall the morphology of the dust column density matches well the \hncofznt\, integrated 
%intensity, they are anti-correlated toward the clump's centre:  the maximum of the dust continuum lies at
%the same position as a minimum in the molecular line emission.  This anti-correlation is more apparent when we plot the ratio of the 
%dust-derived column density to the HNCO integrated intensity and 70\,\um\, extinction (Fig.~\ref{columns}, lower panels) along the 
%cloud's major axis. While the HNCO integrated intensity follows well the dust derived column density toward the clump's edges, 
%it is clearly anti-correlated with the column density toward its center. 

The \Herschel\, dust emission indicates a peak \hh\, column density of 3.3 $\times$ 10$^{23}$\,\cms\, and a total mass for the clump 
of 1.3$\times$ 10$^{5}$\,\Msun. The lack of molecular line emission  (particularly from the isotopologues, e.g. \hntcnt)  at the peak of the 
dust column density is therefore surprising.  Indeed, none of the molecular lines peak toward the dust column density peak.
If the molecular line emission is optically thin, then the most plausible explanation 
is that there is severe chemical gas depletion in the clump's cold interior. Depletion arises when the gas 
and dust have a low temperature and high column density (T$_{gas} \sim$ T$_{dust} \sim$ 10 K, N(\hh) $>$ 10$^{22}$\,\cms , e.g., \citealp{Redman02}), 
and the molecular gas freezes onto the dust grains in the form of ice.  
In these regions, despite a large mass of dust, the usual molecular markers of the gas mass are absent because these
molecular species are frozen and hence do not emit in molecular lines.  
 While the dust conditions inferred for the center of
\cloud\, have precisely the properties required for chemical depletion, the gas temperature may be slightly higher (a few 10\,K). 
Recent modelling of the gas and dust temperature distribution within \cloud, shows that for densities of 10${^4}$\,\cmc, the gas 
temperature is roughly 70\,K, while the dust temperature may be closer to 20\,K \citep{Clark13}. However, as the density increases 
(presumably toward the clump's centre) and the dust temperature decreases ($\sim$ 10\,K), then the gas and dust temperatures 
should become more strongly coupled and the gas temperature will be closer to the dust temperature.
Detailed radiative transfer 
modelling is clearly needed to determine the exact gas temperature distribution within the clump.
Nevertheless, based on these observations we suggest  that \cloud\, is centrally condensed with a cold, 
dense interior in which the gas is depleted. The lack of molecular line emission toward the clump center therefore simply reflects the absence of these
molecular species in the gas phase due to depletion.

\subsection{Velocity gradient, complex gas kinematics}
\label{spectra_section}
 
\begin{sidewaysfigure}
\centering
\vspace{0cm}
\includegraphics[angle=-90,width=0.95\textwidth,clip=true,trim=10mm 30mm 110mm 40mm]{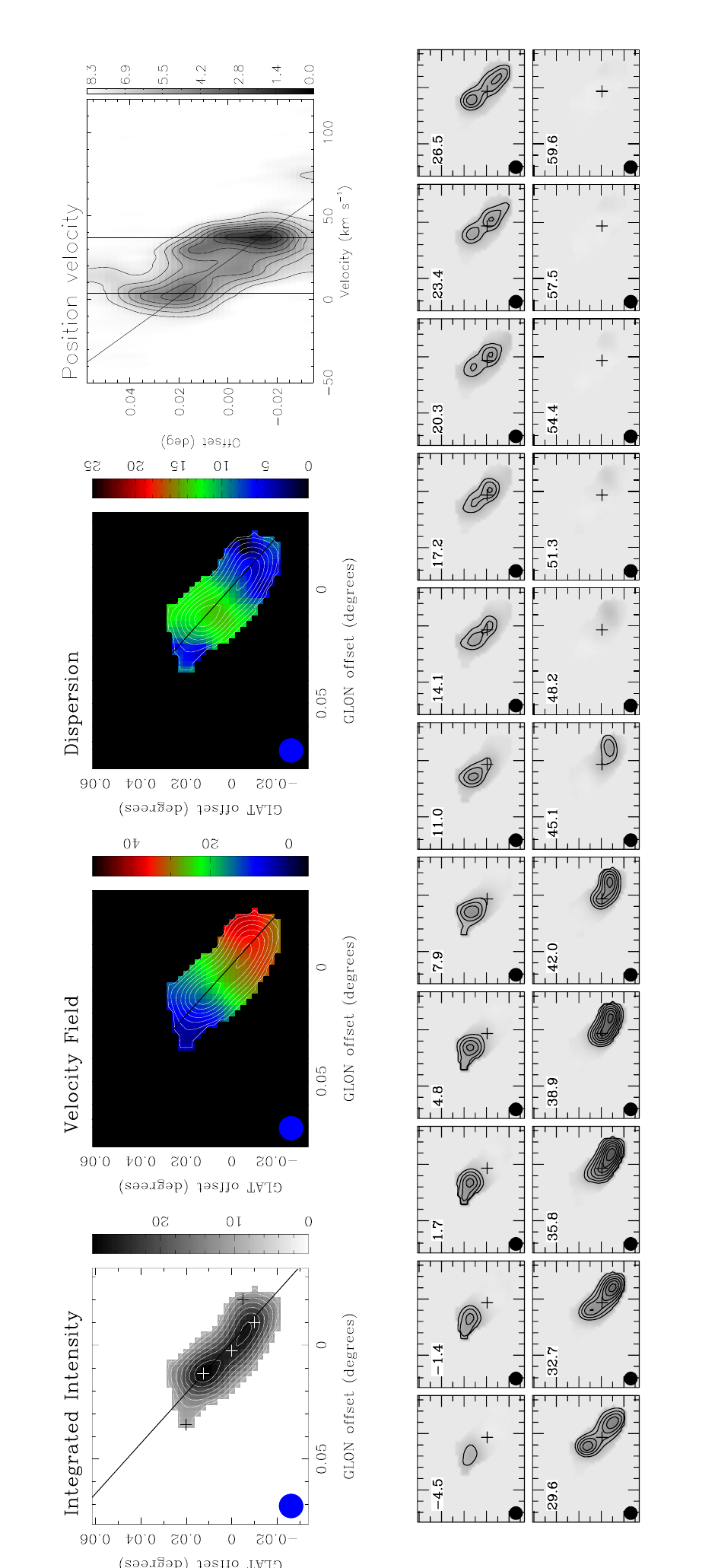}
\caption{\label{hnco_mom_chan} Moment maps, position-velocity diagram, and channel maps for \hncofz. {\it {Upper row, from left to right:}} 
      Integrated intensity map (M$_{0}$, units of \Kkms),  intensity weighted velocity field (M$_{1}$, units of \kms),  intensity weighted dispersion 
      (M$_{2}$, units of \kms), and position-velocity diagram showing the emission along the major axis of \cloud\, (the emission was averaged 
      over the clump's minor axis; contour levels are from 10 to 90\% of the peak; units of K).  Overlaid on the moment maps are contours of the 
      integrated intensity (levels are from 10 to 90\% of the peak in steps of 10\%) and a solid diagonal line marking the major axis of the clump. 
      The small crosses on the integrated intensity image mark the 5 positions a which spectra were extracted across the clump (P1 to P5, from left to right). 
      Overlaid on the position-velocity diagram are solid vertical lines marking the velocities of the two components identified in the \hncofznt\, 
      emission, the diagonal line shows the slope of the velocity gradient across \cloud. Each panel of the channel map shows the emission 
      averaged over $\sim$6\,\kms\, around the listed velocity (contour levels are from 10 to 90\% of the peak in steps of 10\%). The small cross 
      marks the peak in both the dust continuum emission and column density. For the moment and channel maps the beam size is shown in the lower left corner.}
\end{sidewaysfigure}

\begin{sidewaysfigure}
\centering
\includegraphics[angle=-90,width=0.95\textwidth,clip=true,trim=10mm 30mm 110mm 40mm]{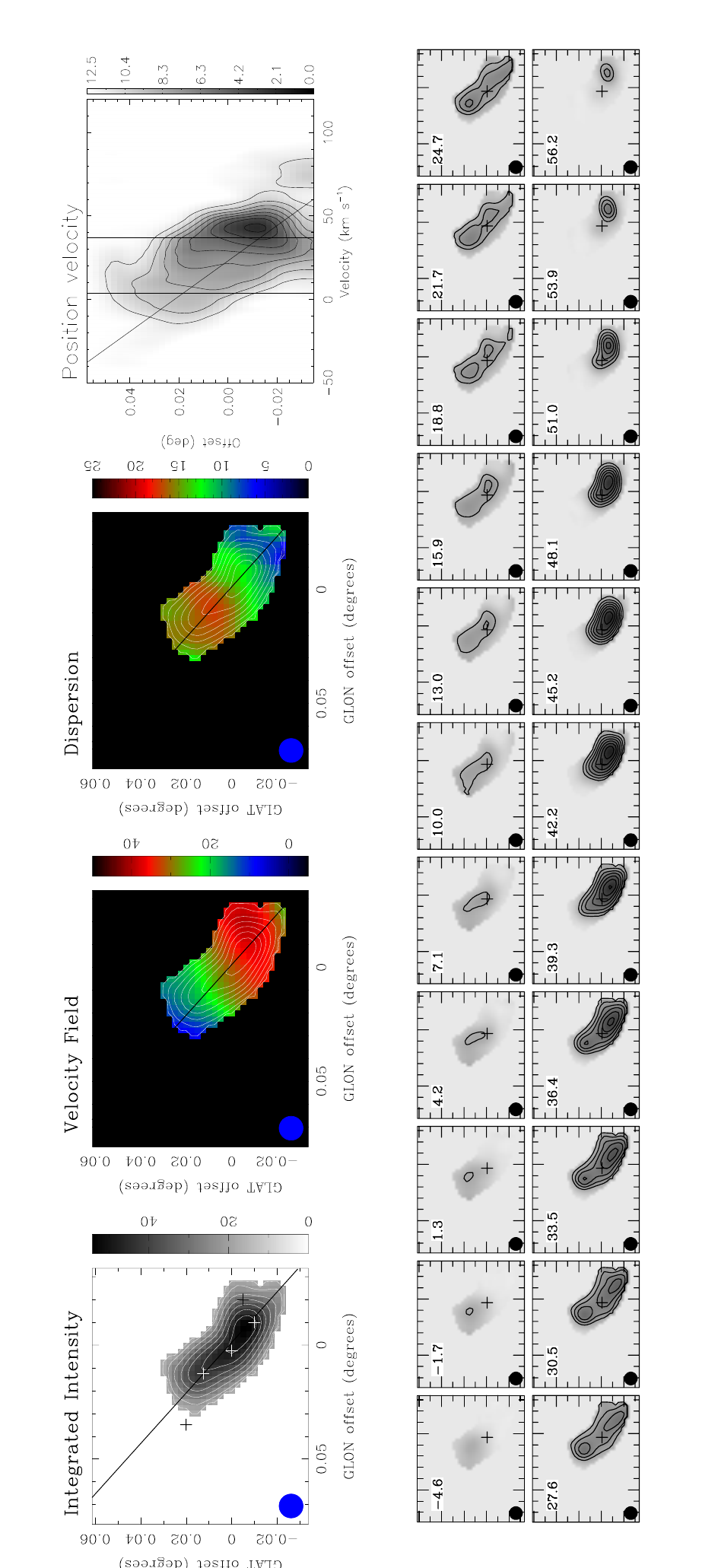}
\caption{\label{nthpoz}Moment maps, position-velocity diagram, and channel maps for \nthp. {\it {Upper row, from left to right:}} Integrated intensity map (M$_{0}$, units of \Kkms),  
intensity weighted velocity field (M$_{1}$, units of \kms),  intensity weighted dispersion (M$_{2}$, units of \kms), and position-velocity diagram showing the 
 emission along the major axis of \cloud\, (the emission was averaged over the clump's minor axis; contour levels are from 10 to 90\% of the peak; units of K).  Overlaid on 
 the moment maps are contours of the integrated intensity (levels are from 10 to 90\% of the peak in steps of 10\%) and a solid diagonal line marking 
the major axis of the clump.  The small crosses on the integrated intensity image mark the 5 positions a which spectra were extracted across the clump (P1 to P5, from left to right). Overlaid on the position-velocity diagram are solid vertical lines marking 
the velocities of the two components identified in the \hncofznt\, emission, the diagonal line shows the slope of the velocity gradient across \cloud.
Each panel of the channel map
shows the emission averaged over $\sim$6\,\kms\, around the listed velocity (contour levels are from 10 to 90\% of the peak in steps of 10\%). The small cross marks the peak in both
the dust continuum emission and column density. For the moment and channel maps the beam size is shown in the lower left corner.}
\end{sidewaysfigure}

\begin{sidewaysfigure}
\centering
%\nthpnt (3--2)
\includegraphics[angle=-90,width=0.95\textwidth,clip=true,trim=10mm 30mm 110mm 40mm]{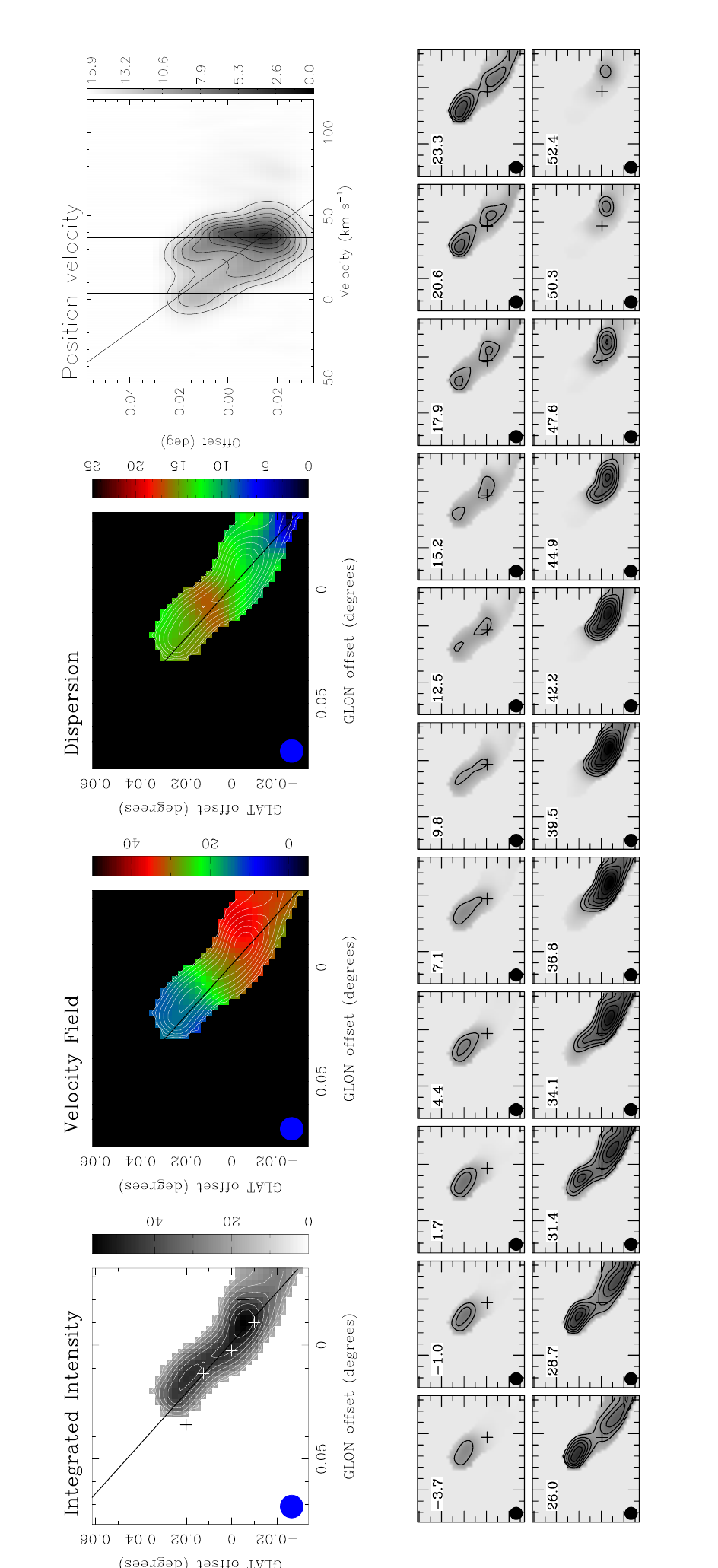}
\caption{\label{nthptt}Moment maps, position-velocity diagram, and channel maps for \nthpnt\, (3--2). {\it {Upper row, from left to right:}} Integrated intensity map (M$_{0}$, units of \Kkms),  
intensity weighted velocity field (M$_{1}$, units of \kms),  intensity weighted dispersion (M$_{2}$, units of \kms), and position-velocity diagram showing the 
 emission along the major axis of \cloud\, (the emission was averaged over the clump's minor axis; contour levels are from 10 to 90\% of the peak; units of K).  Overlaid on 
 the moment maps are contours of the integrated intensity (levels are from 10 to 90\% of the peak in steps of 10\%) and a solid diagonal line marking 
the major axis of the clump.  The small crosses on the integrated intensity image mark the 5 positions a which spectra were extracted across the clump (P1 to P5, from left to right). Overlaid on the position-velocity diagram are solid vertical lines marking 
the velocities of the two components identified in the \hncofznt\, emission, the diagonal line shows the slope of the velocity gradient across \cloud.
Each panel of the channel map
shows the emission averaged over $\sim$6\,\kms\, around the listed velocity (contour levels are from 10 to 90\% of the peak in steps of 10\%). The small cross marks the peak in both
the dust continuum emission and column density. For the moment and channel maps the beam size is shown in the lower left corner.}
\end{sidewaysfigure}

\begin{sidewaysfigure}
\centering
%\nthpnt (4--3)
\includegraphics[angle=-90,width=0.95\textwidth,clip=true,trim=10mm 30mm 110mm 40mm]{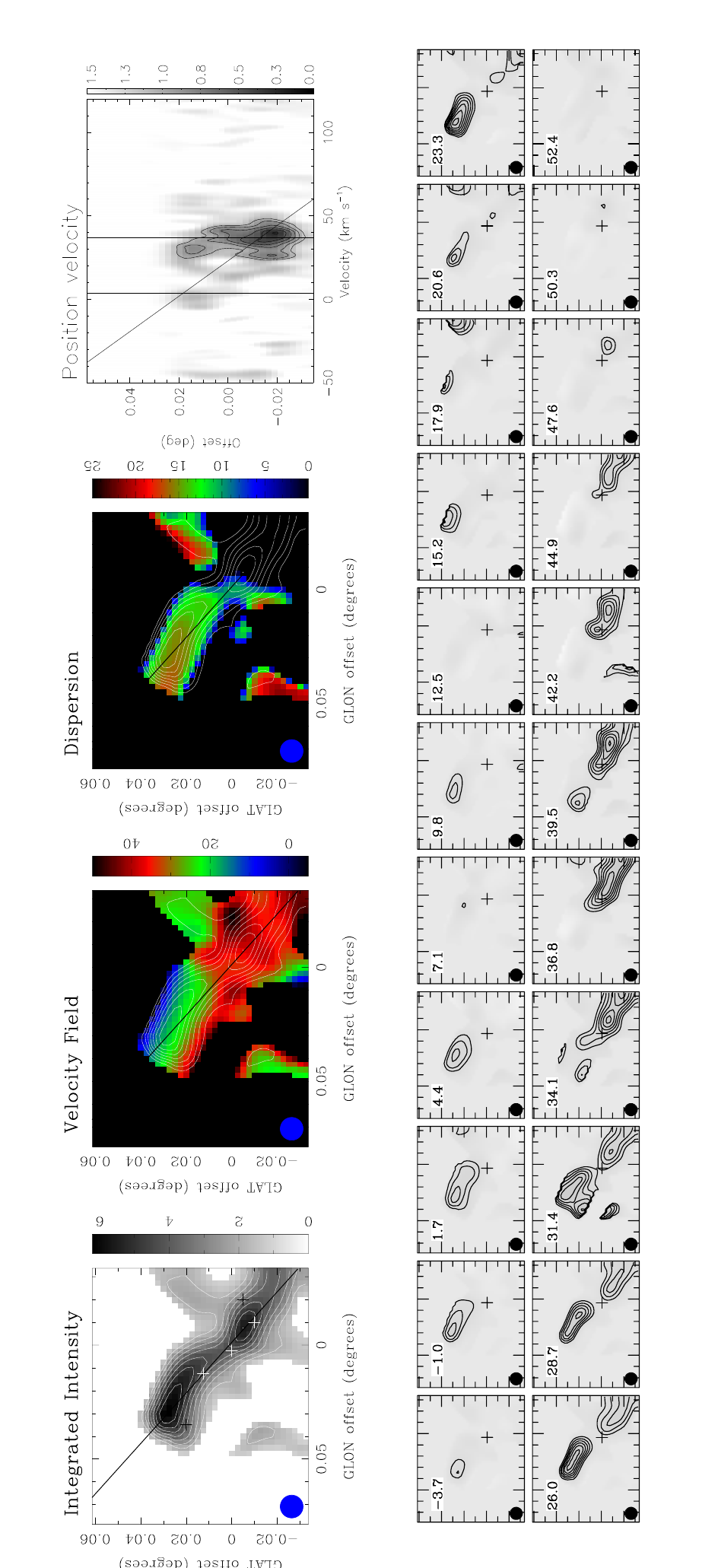}
\caption{\label{nthpft}Moment maps, position-velocity diagram, and channel maps for \nthpnt\, (4--3). {\it {Upper row, from left to right:}} Integrated intensity map (M$_{0}$, units of \Kkms),  
intensity weighted velocity field (M$_{1}$, units of \kms),  intensity weighted dispersion (M$_{2}$, units of \kms), and position-velocity diagram showing the 
 emission along the major axis of \cloud\, (the emission was averaged over the clump's minor axis; contour levels are from 10 to 90\% of the peak; units of K).  Overlaid on 
 the moment maps are contours of the integrated intensity (levels are from 10 to 90\% of the peak in steps of 10\%) and a solid diagonal line marking 
the major axis of the clump.  The small crosses on the integrated intensity image mark the 5 positions a which spectra were extracted across the clump (P1 to P5, from left to right). Overlaid on the position-velocity diagram are solid vertical lines marking 
the velocities of the two components identified in the \hncofznt\, emission, the diagonal line shows the slope of the velocity gradient across \cloud.
Each panel of the channel map
shows the emission averaged over $\sim$6\,\kms\, around the listed velocity (contour levels are from 10 to 90\% of the peak in steps of 10\%). The small cross marks the peak in both
the dust continuum emission and column density. For the moment and channel maps the beam size is shown in the lower left corner.}
\end{sidewaysfigure}

\begin{sidewaysfigure}
\centering
%\hcn
\includegraphics[angle=-90,width=0.95\textwidth,clip=true,trim=10mm 30mm 110mm 40mm]{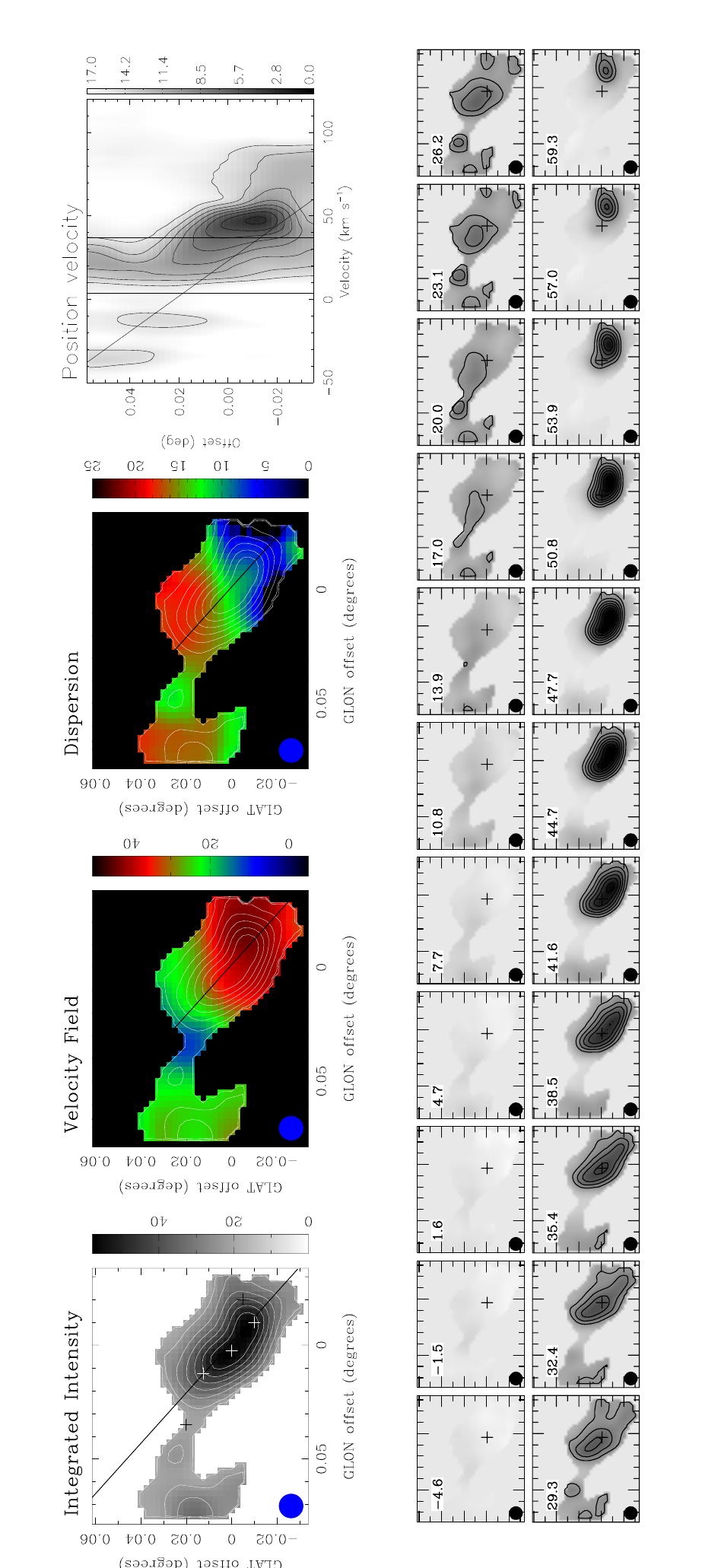}
\caption{\label{hcnoz}Moment maps, position-velocity diagram, and channel maps for \hcn. {\it {Upper row, from left to right:}} Integrated intensity map (M$_{0}$, units of \Kkms),  
intensity weighted velocity field (M$_{1}$, units of \kms),  intensity weighted dispersion (M$_{2}$, units of \kms), and position-velocity diagram showing the 
 emission along the major axis of \cloud\, (the emission was averaged over the clump's minor axis; contour levels are from 10 to 90\% of the peak; units of K).  Overlaid on 
 the moment maps are contours of the integrated intensity (levels are from 10 to 90\% of the peak in steps of 10\%) and a solid diagonal line marking 
the major axis of the clump.  The small crosses on the integrated intensity image mark the 5 positions a which spectra were extracted across the clump (P1 to P5, from left to right). Overlaid on the position-velocity diagram are solid vertical lines marking 
the velocities of the two components identified in the \hncofznt\, emission, the diagonal line shows the slope of the velocity gradient across \cloud.
Each panel of the channel map
shows the emission averaged over $\sim$6\,\kms\, around the listed velocity (contour levels are from 10 to 90\% of the peak in steps of 10\%). The small cross marks the peak in both
the dust continuum emission and column density. For the moment and channel maps the beam size is shown in the lower left corner.}
\end{sidewaysfigure}

\begin{sidewaysfigure}
\centering
%\hcnnt\, (3--2)
\includegraphics[angle=-90,width=0.95\textwidth,clip=true,trim=10mm 30mm 110mm 40mm]{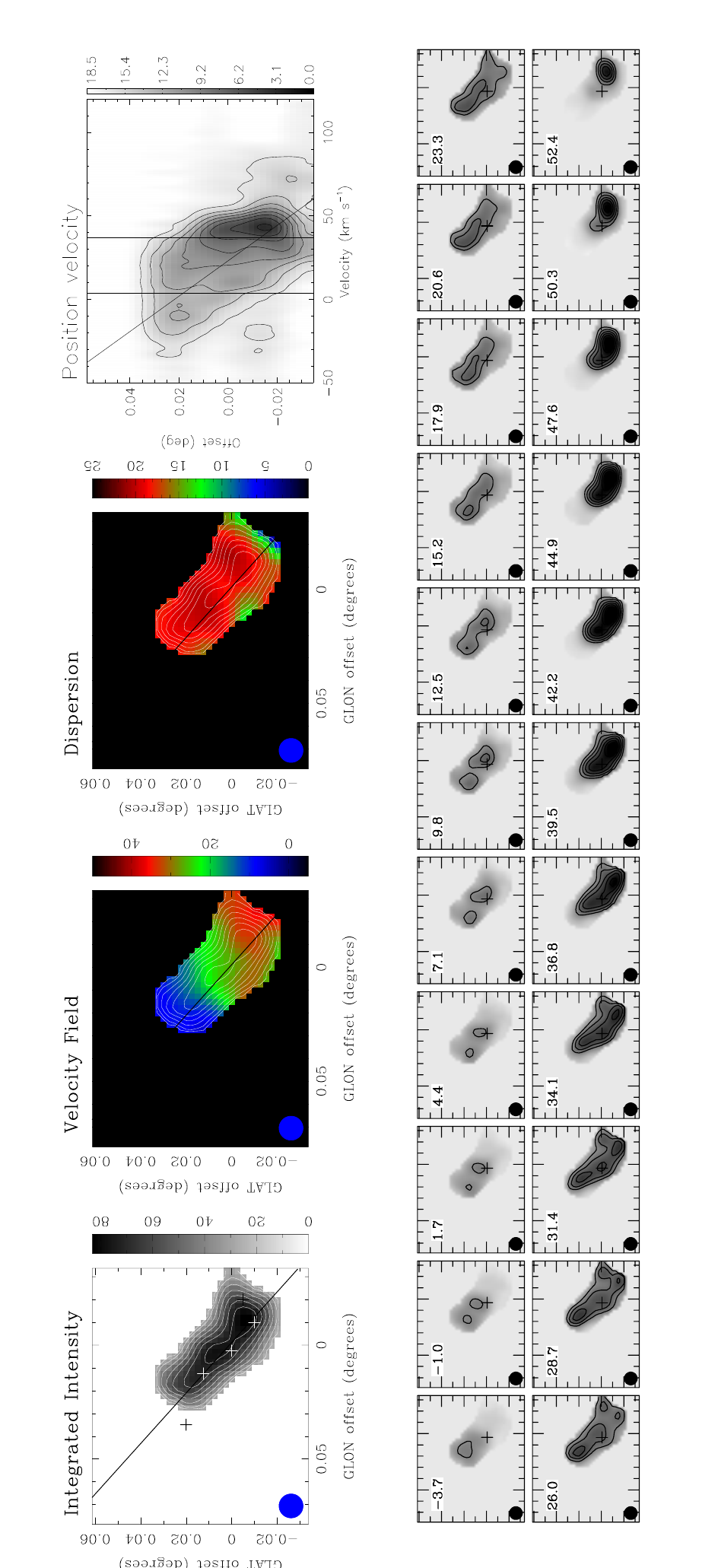}
\caption{\label{hcntt}Moment maps, position-velocity diagram, and channel maps for \hcnnt\, (3--2). {\it {Upper row, from left to right:}} Integrated intensity map (M$_{0}$, units of \Kkms),  
intensity weighted velocity field (M$_{1}$, units of \kms),  intensity weighted dispersion (M$_{2}$, units of \kms), and position-velocity diagram showing the 
 emission along the major axis of \cloud\, (the emission was averaged over the clump's minor axis; contour levels are from 10 to 90\% of the peak; units of K).  Overlaid on 
 the moment maps are contours of the integrated intensity (levels are from 10 to 90\% of the peak in steps of 10\%) and a solid diagonal line marking 
the major axis of the clump.  The small crosses on the integrated intensity image mark the 5 positions a which spectra were extracted across the clump (P1 to P5, from left to right). Overlaid on the position-velocity diagram are solid vertical lines marking 
the velocities of the two components identified in the \hncofznt\, emission, the diagonal line shows the slope of the velocity gradient across \cloud.
Each panel of the channel map
shows the emission averaged over $\sim$6\,\kms\, around the listed velocity (contour levels are from 10 to 90\% of the peak in steps of 10\%). The small cross marks the peak in both
the dust continuum emission and column density. For the moment and channel maps the beam size is shown in the lower left corner.}
\end{sidewaysfigure}

\begin{sidewaysfigure}
\centering
 %\hcnnt\, (4--3)
 \includegraphics[angle=-90,width=0.95\textwidth,clip=true,trim=10mm 30mm 110mm 40mm]{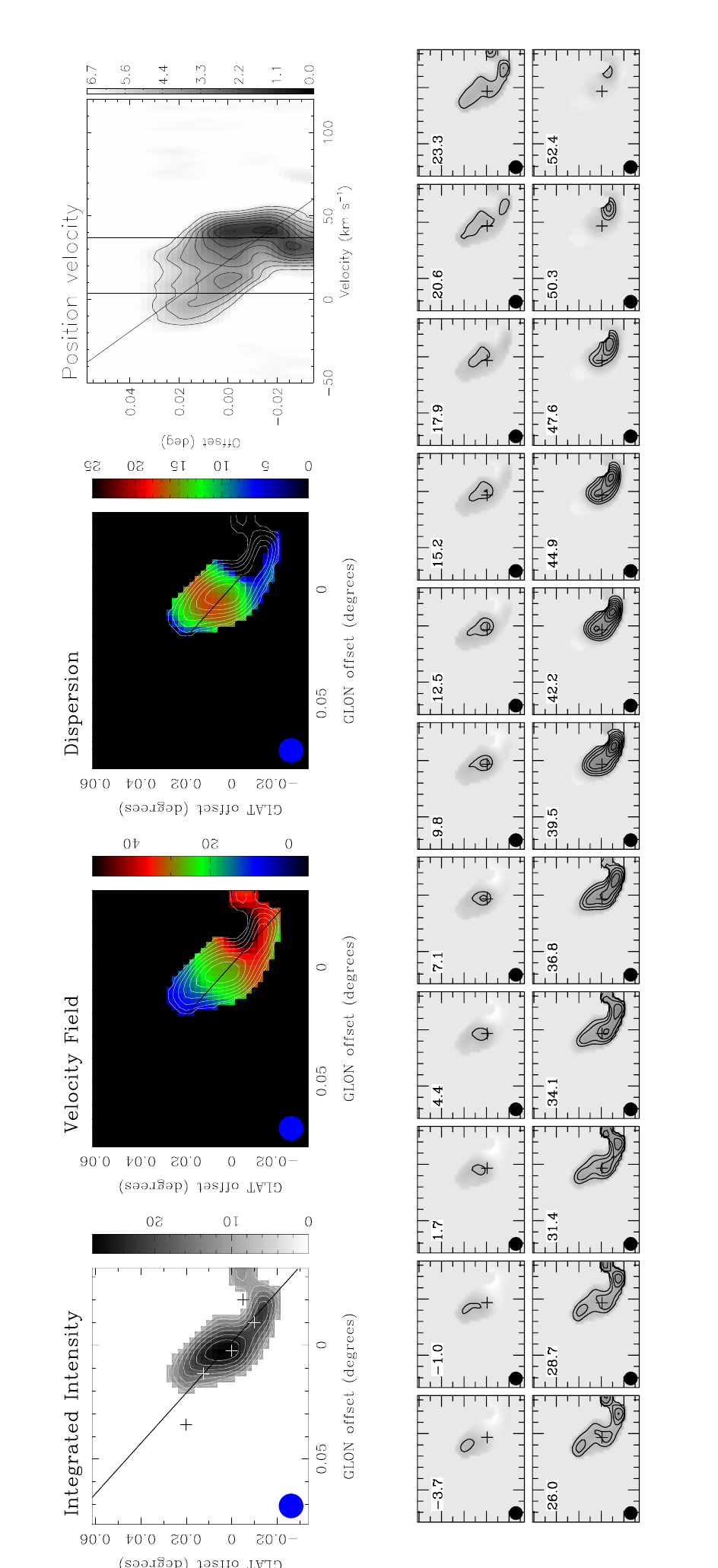}
\caption{\label{hcnft}Moment maps, position-velocity diagram, and channel maps for \hcnnt\, (4--3). {\it {Upper row, from left to right:}} Integrated intensity map (M$_{0}$, units of \Kkms),  
intensity weighted velocity field (M$_{1}$, units of \kms),  intensity weighted dispersion (M$_{2}$, units of \kms), and position-velocity diagram showing the 
 emission along the major axis of \cloud\, (the emission was averaged over the clump's minor axis; contour levels are from 10 to 90\% of the peak; units of K).  Overlaid on 
 the moment maps are contours of the integrated intensity (levels are from 10 to 90\% of the peak in steps of 10\%) and a solid diagonal line marking 
the major axis of the clump.  The small crosses on the integrated intensity image mark the 5 positions a which spectra were extracted across the clump (P1 to P5, from left to right). Overlaid on the position-velocity diagram are solid vertical lines marking 
the velocities of the two components identified in the \hncofznt\, emission, the diagonal line shows the slope of the velocity gradient across \cloud.
Each panel of the channel map
shows the emission averaged over $\sim$6\,\kms\, around the listed velocity (contour levels are from 10 to 90\% of the peak in steps of 10\%). The small cross marks the peak in both
the dust continuum emission and column density. For the moment and channel maps the beam size is shown in the lower left corner.}
\end{sidewaysfigure}

\begin{sidewaysfigure}
\centering
%\hnc
\includegraphics[angle=-90,width=0.95\textwidth,clip=true,trim=10mm 30mm 110mm 40mm]{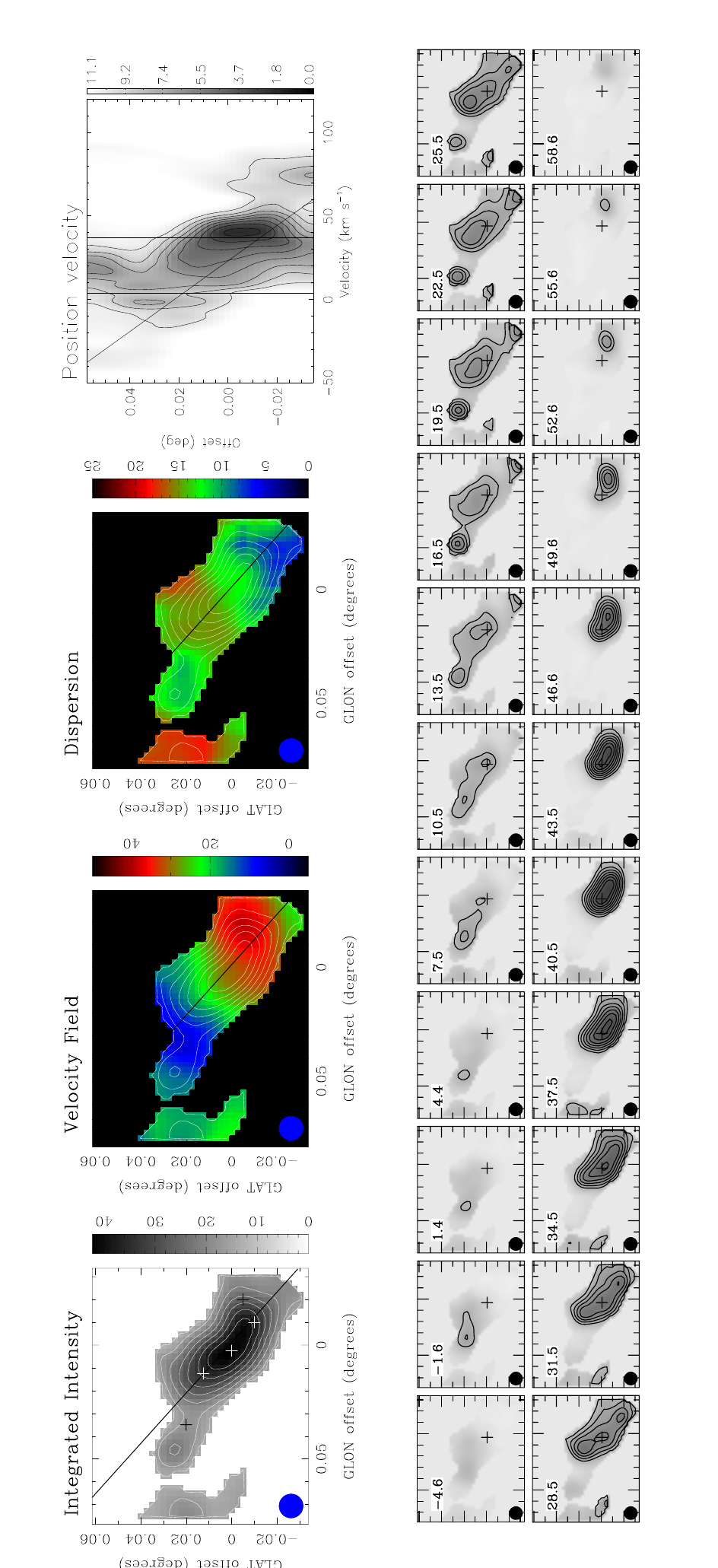}
\caption{\label{hncoz}Moment maps, position-velocity diagram, and channel maps for \hnc. {\it {Upper row, from left to right:}} Integrated intensity map (M$_{0}$, units of \Kkms),  
intensity weighted velocity field (M$_{1}$, units of \kms),  intensity weighted dispersion (M$_{2}$, units of \kms), and position-velocity diagram showing the 
 emission along the major axis of \cloud\, (the emission was averaged over the clump's minor axis; contour levels are from 10 to 90\% of the peak; units of K).  Overlaid on 
 the moment maps are contours of the integrated intensity (levels are from 10 to 90\% of the peak in steps of 10\%) and a solid diagonal line marking 
the major axis of the clump.  The small crosses on the integrated intensity image mark the 5 positions a which spectra were extracted across the clump (P1 to P5, from left to right). Overlaid on the position-velocity diagram are solid vertical lines marking 
the velocities of the two components identified in the \hncofznt\, emission, the diagonal line shows the slope of the velocity gradient across \cloud.
Each panel of the channel map
shows the emission averaged over $\sim$6\,\kms\, around the listed velocity (contour levels are from 10 to 90\% of the peak in steps of 10\%). The small cross marks the peak in both
the dust continuum emission and column density. For the moment and channel maps the beam size is shown in the lower left corner.}
\end{sidewaysfigure}

\begin{sidewaysfigure}
\centering
%\hncnt\,(3--2)
\includegraphics[angle=-90,width=0.95\textwidth,clip=true,trim=10mm 30mm 110mm 40mm]{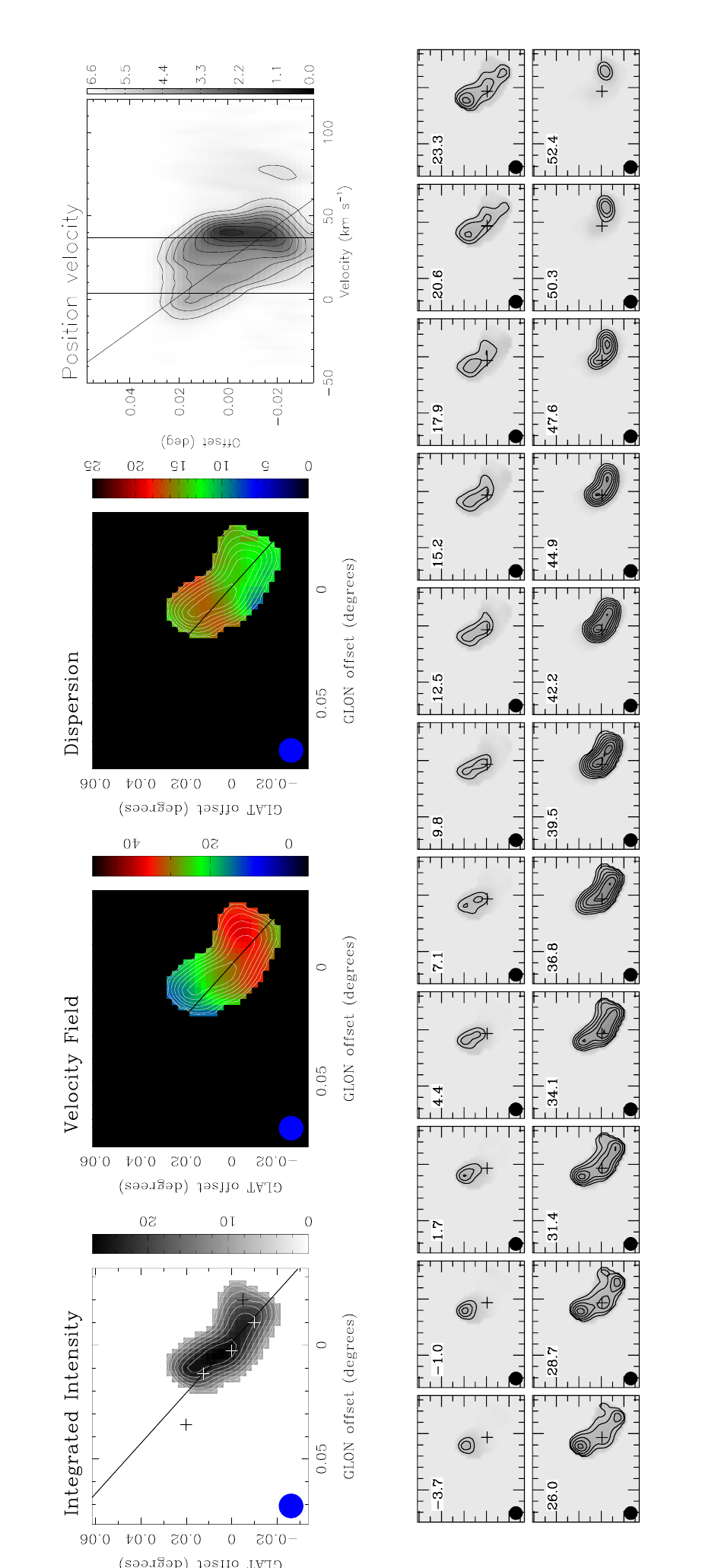}
\caption{\label{hnctt}Moment maps, position-velocity diagram, and channel maps for \hncnt\, (3--2). {\it {Upper row, from left to right:}} Integrated intensity map (M$_{0}$, units of \Kkms),  
intensity weighted velocity field (M$_{1}$, units of \kms),  intensity weighted dispersion (M$_{2}$, units of \kms), and position-velocity diagram showing the 
 emission along the major axis of \cloud\, (the emission was averaged over the clump's minor axis; contour levels are from 10 to 90\% of the peak; units of K).  Overlaid on 
 the moment maps are contours of the integrated intensity (levels are from 10 to 90\% of the peak in steps of 10\%) and a solid diagonal line marking 
the major axis of the clump.  The small crosses on the integrated intensity image mark the 5 positions a which spectra were extracted across the clump (P1 to P5, from left to right). Overlaid on the position-velocity diagram are solid vertical lines marking 
the velocities of the two components identified in the \hncofznt\, emission, the diagonal line shows the slope of the velocity gradient across \cloud.
Each panel of the channel map
shows the emission averaged over $\sim$6\,\kms\, around the listed velocity (contour levels are from 10 to 90\% of the peak in steps of 10\%). The small cross marks the peak in both
the dust continuum emission and column density. For the moment and channel maps the beam size is shown in the lower left corner.}
\end{sidewaysfigure}

\begin{sidewaysfigure}
\centering
%\hncnt\,(4--3)
\includegraphics[angle=-90,width=0.95\textwidth,clip=true,trim=10mm 30mm 110mm 40mm]{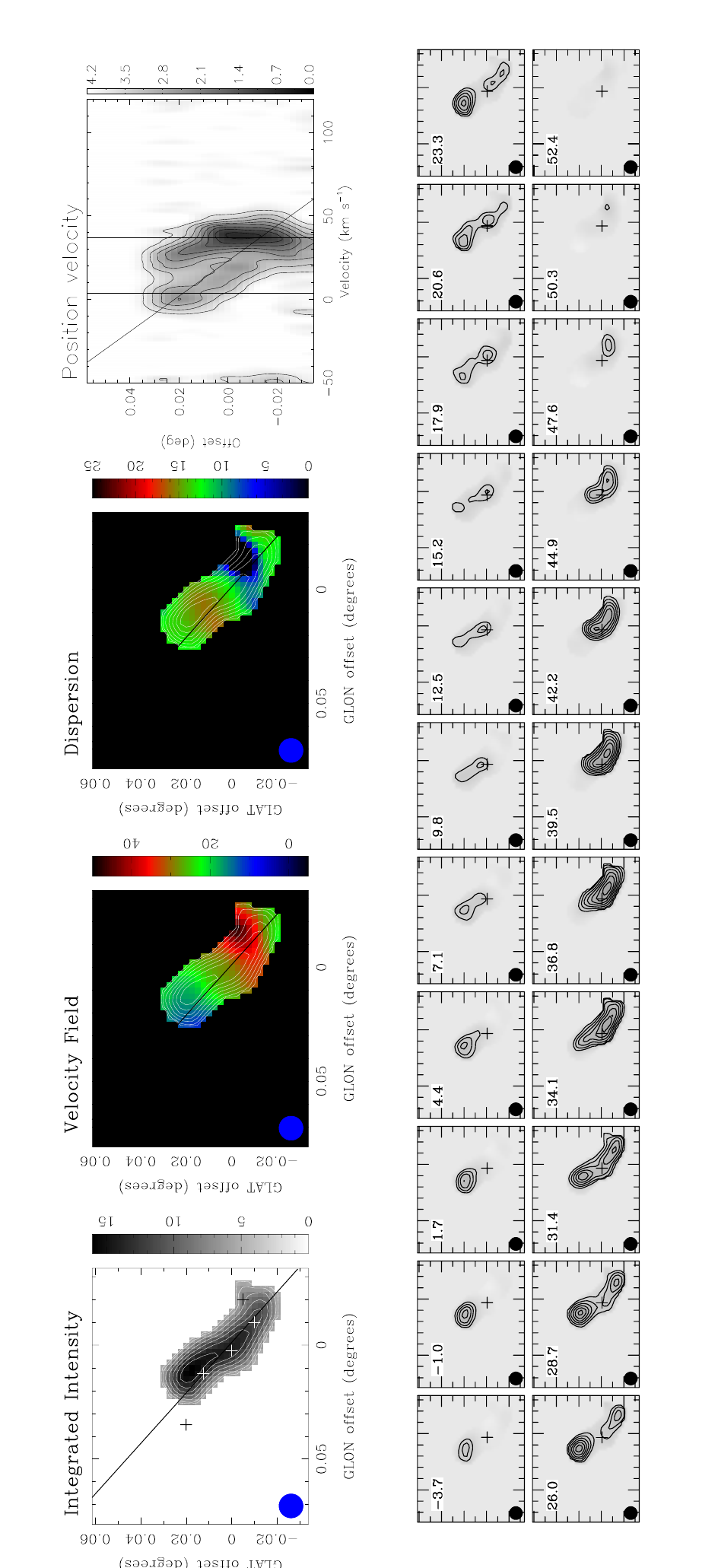}
\caption{\label{hncft}Moment maps, position-velocity diagram, and channel maps for \hncnt\, (4--3). {\it {Upper row, from left to right:}} Integrated intensity map (M$_{0}$, units of \Kkms),  
intensity weighted velocity field (M$_{1}$, units of \kms),  intensity weighted dispersion (M$_{2}$, units of \kms), and position-velocity diagram showing the 
 emission along the major axis of \cloud\, (the emission was averaged over the clump's minor axis; contour levels are from 10 to 90\% of the peak; units of K).  Overlaid on 
 the moment maps are contours of the integrated intensity (levels are from 10 to 90\% of the peak in steps of 10\%) and a solid diagonal line marking 
the major axis of the clump.  The small crosses on the integrated intensity image mark the 5 positions a which spectra were extracted across the clump (P1 to P5, from left to right). Overlaid on the position-velocity diagram are solid vertical lines marking 
the velocities of the two components identified in the \hncofznt\, emission, the diagonal line shows the slope of the velocity gradient across \cloud.
Each panel of the channel map
shows the emission averaged over $\sim$6\,\kms\, around the listed velocity (contour levels are from 10 to 90\% of the peak in steps of 10\%). The small cross marks the peak in both
the dust continuum emission and column density. For the moment and channel maps the beam size is shown in the lower left corner.}
\end{sidewaysfigure}

\begin{sidewaysfigure}
\centering
 %\hntc
 \includegraphics[angle=-90,width=0.95\textwidth,clip=true,trim=10mm 30mm 110mm 40mm]{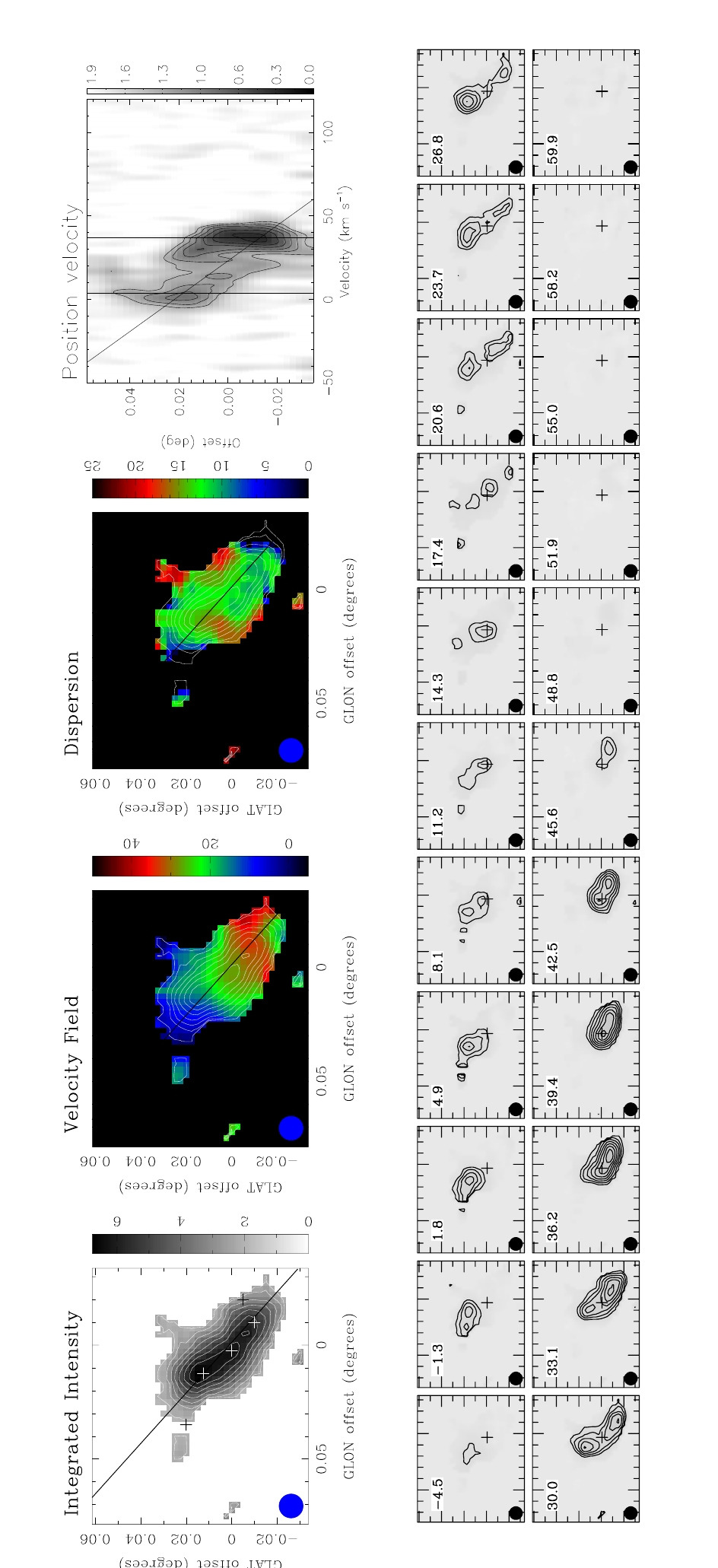}
\caption{\label{hntc}Moment maps, position-velocity diagram, and channel maps for \hntc. {\it {Upper row, from left to right:}} Integrated intensity map (M$_{0}$, units of \Kkms),  
intensity weighted velocity field (M$_{1}$, units of \kms),  intensity weighted dispersion (M$_{2}$, units of \kms), and position-velocity diagram showing the 
 emission along the major axis of \cloud\, (the emission was averaged over the clump's minor axis; contour levels are from 10 to 90\% of the peak; units of K).  Overlaid on 
 the moment maps are contours of the integrated intensity (levels are from 10 to 90\% of the peak in steps of 10\%) and a solid diagonal line marking 
the major axis of the clump.  The small crosses on the integrated intensity image mark the 5 positions a which spectra were extracted across the clump (P1 to P5, from left to right). Overlaid on the position-velocity diagram are solid vertical lines marking 
the velocities of the two components identified in the \hncofznt\, emission, the diagonal line shows the slope of the velocity gradient across \cloud.
Each panel of the channel map
shows the emission averaged over $\sim$6\,\kms\, around the listed velocity (contour levels are from 10 to 90\% of the peak in steps of 10\%). The small cross marks the peak in both
the dust continuum emission and column density. For the moment and channel maps the beam size is shown in the lower left corner.}
\end{sidewaysfigure}

\begin{sidewaysfigure}
\centering
% \hcop
\includegraphics[angle=-90,width=0.95\textwidth,clip=true,trim=10mm 30mm 110mm 40mm]{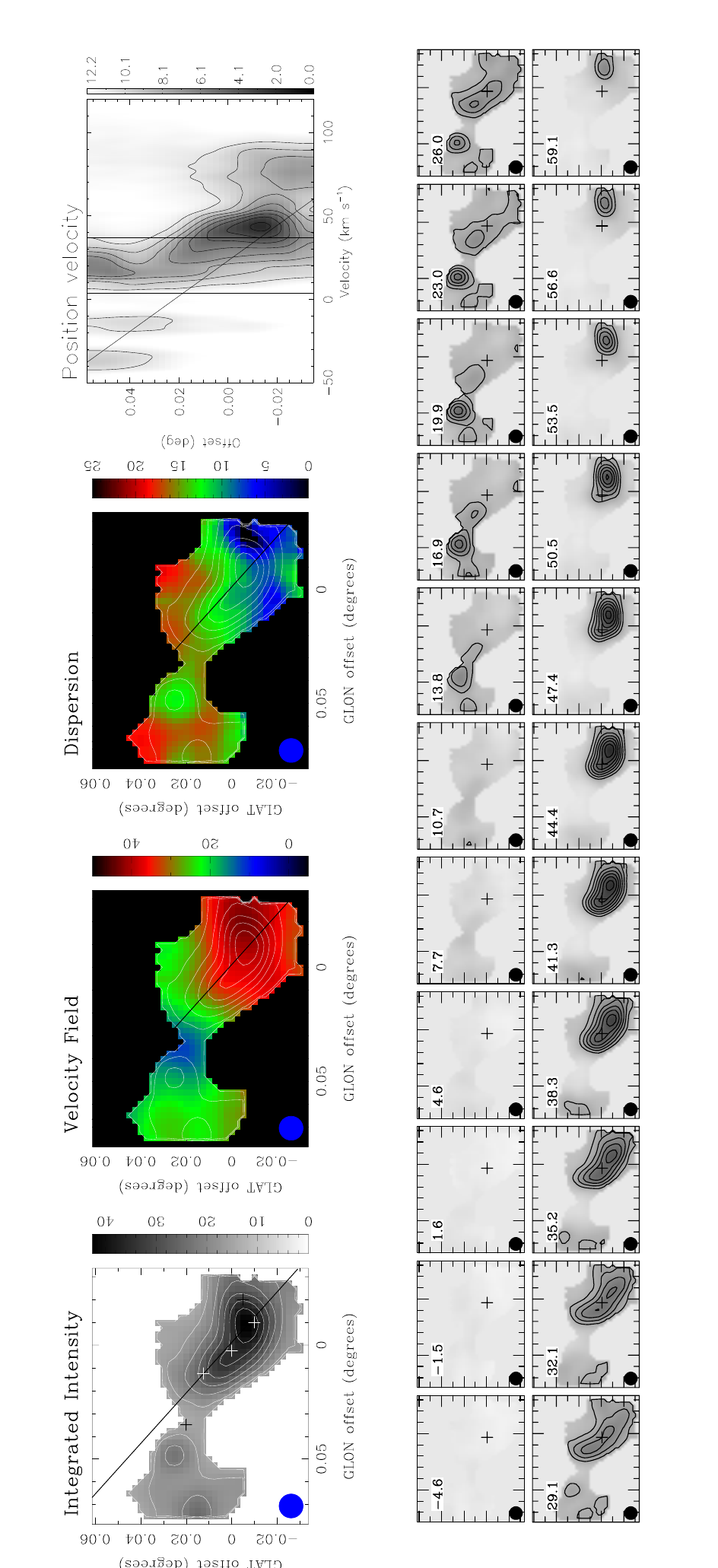}
\caption{\label{hcopoz}Moment maps, position-velocity diagram, and channel maps for \hcop. {\it {Upper row, from left to right:}} Integrated intensity map (M$_{0}$, units of \Kkms),  
intensity weighted velocity field (M$_{1}$, units of \kms),  intensity weighted dispersion (M$_{2}$, units of \kms), and position-velocity diagram showing the 
 emission along the major axis of \cloud\, (the emission was averaged over the clump's minor axis; contour levels are from 10 to 90\% of the peak; units of K).  Overlaid on 
 the moment maps are contours of the integrated intensity (levels are from 10 to 90\% of the peak in steps of 10\%) and a solid diagonal line marking 
the major axis of the clump.  The small crosses on the integrated intensity image mark the 5 positions a which spectra were extracted across the clump (P1 to P5, from left to right). Overlaid on the position-velocity diagram are solid vertical lines marking 
the velocities of the two components identified in the \hncofznt\, emission, the diagonal line shows the slope of the velocity gradient across \cloud.
Each panel of the channel map
shows the emission averaged over $\sim$6\,\kms\, around the listed velocity (contour levels are from 10 to 90\% of the peak in steps of 10\%). The small cross marks the peak in both
the dust continuum emission and column density. For the moment and channel maps the beam size is shown in the lower left corner.}
 \end{sidewaysfigure}

\begin{sidewaysfigure}
\centering
%\hcopnt\,(3--2)
\includegraphics[angle=-90,width=0.95\textwidth,clip=true,trim=10mm 30mm 110mm 40mm]{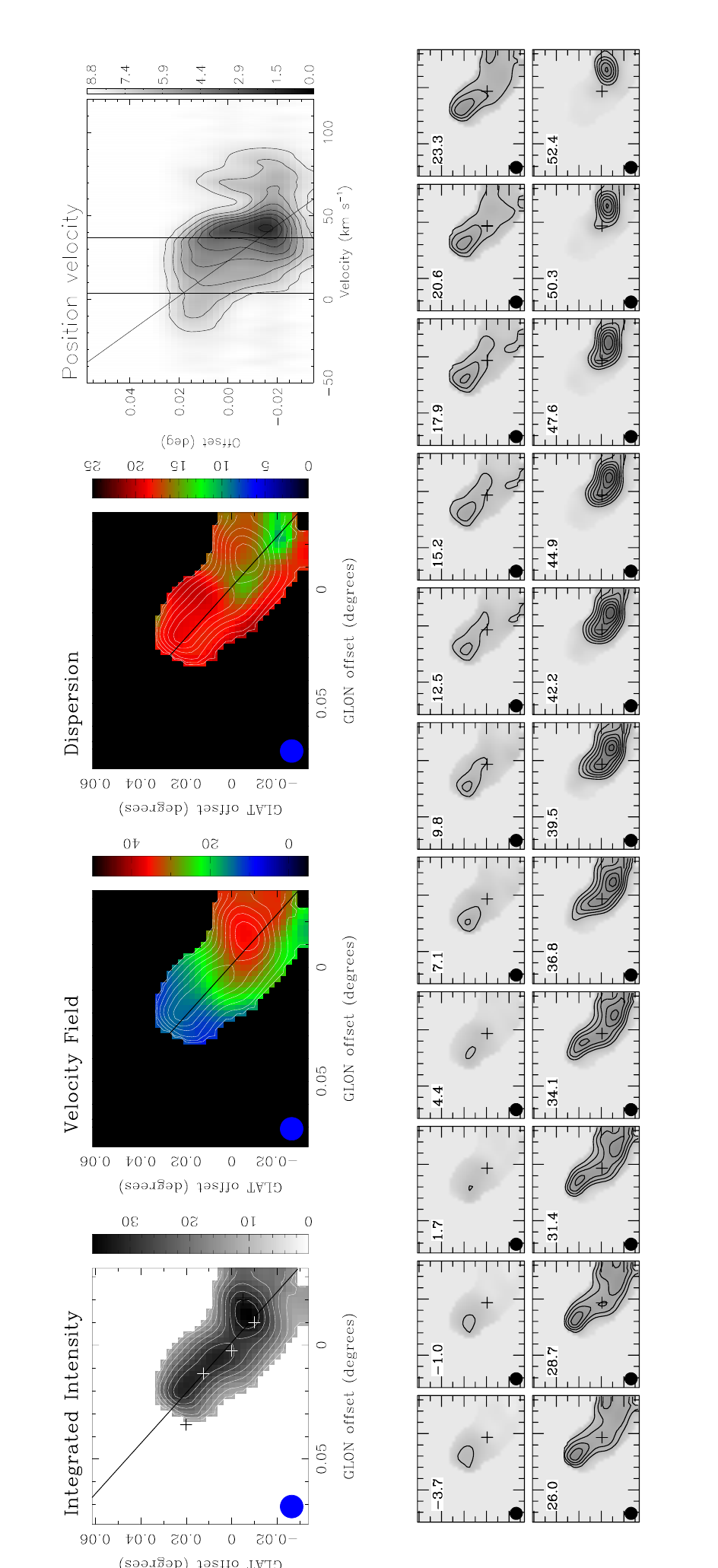}
\caption{\label{hcoptt}Moment maps, position-velocity diagram, and channel maps for \hcopnt\, (3--2). {\it {Upper row, from left to right:}} Integrated intensity map (M$_{0}$, units of \Kkms),  
intensity weighted velocity field (M$_{1}$, units of \kms),  intensity weighted dispersion (M$_{2}$, units of \kms), and position-velocity diagram showing the 
 emission along the major axis of \cloud\, (the emission was averaged over the clump's minor axis; contour levels are from 10 to 90\% of the peak; units of K).  Overlaid on 
 the moment maps are contours of the integrated intensity (levels are from 10 to 90\% of the peak in steps of 10\%) and a solid diagonal line marking 
the major axis of the clump.  The small crosses on the integrated intensity image mark the 5 positions a which spectra were extracted across the clump (P1 to P5, from left to right). Overlaid on the position-velocity diagram are solid vertical lines marking 
the velocities of the two components identified in the \hncofznt\, emission, the diagonal line shows the slope of the velocity gradient across \cloud.
Each panel of the channel map
shows the emission averaged over $\sim$6\,\kms\, around the listed velocity (contour levels are from 10 to 90\% of the peak in steps of 10\%). The small cross marks the peak in both
the dust continuum emission and column density. For the moment and channel maps the beam size is shown in the lower left corner.}
\end{sidewaysfigure}

\begin{sidewaysfigure}
\centering
%\hcopnt\,(4--3)
\includegraphics[angle=-90,width=0.95\textwidth,clip=true,trim=10mm 30mm 110mm 40mm]{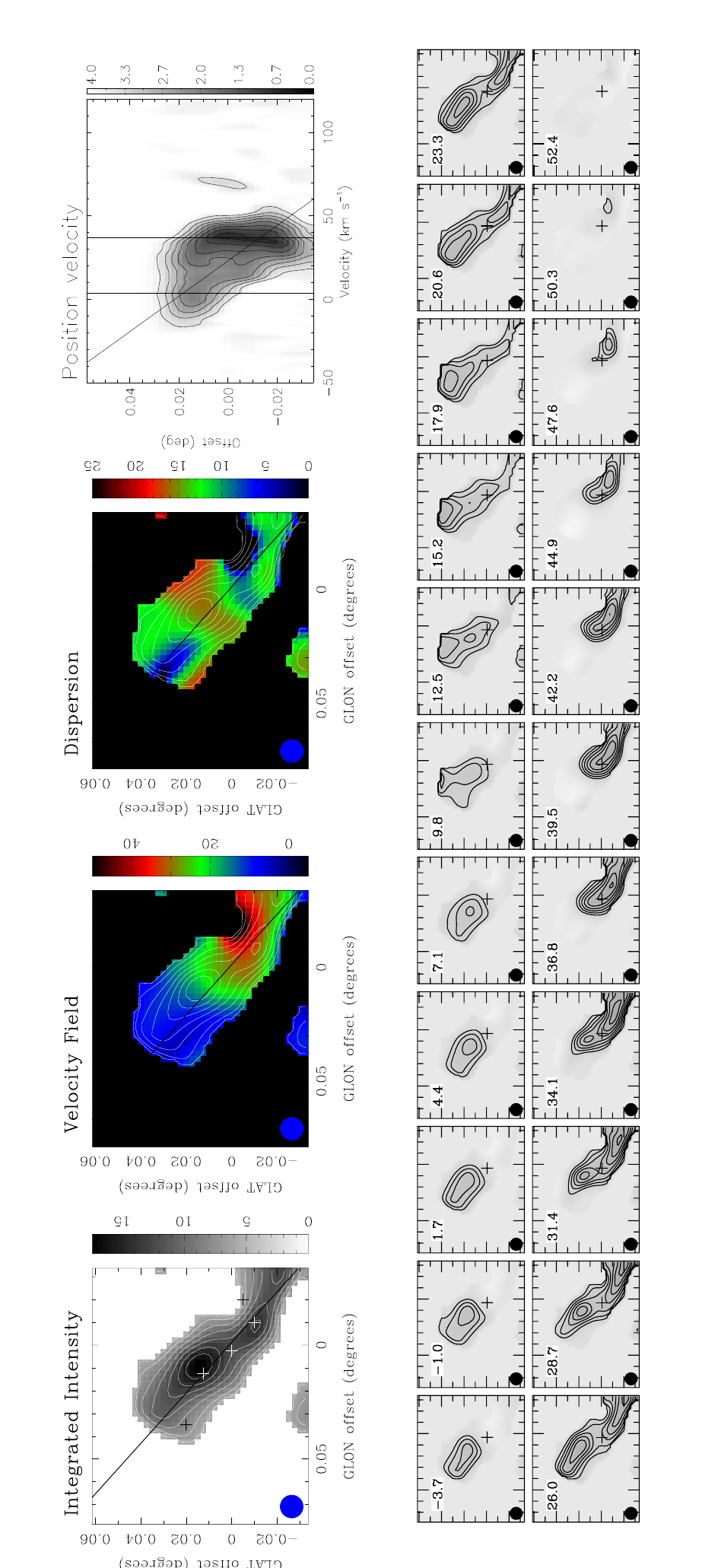}
\caption{\label{hcopft}Moment maps, position-velocity diagram, and channel maps for \hcopnt\, (4--3). {\it {Upper row, from left to right:}} Integrated intensity map (M$_{0}$, units of \Kkms),  
intensity weighted velocity field (M$_{1}$, units of \kms),  intensity weighted dispersion (M$_{2}$, units of \kms), and position-velocity diagram showing the 
 emission along the major axis of \cloud\, (the emission was averaged over the clump's minor axis; contour levels are from 10 to 90\% of the peak; units of K).  Overlaid on 
 the moment maps are contours of the integrated intensity (levels are from 10 to 90\% of the peak in steps of 10\%) and a solid diagonal line marking 
the major axis of the clump.  The small crosses on the integrated intensity image mark the 5 positions a which spectra were extracted across the clump (P1 to P5, from left to right). Overlaid on the position-velocity diagram are solid vertical lines marking 
the velocities of the two components identified in the \hncofznt\, emission, the diagonal line shows the slope of the velocity gradient across \cloud.
Each panel of the channel map
shows the emission averaged over $\sim$6\,\kms\, around the listed velocity (contour levels are from 10 to 90\% of the peak in steps of 10\%). The small cross marks the peak in both
the dust continuum emission and column density. For the moment and channel maps the beam size is shown in the lower left corner.}
\end{sidewaysfigure}

\begin{sidewaysfigure}
\centering
%\htcop
\includegraphics[angle=-90,width=0.95\textwidth,clip=true,trim=10mm 30mm 110mm 40mm]{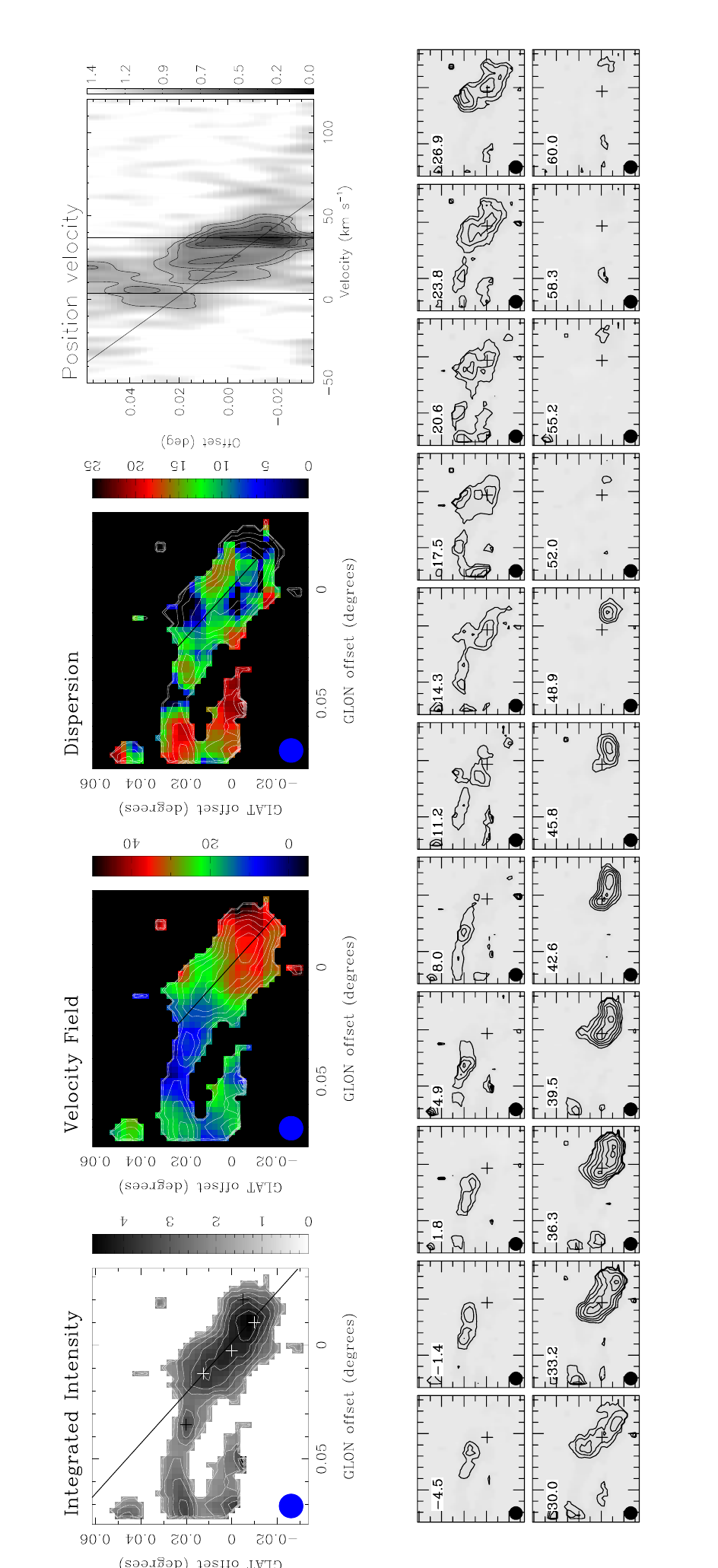}
\caption{\label{htcop}Moment maps, position-velocity diagram, and channel maps for \htcop. {\it {Upper row, from left to right:}} Integrated intensity map (M$_{0}$, units of \Kkms),  
intensity weighted velocity field (M$_{1}$, units of \kms),  intensity weighted dispersion (M$_{2}$, units of \kms), and position-velocity diagram showing the 
 emission along the major axis of \cloud\, (the emission was averaged over the clump's minor axis; contour levels are from 10 to 90\% of the peak; units of K).  Overlaid on 
 the moment maps are contours of the integrated intensity (levels are from 10 to 90\% of the peak in steps of 10\%) and a solid diagonal line marking 
the major axis of the clump.  The small crosses on the integrated intensity image mark the 5 positions a which spectra were extracted across the clump (P1 to P5, from left to right). Overlaid on the position-velocity diagram are solid vertical lines marking 
the velocities of the two components identified in the \hncofznt\, emission, the diagonal line shows the slope of the velocity gradient across \cloud.
Each panel of the channel map
shows the emission averaged over $\sim$6\,\kms\, around the listed velocity (contour levels are from 10 to 90\% of the peak in steps of 10\%). The small cross marks the peak in both
the dust continuum emission and column density. For the moment and channel maps the beam size is shown in the lower left corner.}
\end{sidewaysfigure}

\begin{sidewaysfigure}
\centering
%\tcs
\includegraphics[angle=-90,width=0.95\textwidth,clip=true,trim=10mm 30mm 110mm 40mm]{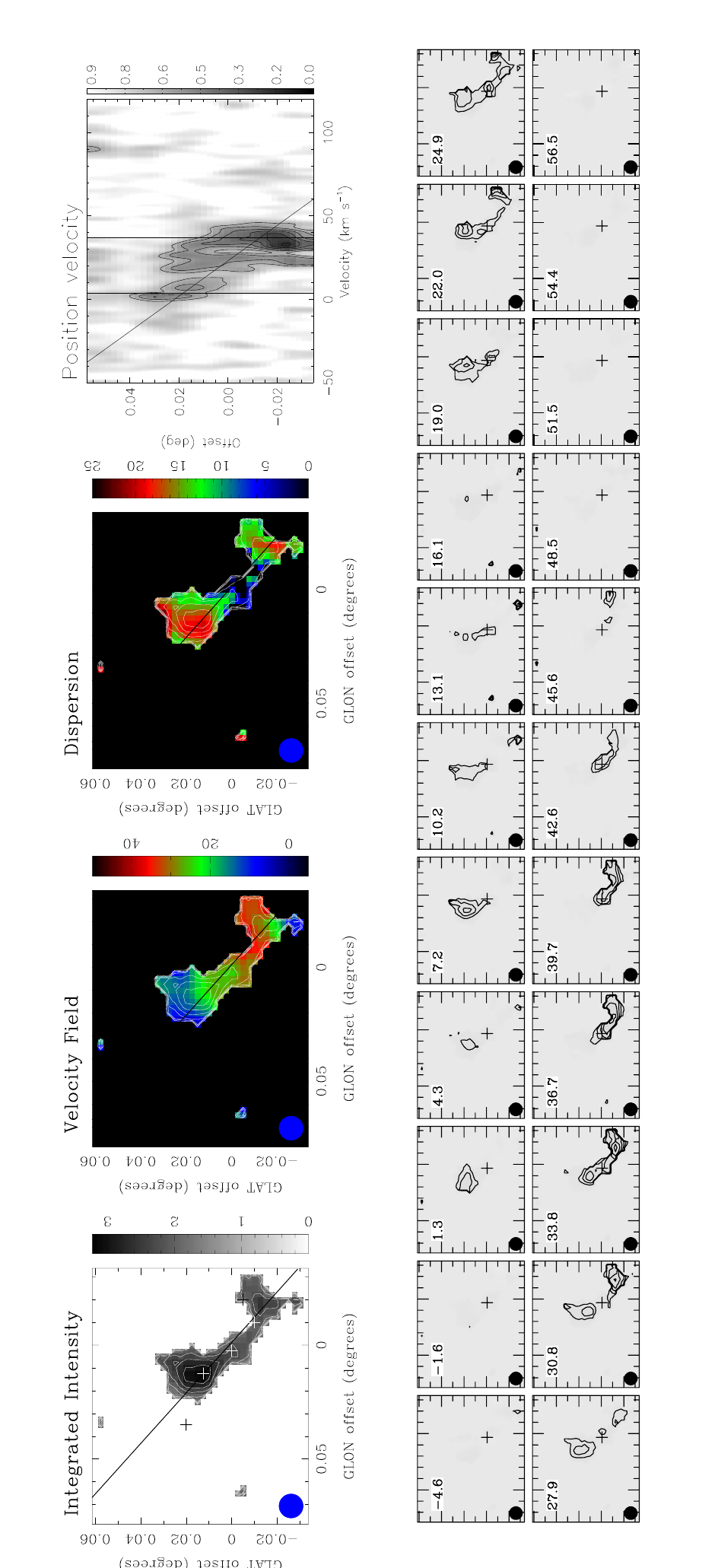}
\caption{\label{tcs}Moment maps, position-velocity diagram, and channel maps for \tcs. {\it {Upper row, from left to right:}} Integrated intensity map (M$_{0}$, units of \Kkms),  
intensity weighted velocity field (M$_{1}$, units of \kms),  intensity weighted dispersion (M$_{2}$, units of \kms), and position-velocity diagram showing the 
 emission along the major axis of \cloud\, (the emission was averaged over the clump's minor axis; contour levels are from 10 to 90\% of the peak; units of K).  Overlaid on 
 the moment maps are contours of the integrated intensity (levels are from 10 to 90\% of the peak in steps of 10\%) and a solid diagonal line marking 
the major axis of the clump.  The small crosses on the integrated intensity image mark the 5 positions a which spectra were extracted across the clump (P1 to P5, from left to right). Overlaid on the position-velocity diagram are solid vertical lines marking 
the velocities of the two components identified in the \hncofznt\, emission, the diagonal line shows the slope of the velocity gradient across \cloud.
Each panel of the channel map
shows the emission averaged over $\sim$6\,\kms\, around the listed velocity (contour levels are from 10 to 90\% of the peak in steps of 10\%). The small cross marks the peak in both
the dust continuum emission and column density. For the moment and channel maps the beam size is shown in the lower left corner.}
\end{sidewaysfigure}

\begin{sidewaysfigure}
\centering
%\sio
\includegraphics[angle=-90,width=0.95\textwidth,clip=true,trim=10mm 30mm 110mm 40mm]{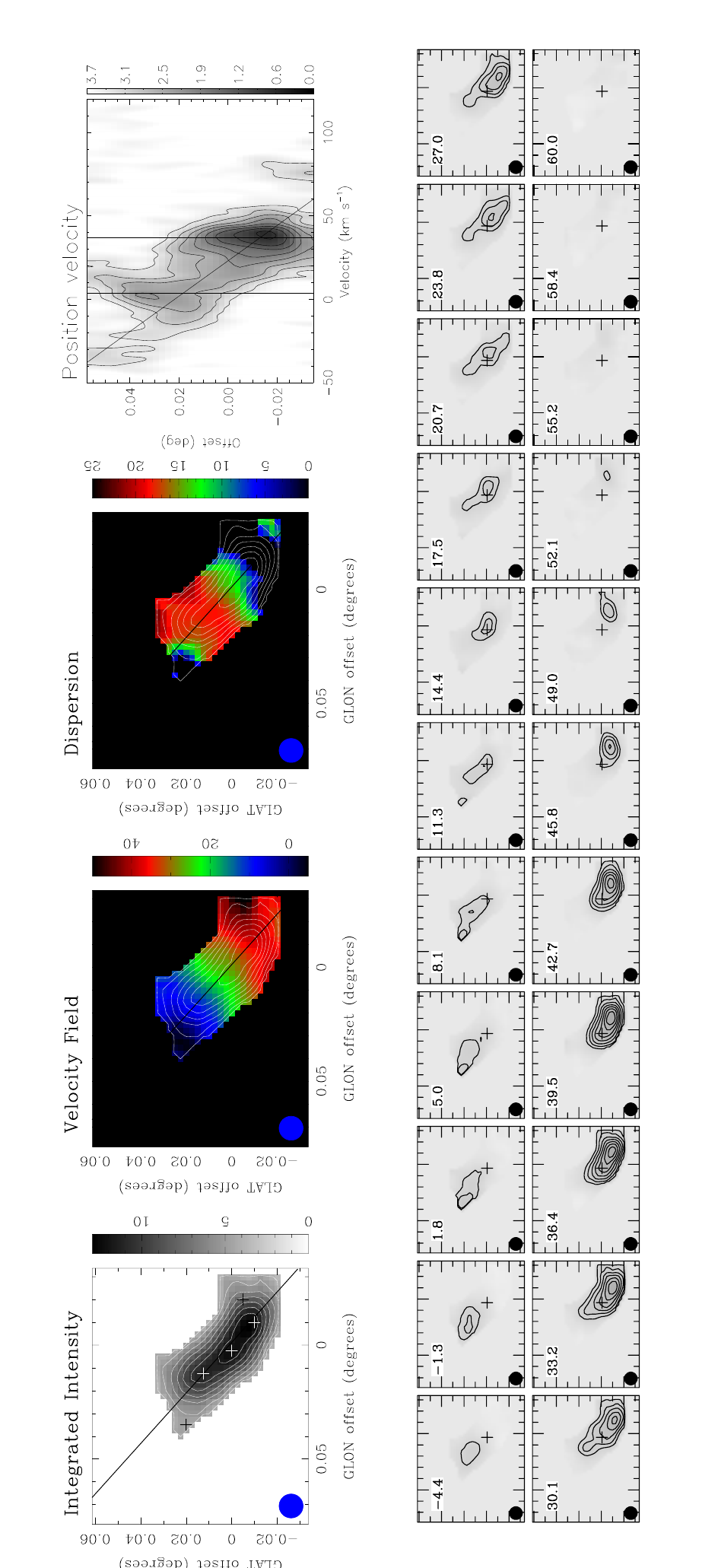}
\caption{\label{sioto}Moment maps, position-velocity diagram, and channel maps for \sio. {\it {Upper row, from left to right:}} Integrated intensity map (M$_{0}$, units of \Kkms),  
intensity weighted velocity field (M$_{1}$, units of \kms),  intensity weighted dispersion (M$_{2}$, units of \kms), and position-velocity diagram showing the 
 emission along the major axis of \cloud\, (the emission was averaged over the clump's minor axis; contour levels are from 10 to 90\% of the peak; units of K).  Overlaid on 
 the moment maps are contours of the integrated intensity (levels are from 10 to 90\% of the peak in steps of 10\%) and a solid diagonal line marking 
the major axis of the clump.  The small crosses on the integrated intensity image mark the 5 positions a which spectra were extracted across the clump (P1 to P5, from left to right). Overlaid on the position-velocity diagram are solid vertical lines marking 
the velocities of the two components identified in the \hncofznt\, emission, the diagonal line shows the slope of the velocity gradient across \cloud.
Each panel of the channel map
shows the emission averaged over $\sim$6\,\kms\, around the listed velocity (contour levels are from 10 to 90\% of the peak in steps of 10\%). The small cross marks the peak in both
the dust continuum emission and column density. For the moment and channel maps the beam size is shown in the lower left corner.}
\end{sidewaysfigure}

\begin{sidewaysfigure}
\centering
%\siont\,(5--4)
\includegraphics[angle=-90,width=0.95\textwidth,clip=true,trim=10mm 30mm 110mm 40mm]{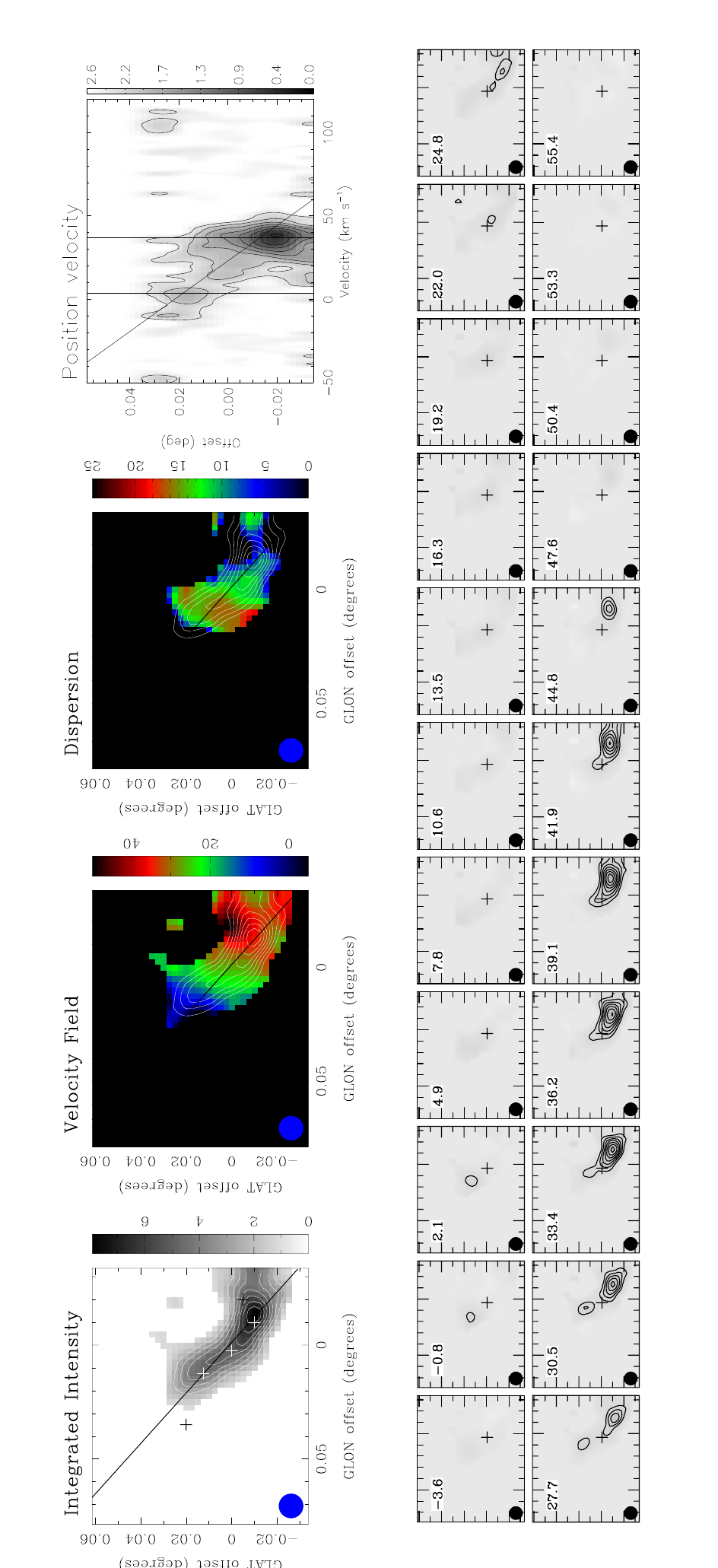}
\caption{\label{sioff}Moment maps, position-velocity diagram, and channel maps for \siont\, (5--4). {\it {Upper row, from left to right:}} Integrated intensity map (M$_{0}$, units of \Kkms),  
intensity weighted velocity field (M$_{1}$, units of \kms),  intensity weighted dispersion (M$_{2}$, units of \kms), and position-velocity diagram showing the 
 emission along the major axis of \cloud\, (the emission was averaged over the clump's minor axis; contour levels are from 10 to 90\% of the peak; units of K).  Overlaid on 
 the moment maps are contours of the integrated intensity (levels are from 10 to 90\% of the peak in steps of 10\%) and a solid diagonal line marking 
the major axis of the clump.  The small crosses on the integrated intensity image mark the 5 positions a which spectra were extracted across the clump (P1 to P5, from left to right). Overlaid on the position-velocity diagram are solid vertical lines marking 
the velocities of the two components identified in the \hncofznt\, emission, the diagonal line shows the slope of the velocity gradient across \cloud.
Each panel of the channel map
shows the emission averaged over $\sim$6\,\kms\, around the listed velocity (contour levels are from 10 to 90\% of the peak in steps of 10\%). The small cross marks the peak in both
the dust continuum emission and column density. For the moment and channel maps the beam size is shown in the lower left corner.}
\end{sidewaysfigure}

\begin{sidewaysfigure}
\centering
%\hctn
\includegraphics[angle=-90,width=0.95\textwidth,clip=true,trim=10mm 30mm 110mm 40mm]{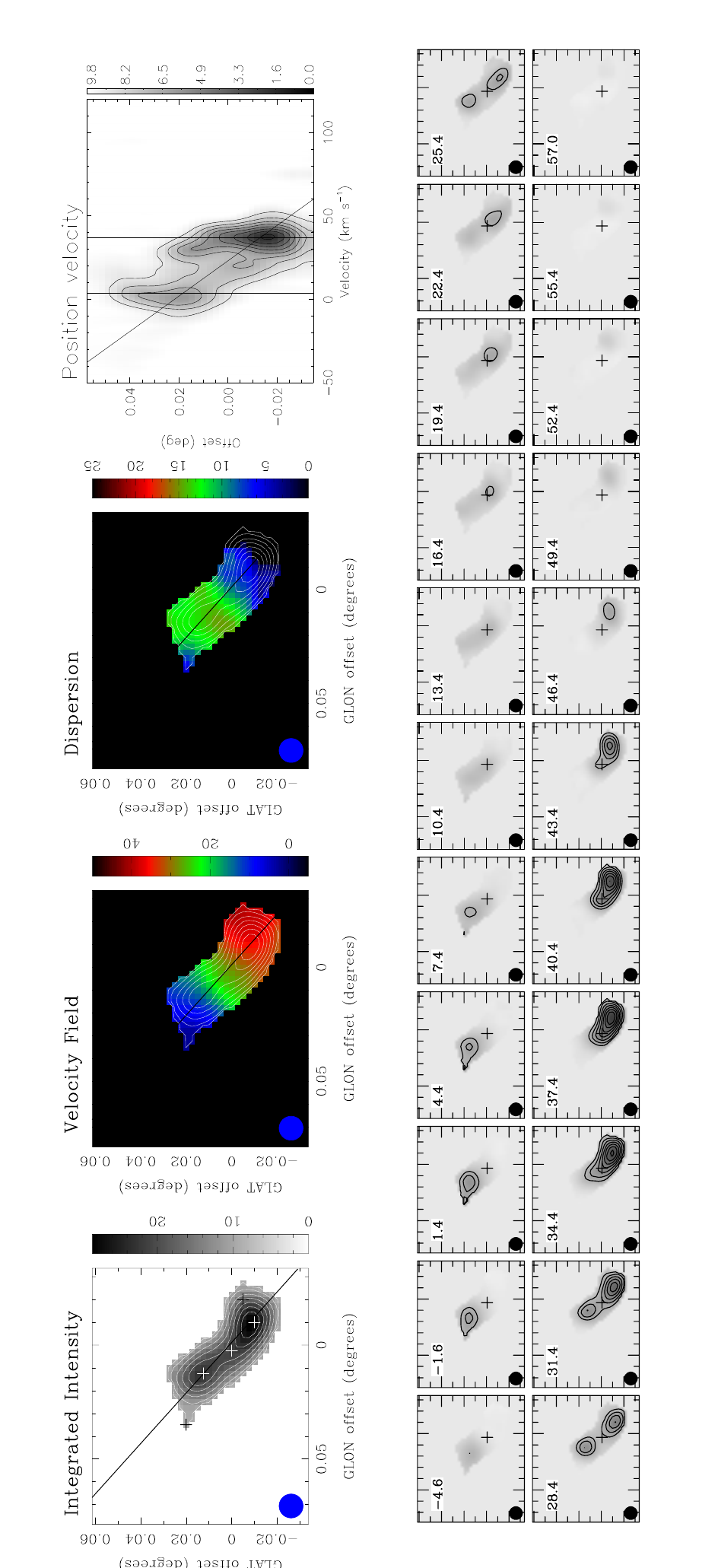}
\caption{\label{hctn}Moment maps, position-velocity diagram, and channel maps for \hctn. {\it {Upper row, from left to right:}} Integrated intensity map (M$_{0}$, units of \Kkms),  
intensity weighted velocity field (M$_{1}$, units of \kms),  intensity weighted dispersion (M$_{2}$, units of \kms), and position-velocity diagram showing the 
 emission along the major axis of \cloud\, (the emission was averaged over the clump's minor axis; contour levels are from 10 to 90\% of the peak; units of K).  Overlaid on 
 the moment maps are contours of the integrated intensity (levels are from 10 to 90\% of the peak in steps of 10\%) and a solid diagonal line marking 
the major axis of the clump.  The small crosses on the integrated intensity image mark the 5 positions a which spectra were extracted across the clump (P1 to P5, from left to right). Overlaid on the position-velocity diagram are solid vertical lines marking 
the velocities of the two components identified in the \hncofznt\, emission, the diagonal line shows the slope of the velocity gradient across \cloud.
Each panel of the channel map
shows the emission averaged over $\sim$6\,\kms\, around the listed velocity (contour levels are from 10 to 90\% of the peak in steps of 10\%). The small cross marks the peak in both
the dust continuum emission and column density. For the moment and channel maps the beam size is shown in the lower left corner.}
\end{sidewaysfigure}

\begin{sidewaysfigure}
\centering
%\chtcn
\includegraphics[angle=-90,width=0.95\textwidth,clip=true,trim=10mm 30mm 110mm 40mm]{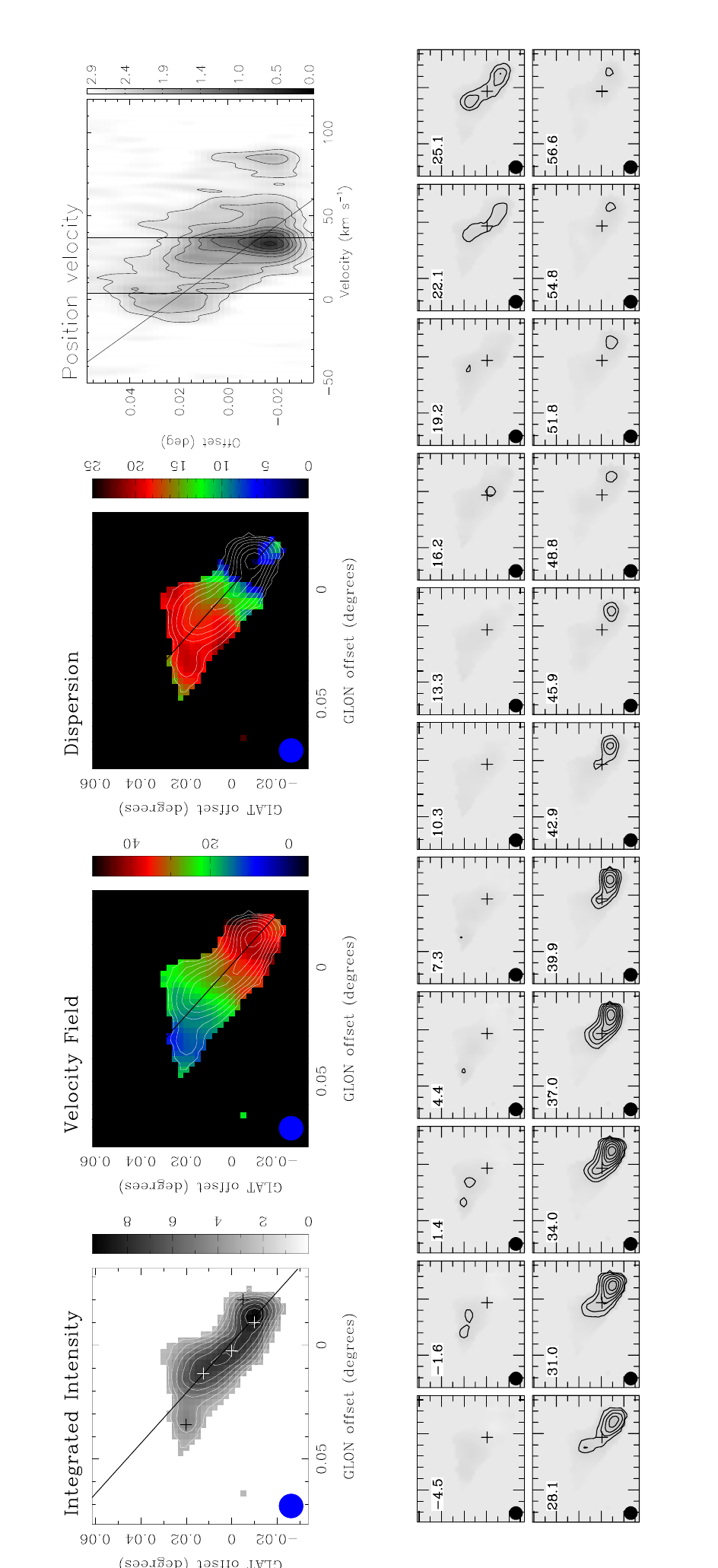}
\caption{\label{chtcn}Moment maps, position-velocity diagram, and channel maps for \chtcn. {\it {Upper row, from left to right:}} Integrated intensity map (M$_{0}$, units of \Kkms),  
intensity weighted velocity field (M$_{1}$, units of \kms),  intensity weighted dispersion (M$_{2}$, units of \kms), and position-velocity diagram showing the 
 emission along the major axis of \cloud\, (the emission was averaged over the clump's minor axis; contour levels are from 10 to 90\% of the peak; units of K).  Overlaid on 
 the moment maps are contours of the integrated intensity (levels are from 10 to 90\% of the peak in steps of 10\%) and a solid diagonal line marking 
the major axis of the clump.  The small crosses on the integrated intensity image mark the 5 positions a which spectra were extracted across the clump (P1 to P5, from left to right). Overlaid on the position-velocity diagram are solid vertical lines marking 
the velocities of the two components identified in the \hncofznt\, emission, the diagonal line shows the slope of the velocity gradient across \cloud.
Each panel of the channel map
shows the emission averaged over $\sim$6\,\kms\, around the listed velocity (contour levels are from 10 to 90\% of the peak in steps of 10\%). The small cross marks the peak in both
the dust continuum emission and column density. For the moment and channel maps the beam size is shown in the lower left corner.}
\end{sidewaysfigure}

\begin{sidewaysfigure}
\centering
%\
\includegraphics[angle=-90,width=0.95\textwidth,clip=true,trim=10mm 30mm 110mm 40mm]{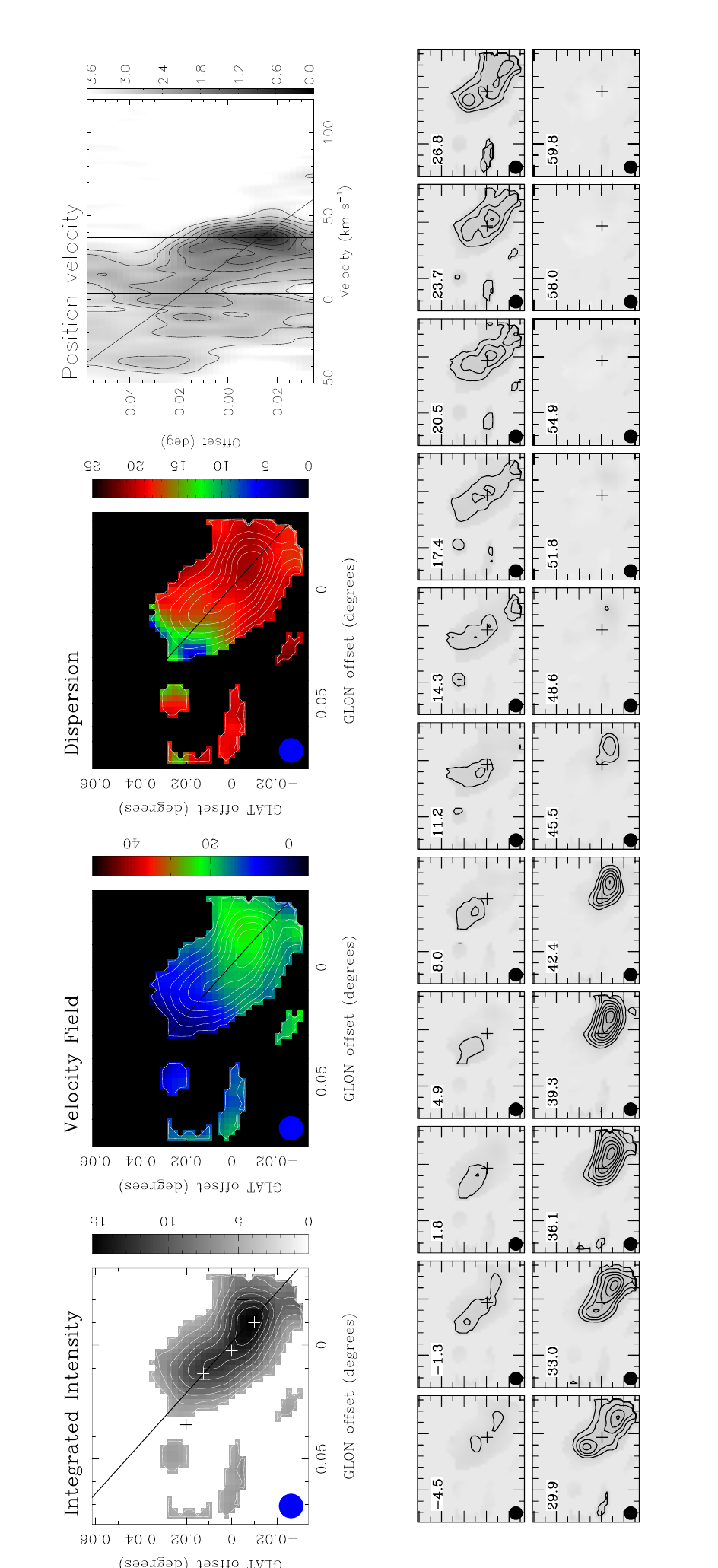}
\caption{\label{cth}Moment maps, position-velocity diagram, and channel maps for \cch. {\it {Upper row, from left to right:}} Integrated intensity map (M$_{0}$, units of \Kkms),  
intensity weighted velocity field (M$_{1}$, units of \kms),  intensity weighted dispersion (M$_{2}$, units of \kms), and position-velocity diagram showing the 
 emission along the major axis of \cloud\, (the emission was averaged over the clump's minor axis; contour levels are from 10 to 90\% of the peak; units of K).  Overlaid on 
 the moment maps are contours of the integrated intensity (levels are from 10 to 90\% of the peak in steps of 10\%) and a solid diagonal line marking 
the major axis of the clump.  The small crosses on the integrated intensity image mark the 5 positions a which spectra were extracted across the clump (P1 to P5, from left to right). Overlaid on the position-velocity diagram are solid vertical lines marking 
the velocities of the two components identified in the \hncofznt\, emission, the diagonal line shows the slope of the velocity gradient across \cloud.
Each panel of the channel map
shows the emission averaged over $\sim$6\,\kms\, around the listed velocity (contour levels are from 10 to 90\% of the peak in steps of 10\%). The small cross marks the peak in both
the dust continuum emission and column density. For the moment and channel maps the beam size is shown in the lower left corner.}
\end{sidewaysfigure}

Figure~\ref{hnco_mom_chan} shows, for \hncofznt, the three moment maps, the channel maps, 
and position-velocity ($pv$) diagram (similar figures for all other detected molecules transitions 
are included on-line, Figs.~\ref{nthpoz}--\ref{cth}).  If the emission arises from a single clump, 
then a velocity gradient is clearly evident from the intensity weighted velocity field.  Included on the 
position-velocity diagrams is a solid diagonal line marking the HNCO velocity gradient 
seen across \cloud\, ($\sim$18.6\,\kms\,pc$^{-1}$). 

\begin{figure}
\hspace{-1.0cm}
\includegraphics[width=1.1\textwidth,clip=true,trim=1mm 1mm 1mm 1mm]{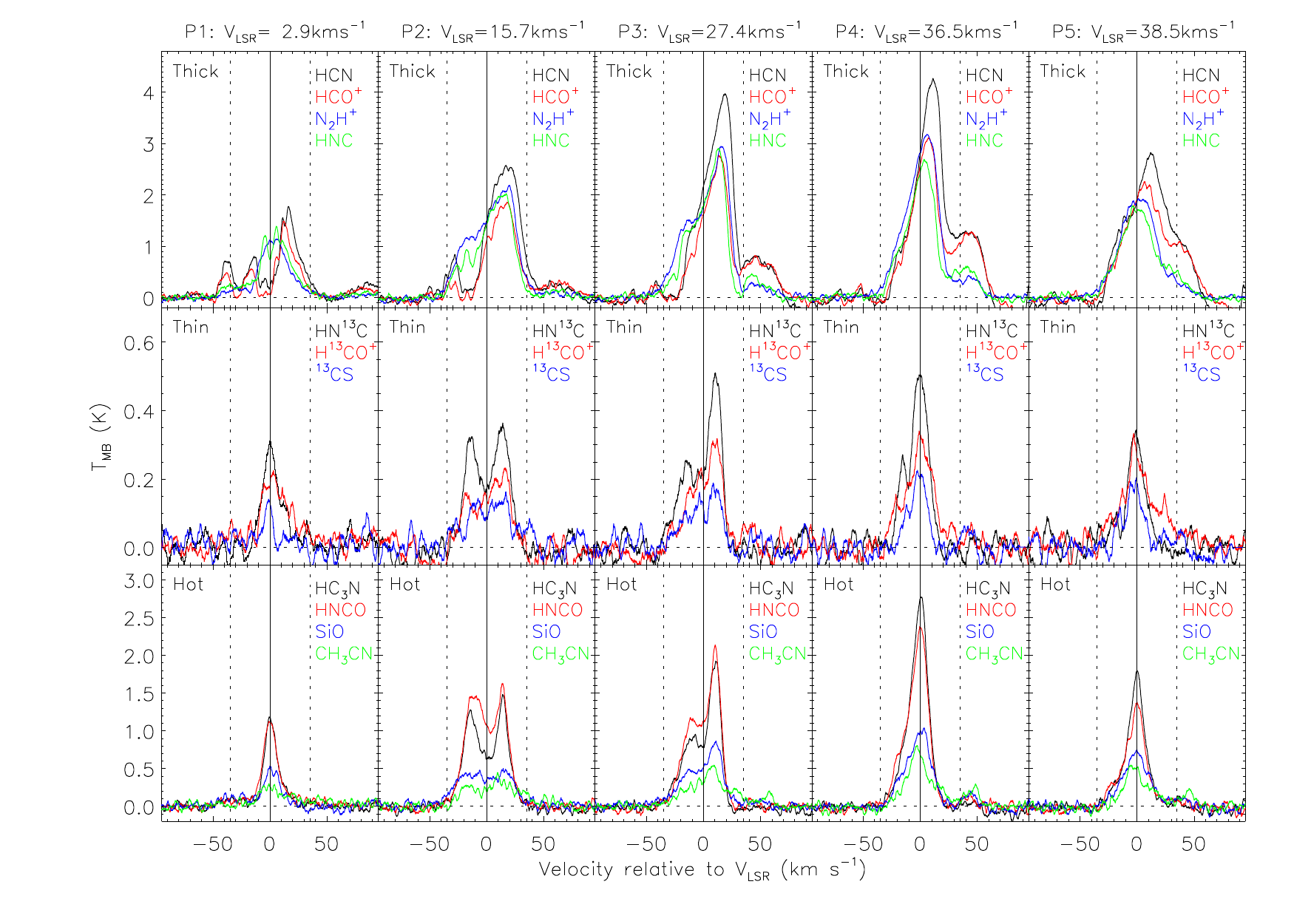}
\caption{\label{spectra-all} Spectra toward 5 positions across \cloud. The spectra were extracted 
	at the positions marked on the integrated intensity image shown in Figure~\ref{hnco_mom_chan} 
	and show details of the kinematics across the clump.  
	The molecules were grouped based on whether they typically trace emission that is likely to be 
	optically thick (upper panels, HCN, \hcopnt, \nthpnt\, and HNC), optically thin (middle panels, 
	\hntcnt, \htcopnt, and \tcsnt), or from hot/shocked gas (lower panels, \hctnnt, \hncofznt, SiO, 
	and \chtcnnt). The solid vertical line marks the derived \vlsr\, of the emission at that location, 
	derived from the \hncofznt\, intensity weighted velocity field (the \vlsr\, is marked at the top of each 
	column). The vertical dotted lines delineate the velocity range we attribute to the emission from 
	\cloud\, ($\pm$ 35\,\kms\, around the derived \vlsr). We find that the spectra from the optically thin 
	tracers and those from the hot/shocked gas have similar profiles. In contrast, spectra from the 
	optically thick gas tracers show that toward all positions the emission peaks red-ward of the other 
	species and of the derived \vlsr.}
\end{figure}

Figure~\ref{spectra-all} shows spectra at 5 positions across \cloud. The spectra were extracted 
at the positions marked on the integrated intensity image show in Figure~\ref{hnco_mom_chan} and 
were selected to show details of the kinematics across the clump.  We group the molecules into 
three separate panels based on whether they typically trace emission that is likely to be optically 
thick (\hcnnt, \hcopnt, \nthpnt, \hncnt; upper panels), optically thin (\hntcnt, \htcopnt, \tcsnt; middle panels), 
or from the higher excitation energy `hot core' molecules (\hctnnt, \hncofznt, \siont, \chtcnnt; lower panels).
Although the line profiles within each of these groups are quite similar, the profiles differ in
a systematic way between the optically thick lines and the other two optically thin and hot gas categories.  

To emphasize their systematic differences, we remove the clump's velocity gradient 
using the velocity determined from the intensity-weighted velocity field of \hncofznt\, (we select this 
transition as it shows well the velocity gradient and the emission is detected with a high signal to noise).
Plotting the spectra in this way simplifies their interpretation.

For each position, we plot the spectra on a velocity axis  relative to the derived \vlsr\, at 
that position. The derived \vlsr\, for each position is shown at the top of each column of Figure~\ref{spectra-all} 
and is marked by the solid vertical line in each spectrum at the velocity offset of 0\,\kms.  The 
vertical dotted lines delineate the velocity range we attribute to the emission from \cloud\, ($\pm$ 35\,\kms\, 
around the derived \vlsr). Evident in the \hcop\, spectra are two other molecular clouds along the line of 
sight (at \vlsr\, of $-$40 and 75\,\kms). These are unrelated clouds\footnote{While the 75\,\kms\, cloud is clearly 
detected in \hcop\, and \sio, we believe it is unrelated to 
\cloud. In the `elliptical ring' model presented recently by \cite{Molinari11} based on molecular line velocities, 
it is thought that the 75\,\kms\, cloud lies on the back side of the ring, compared to \cloud\, which lies on the front side. Moreover, the location 
of these two clouds lies close to the `cross over' point of the ring, where the front and back sides cross our line of sight.
The fact that \cloud\, is IR-dark while the 75\,\kms\, cloud is not, also argues that they are unrelated and probably 
at different physical locations along the line of sight.} and as such we exclude their 
emission from all subsequent analyses.  

Toward all selected positions, the spectra from the optically thin tracers and those 
from the hot/shocked gas have similar profiles: a single velocity component is seen toward the 
edge of the cloud (P1, P4 and P5), while toward its centre (P2 and P3) there are two apparent velocity 
components. In contrast, spectra from the optically thick gas tracers show that toward all positions 
the emission peaks red-ward of the other species and of the derived \vlsr.   The consistent pattern of 
optically thick lines peaking at more positive velocities than the optically thin or hot core 
lines persists regardless of the local systemic velocity of the cloud.
In other words, the redshift of the optically thick lines with respect to the optically thin and hot core lines is
independent of the overall velocity gradient of the cloud.  

The \hcopnt\, and \hcnnt\, profiles toward P1 and P2 show additional blue-shifted absorption at \vlsr\ 
near 0\,\kms.  Because all gas in circular Galactic orbits has a radial velocity of 0\,\kms\, toward the Galactic 
Center, any cold gas in the intervening 8.5\,kpc will absorb at this velocity.  Indeed,  a 0\,\kms\, absorption 
feature is widely seen throughout the CMZ and is most obvious from the molecules that are widespread, 
with low excitation energies (e.g. \hcopnt\, and \hcnnt; \citealp{Jones12}).

\begin{figure}
\hspace{-1.0cm}
\includegraphics[width=1.1\textwidth,clip=true,trim=1mm 60mm 1mm 1mm]{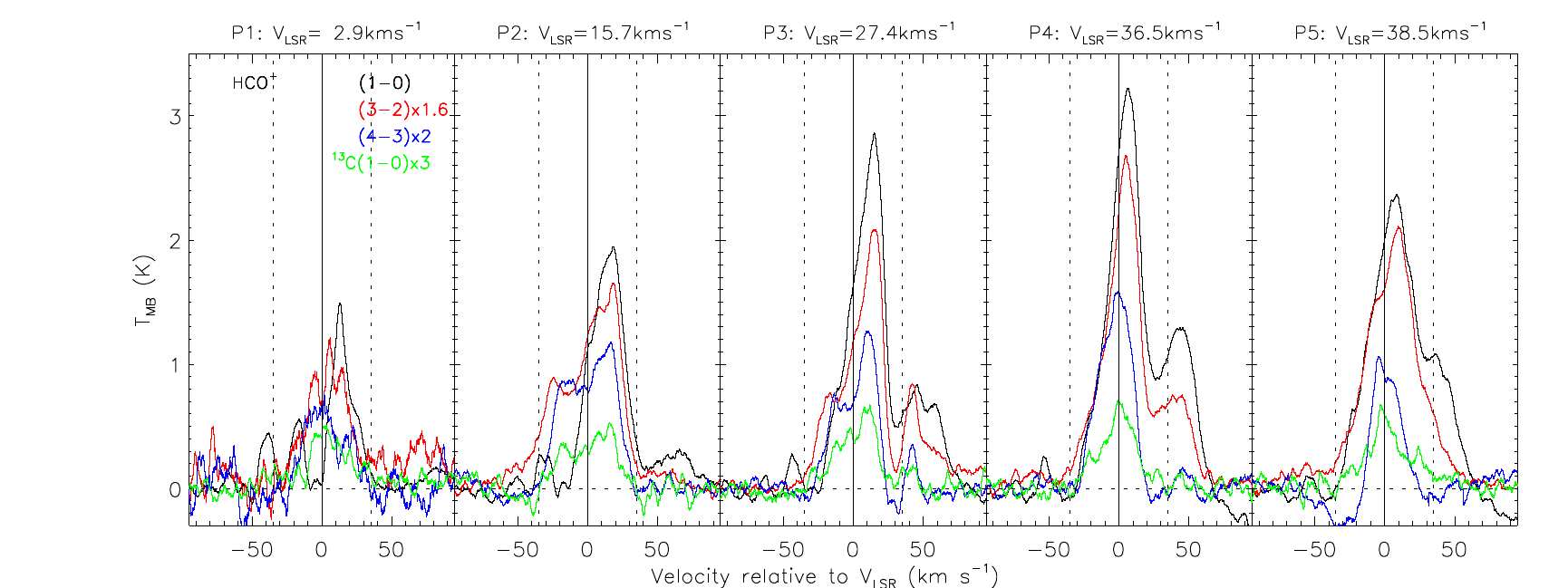}
\caption{\label{spectra-hcop} \hcopnt\, spectra from different transitions toward 5 positions across \cloud. 
	The spectra were extracted 	at the positions marked on the integrated intensity image shown in Figure~\ref{hnco_mom_chan}. The solid vertical line marks 
	the derived \vlsr\, of the emission at that location, derived from the \hncofznt\, intensity weighted velocity 
	field (the \vlsr\, is marked at the top of each column). The vertical dotted lines delineate the velocity 
	range we attribute to the emission from \cloud\, ($\pm$ 35\,\kms\, around the derived \vlsr).  Because 
	the molecular transitions are tracers of material at different critical densities, their apparent shift to more central  
	velocities in the higher density gas, is indicative of a density and velocity gradient of material that is centrally condensed.}
\end{figure}

Figure~\ref{spectra-hcop} compares the emission profiles from several different transitions of the same 
molecular species. We select \hcopnt\, as we have observed it in four different transitions and it does not have 
hyperfine components. Because the molecular transitions are tracers of material at different critical densities 
and excitation energies, their relative extent, location and velocity profiles can be used to trace density and 
temperature gradients within the cloud. The emission from the different transitions peak at different velocities: 
transitions with higher critical densities peak closer to the local systemic velocity derived from the optically thin lines.

\subsection{Small-scale structure}

Given an effective radius of 2.9\,pc and a dust-derived mass of $\sim$ 1.3 $\times$ 10$^{5}$\,\Msun, the 
average H$_2$ volume density of \cloud\, is $\sim$ 2$\times$10$^{4}$\,\cmc. For this density, the estimated 
free-fall time is $\sim$ 0.2\,Myrs. Because this is comparable to the turbulent crossing time of $\sim$ 0.4\,Myrs, 
the clump should be undergoing gravitational collapse and fragmentation.  Indeed, 
the 450\,\um\, dust continuum emission shows substructure down to the limit of the angular 
resolution of the data (7.5\arcsec; 0.3\,pc; \citealp{Longmore12}). 

With a kinetic temperature of $\sim$70 K and a volume density of $\sim$ 10$^{6}$\,\cmc\, RADEX models can 
be used to generate the expected line intensities. The input column density of HNC (10$^{15}$\,\cms) was 
estimated from the dust-derived column density measured toward  \cloud\, and assuming an abundance of 
HNC relative to \hh\, of 3$\times 10^{-9}$ \citep{Schilke92}. For this 
temperature and density, the predicted HNC (1--0), (3--2), and (4--3) line brightness temperatures are $\sim$48, $\sim$34, and 
$\sim$27 K respectively for homogeneous gas that fills the beam. These intensities are much higher 
than what is measured toward \cloud, which suggests 
that the gas could be highly clumped on small scales and that the smaller observed intensities result from low beam-filling factors.
Indeed, the channel 
maps clearly show that the emission has structure in both position and velocity (e.g.\,Fig.~\ref{hnco_mom_chan}).
While the angular resolution of our current data is insufficient to 
pinpoint the individual small-scale fragments, the data are consistent with a clump that is highly sub-structured.

The ratio of the predicted to measured line intensities can be used to
estimate the beam filling factor. We find values of $\sim$ 0.032, $\sim$ 0.025, and $\sim$ 0.019 for the 
HNC  (1--0), (3--2), and (4--3) lines respectively. The general trend of a higher filling factor for the lower J 
transitions is expected as the emission from the lower J transitions traces less dense, colder material that is presumably 
more widespread within the beam.

%While the filling factors for the (3--2) and (4--3) transitions are consistent with 
%this trend, they are not wildly different. Thus, our assumption of similar filling factors in these two transitions 
%when determining the gas densities from the line ratios is justified. 

The inferred clumping of the molecular emission, however, depends on the validity of the RADEX model assumption of uniform
gas properties throughout the cloud.  If multiple components of different temperatures and
densities exist along the line of sight, then the model
results may be invalid because the radiative transfer and optical depth effects are not properly accounted for.
For example, if the optically thick (1--0) lines are suppressed through self-absorption, RADEX would overpredict the
densities and brightness temperatures due to artificially high (3 --2)/(1 --0) line ratios.

Higher angular resolution observations are clearly needed to determine the
small-scale structure of \cloud. Despite their significant improvement in angular resolution, recent CARMA 
and SMA observations of this clump show a lack of dense gas on spatial scales of $\sim$ 2\arcsec, which 
led \cite{Kauffmann13} to speculate that the clump has insufficient dense material to form stars.  This result
is puzzling if the gas in fact is highly clumped with high brightness temperatures as the RADEX modelling suggests.    
Our recent ALMA cycle 0 observations of the 90\,GHz line and continuum emission toward \cloud\, show many 
compact fragments in dense gas tracers (1.7\arcsec; \citealp{Rathborne-alma}). The latter detection of the small-scale dense 
gas is a result of the improvement in sensitivity provided by ALMA and will be discussed in future work.

%%%%%%%%%%%%%%%%%%%%%%%%%%%%%%%%%%%%%%%%%%%%%%%%%%%%%%%%%
\section{Discussion}
%%%%%%%%%%%%%%%%%%%%%%%%%%%%%%%%%%%%%%%%%%%%%%%%%%%%%%%%%

The presence of cold dust, `hot core' chemistry, and complicated kinematics within \cloud\, is intriguing 
and may provide clues as to how this clump formed and whether it is on the verge of collapse. While the presence of cold dust, 
complex chemistry, and broad line-widths is well documented in the CMZ (e.g., \citealp{Wilson82,Martin-Pintado01}), 
\cloud\, is extreme compared to more typical clumps in the Galactic disk as it has a high-density, shows little 
star formation, and has sufficient mass to form a YMC through direct collapse. 

In this section we discuss the formation of \cloud\, and 
posit two scenarios that may explain the presence of the hot gas within \cloud\, with very 
different predictions for its distribution and kinematics. 

%While recent observations of the 
%dense gas on small scales suggests that \cloud\, may have little star formation potential \citep{Kauffmann13}, 
%three compact \hii\, regions consistent with being powered by B0.5 ZAMS stars have been detected in its 
%periphery \citep{Rodriguez13}.
\subsection{The formation of \cloud\, via a close passage to Sgr\,A$^{*}$}

%Although \cloud is the most high-mass clump that is a possible progenitor of a YMC, three additional, less
%high-mass clumps with
%the required characteristics have also been identified in the Central Molecular Zone \citep{Longmore13b}.  
%Despite intensive searches,
%no additional such clumps have
%been found outside of the CMZ.  These results suggest a link between the Galactic Center region and the
%formation of YMC progenitors.  

 \cite{Longmore13b} suggest that the formation of \cloud\, has been triggered by the pericentre 
passage of a gas stream close to the bottom of the Galactic gravitational potential near Sgr\,A$^{*}$, during 
which the gas is stretched in the orbital direction, but compressed in the direction perpendicular to its orbit. It is then argued 
that this compression leads to an accelerated dissipation of turbulent energy and hence triggers 
star formation. In this picture, \cloud\, recently passed pericentre $\sim$ 0.6\,Myrs ago and therefore should be on the verge of 
initiating star formation, whereas clouds like Sgr~B2 that passed pericentre 1--2 Myr ago should be 
exhibiting prevalent star formation. This scenario accounts for many of the observed properties of 
Galactic Center clouds.

%suggest the formation of \cloud\,  has been 
%triggered by the passage of a gas stream close to Sgr\,A$^{*}$.  Their models show that a clump
%passing near the large gravitational potential minimum of the Galaxy will be be compressed in the direction
%perpendicular to its orbit and stretched in the direction along the orbit.  These motions are due to the tidal effects
%induced by the large, concentrated mass of the nuclear bulge stars.  In this model, the compression leads to the triggering of star 
%formation.  For clouds like \cloud\, that have only recently passed  close by to Sgr\,A$^{*}$ ($\sim 0.6$ Myr ago), 
%star formation should just be beginning, and for clouds that passed several Myr earlier (i.e. Sgr B2), star formation should now 
%be much more active.  This model accounts for many of the overall properties of Galactic Center clouds. 

 If this scenario accurately describes \cloud, then its molecular line emission ought to 
show: (1) that the dynamical time scales of its motions are comparable to the time since pericentre passage 
($\sim$0.6 Myr in the model of \citealp{Molinari11}), (2) that it is elongated in the direction of its orbital motion, 
and (3) that there is evidence for bulk radial motions as the cloud is stretched and compressed.
 
Each of these predictions are consistent with the observed properties of \cloud. Firstly, the implied dynamical time 
scale for radial motions is $\sim 0.6$ Myr, in good agreement with the estimated time since closest passage 
to the Galactic Center in the \cite{Molinari11} model (the optically thick lines peak at velocities red-ward of the 
optically thin lines by $\sim 5$ \kms\, and its effective radius is 2.9 pc, leading to a dynamical time scale, 
$t = r/v$, of 0.6 Myr). Second, the morphology of \cloud\, is indeed elongated, in the predicted direction 
along its orbit.  Third, the observations clearly show bulk radial motions, as evidenced by the systematic redshift of 
the optically thick emission with respect to the optically thin and hot gas tracers.

 In the scenario where \cloud\, results from a pericentre passage, preliminary results from numerical simulations of this 
process (Kruijssen et al. in preparation) show that its center (within the tidal radius) will dissipate its turbulent energy at 
an accelerated rate and will therefore collapse, while its diffuse outer envelope (outside the tidal radius) will be stripped. 
Because both the collapsing center and the stripped envelope are characterised by radial motions, it is not clear whether the 
observed optically thick molecular lines are probing the collapsing center or the expanding envelope. 
Nevertheless, in either case, radial motions are clearly seen in the molecular line data presented here. The exact comparison to the 
numerical simulations will be presented in a future paper. 

\subsection{Two clumps colliding} 

One scenario for the presence of hot gas within  \cloud\, is that it has formed recently as a result of clump-clump
collisions. In this scenario, the shocks from the collisions are responsible for heating the gas. Indeed, widespread 
shocks, traced via SiO, are clearly associated with \cloud\, \citep{Lis01,Kauffmann13}. From large-scale CO mapping 
toward \cloud, \cite{Lis98} found evidence for a velocity gradient and complex kinematics within the clump that they 
interpret to be the superposition of two spatially overlapping components that may be interacting. In this collision/interaction 
scenario, the dust is not thermally coupled to the gas and, thus, has not yet had time to be heated. If true, then we 
would expect to see two velocity components in the dense gas tracers and  a central zone  of hot gas at the site of the collision. 
The hot/shocked gas should coincide with the position and velocity of the collision site.  

While our observations show two velocity components in the dense gas tracers toward the centre of the clump 
where presumably the components are interacting, the hot/shocked gas tracers are not isolated to this central 
region ( in position or velocity ). Instead, the hot/shocked gas has similar distribution and kinematics to the optically thin gas tracers and, 
thus, arises from similar regions across the clump. The fact that the emission from the hot/shocked gas is spread 
over the whole clump argues against this simple clump-clump collision scenario as a mechanism for producing the 
hot/shocked gas. There is no evidence for a distinct interaction zone.

\subsection{A centrally condensed clump, with depletion in its cold interior} 

An alternative scenario is one in which the clump is a single, coherent structure, with the hot gas distributed throughout.
When comparing the emission from different transitions of the same molecule, the integrated intensity images and 
position-velocity diagrams show that the (1--0) emission is more extended compared to the emission from the 
(3--2) and (4--3) transitions and isotopologues (for \hcopnt\, see Figs.~\ref{hcopoz}, \ref{hcoptt}, \ref{hcopft} and 
\ref{htcop}). Moreover, their spectra (Fig.~\ref{spectra-hcop}) show that the emission from the different transitions 
also peak at different velocities. Our data are consistent with a gradient in velocity that follows the critical density 
of the transitions: emission from the higher J transitions peak toward a more central velocity. This velocity shift 
combined with a decrease in the size of the emitting region in the higher density gas indicates a density
gradient of material that is centrally condensed.

Moreover, the observed anti-correlation between the dust column density and the molecular integrated intensity 
toward the clump's centre shows that the molecules are absent in the highest density region at the clump's center.  
One possible explanation for this disparity is molecular depletion in its cold interior.  Thus, the absence of emission 
from the optically thin species toward the clump's center and systemic velocity  may simply reflect depletion in the 
cloud's cold, dense interior.  If true, then the two velocity components observed in the both the optically thin and 
hot/shocked gas are not tracing physically distinct clumps, but are instead simply tracing the velocity fields of a 
centrally condensed clump. In this scenario, the apparent presence of two distinct velocity components arises 
from the lack of emission in the center of a large, extended clump with a smooth velocity gradient.

\subsection{Radial motions}

The fact that the observed velocity field of the  heated/shocked gas matches well the optically thin isotopologues 
implies that the emission from all of these  molecules are tracing the same material. In contrast, we find that the 
optically thick gas is always red-shifted with respect to the optically thin/hot gas tracers (e.g. Fig.~\ref{spectra-all}). 
Because the optically thick lines probe the $\tau = 1$ surface, their persistent redshift with respect to the optically 
thin and hot gas tracers demonstrates that the cloud has radial motions and that the gas properties vary along the line 
of sight.  Such asymmetries arise from the fact that the radial motions separate these distinct gas components since 
they have different radial velocities. Such an effect cannot arise from a foreground cloud; because of the large velocity 
gradient, such a foreground cloud would  have to have exactly the same velocity gradient to absorb at precisely the 
right velocity to maintain the constant red-ward asymmetry.

\subsubsection{P Cygni profile: an expanding, centrally condensed clump}

\begin{figure}
\centering
\includegraphics[width=0.95\textwidth,angle=0,clip=true,trim=1mm 1mm 1mm 1mm]{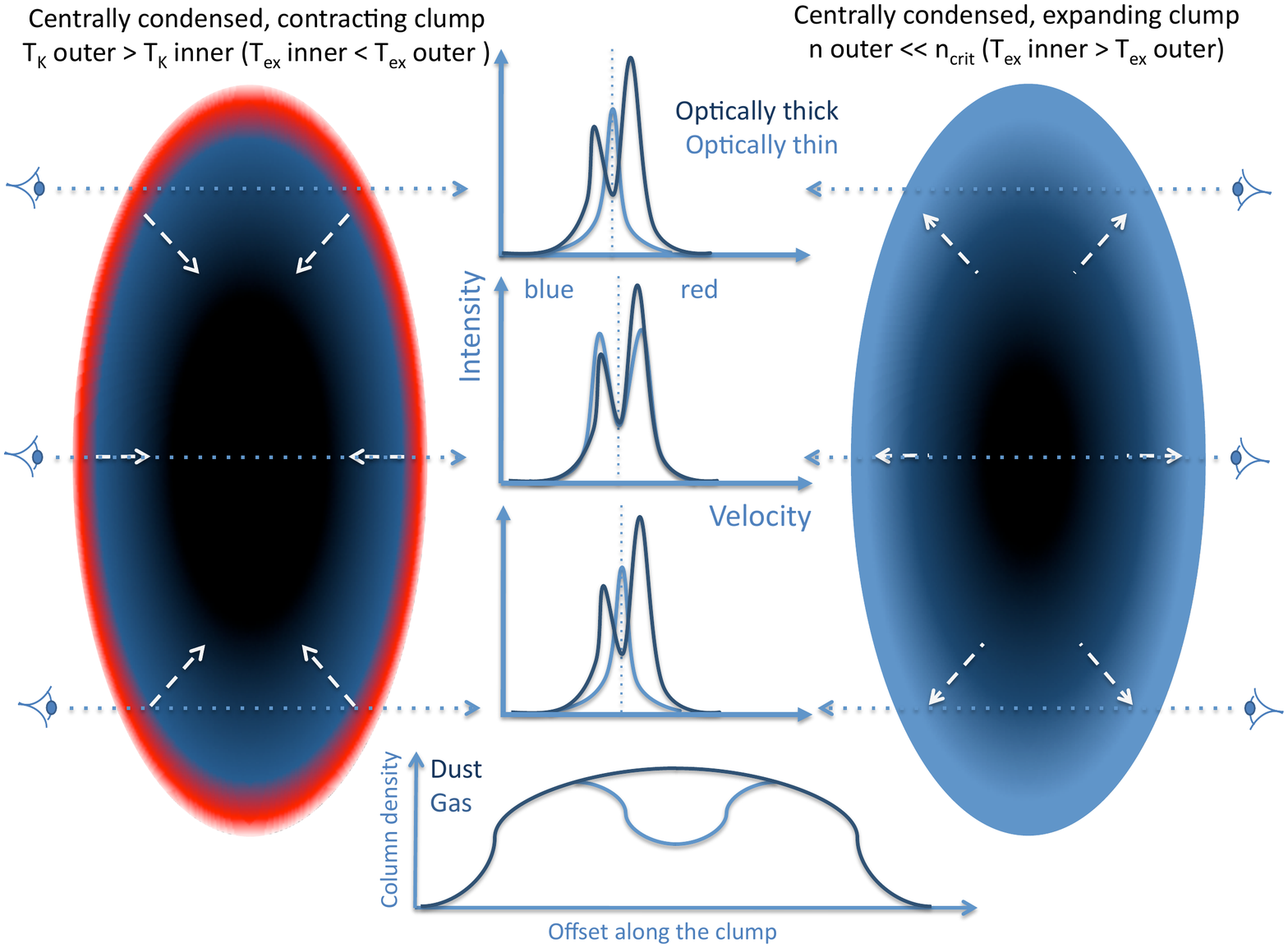}
\caption{\label{schematic}  Two alternative models for \cloud, consistent with both the apparent lack of 
molecular line emission at the central column density peak, and with the observed red-blue asymmetry in the optically 
thick molecular lines: externally heated, collapsing (left, Baked Alaska) vs 
sub-thermal excitation, expanding (right, P Cygni). In both scenarios the clump has a density gradient that increases 
toward its centre (white arrows show the gas motion). For an externally heated collapsing clump, the optically 
thick gas will show a red asymmetry in its line profile: the red velocity component comes from the warmer and 
brighter $\tau$=1 surface at the front of the cloud, while the blue component comes from the cooler and 
fainter $\tau$=1 surface toward the centre. In contrast, if the gas is expanding, then the brighter emission 
at red-shifted velocities arises from the interior of the cloud which has a significantly higher excitation 
temperature than the exterior portion. Given that the observed dust temperature implies that the clump is externally 
heated, this scenario requires sub-thermal excitation in the outer edges (n $\ll$ n$_{crit}$). 
In both cases, emission from an optically thin tracer will peak at the clump's  \vlsr.  However, if the gas is depleted in the clumps' centre,
	then the profile from the optically thin tracer will instead show two velocity components that peak in velocity either side of the 
	cloud's \vlsr. The distribution of dust and gas emission will not be correlated toward the cloud's centre; instead, the 
	gas will be deficient compared to the dust toward the centre.}
\end{figure}

One standard interpretation of this red-ward asymmetry is a `P Cygni'  type line profile due to expanding motions 
(see Fig.~\ref{schematic} for a schematic).  
In this interpretation,  the blue-shifted material arises from the outer surface of the cloud, and the red-shifted 
material from the cloud's interior.  The brighter emission at red-shifted velocities then arises from the fact that 
the interior of the cloud has a significantly higher excitation temperature than the exterior portion 
(T$_{ex}$ inner $>$ T$_{ex}$ outer).  The lower excitation temperature in the exterior layers could be due to colder 
temperatures in the exterior if the gas is thermalized (n $>>$ n$_{crit}$), or to sub-thermal excitation 
(n $<<$ n$_{crit}$).  The lower excitation exterior will not only be fainter than the higher excitation temperature interior, it can also absorb 
line emission from the interior. Combined, these effects lead to substantial red-blue asymmetries 
in the line profiles of optically thick gas.

\subsubsection{Baked Alaska: a collapsing, centrally condensed clump}

An alternative to this `P Cygni' interpretation is a `Baked Alaska' collapse model where the clump is 
externally heated (T$_{K}$ outer $>$ T$_{K}$ inner).  For a collapsing cloud, redshifted optically 
thick emission arises from the cloud's exterior and blue-shifted emission from the cloud's interior.  If 
\cloud\, is indeed collapsing, the brighter redshifted emission arises from the fact that the
exterior layers have a higher excitation temperature than the cloud's interior (see Fig.~\ref{schematic} for 
a schematic).  In this scenario, the gas is thermalized throughout the cloud (n$>>$n$_{crit}$).
This hot-exterior, cold-interior `Baked Alaska' model is consistent with the
%This is in contrast to the typical asymmetry of an optically thick molecular line toward star forming clumps, with a brighter blue velocity 
%component relative to the red which arises when the contracting material is centrally heated.
%The dust temperatures in \cloud\, are in fact higher in the outer portions of the cloud, which is consistent 
%with an externally heated cloud. 
recent observations and SPH modelling of the dust and gas temperature distribution in \cloud: both indicate that 
the clump is externally heated \citep{Lis01,Longmore12,Clark13}. 
Thus, there is some evidence to support the idea that the radial motions 
indicate collapse, a scenario which also accounts nicely for the extremely high mass and density of the cloud.

\subsection{The cluster formation potential of \cloud\, and the implications for the formation of high-mass, bound clusters}

%this is a single clump, that has the potential to form a star cluster (not simply the collision between two smaller clumps)

The fact that \cloud\, may be unique in the Galaxy in terms of its high density, high mass, and 
lack of prevalent star formation may be important for understanding how high-mass, bound clusters are formed. While the 
simplest idea for the formation of a high-mass cluster is through direct collapse of a large, high-mass, dense and 
cold molecular clump progenitor, there may in fact be multiple channels for the formation of high-mass star 
clusters. Rather than the simple collapse of a single molecular clump,  high-mass clusters may form more 
slowly via gradual star formation as gas is continually accreted onto a central potential well from a more 
extend reservoir of material. While plausible, the observed lack of age spreads in young 
high-mass clusters argues against this formation scenario \citep{Clark05,Negueruela10,Kudryavtseva12}.
Massive clusters may also form through mergers of smaller groups 
of stars that have already formed but are part of a common potential well (e.g., \citealp{McMillan07,Allison09}).

The observed gas morphology and kinematics suggest that \cloud\, is indeed a single, coherent high-mass dense, 
clump that may be highly fragmented on small spatial scales. As such, it has the potential to form a 
high-mass cluster. The puzzle, however, is how a dense, 10$^{5}$ \Msun\, 
clump forms without rapidly producing stars. For \cloud, its location in the CMZ may provide 
the clue: the increased turbulence in the CMZ may in fact inhibit star formation below a column density threshold which 
is much higher in the CMZ than in the Galactic plane \citep{Kruijssen13a}.
However, the fact that some of the most massive Galactic YMCs are located outside of the Galactic centre region 
(e.g., Westerlund 1, Glimpse-C01, NGC 3603,  RSGCs) suggests that the unique conditions within the 
CMZ are not essential for their formation. 

High-mass clusters are thought to form quickly, perhaps in less than a few  Myrs. Recent observations of
NGC  3603 and Westerlund 1 suggests that the age spread within the stellar population may be 
$<$0.4 Myr \citep{Clark05,Negueruela10,Kudryavtseva12}. The fact that \cloud\, is one of a handful of clumps  
with sufficient mass to form a YMC may support this fast formation scenario; at any given time the number of 
high-mass cluster precursors within the Galaxy will be limited to just a few \citep{Longmore12}. If true, then perhaps 
we are catching \cloud\, at a very special time, immediately before the formation of the high-mass cluster. 
In this case, we would expect the gas and dust to be highly fragmented, the fragments being the 
precursors to the individual stars or sub-clusters. Recent models do predict that \cloud\, should form a bound cluster 
through hierarchical fragmentation \citep{Kruijssen12b}. While our data show evidence for fragmentation, 
higher-angular resolution data of both the optically thin and shocked gas tracers down to small scales ($<$ 0.1pc)
are clearly needed to definitively determine whether or not  \cloud\, with give rise to a cluster in the future.

%%%%%%%%%%%%%%%%%%%%%%%%%%%%%%%%%%%%%%%%%%%%%%%%%%%%%%%%%
\section{Summary}
%%%%%%%%%%%%%%%%%%%%%%%%%%%%%%%%%%%%%%%%%%%%%%%%%%%%%%%%%

Utilizing molecular line data from the MALT90 survey combined with complementary APEX observations, 
we have investigated the global conditions and kinematics of the gas within \cloud. These data reveal 
a wealth of information because they combine molecular transitions that cover a broad range in critical 
densities and excitation energies and also include a number of key transitions that trace regions of complex 
chemistry and shocked gas.

While \cloud\, appears dark in the mid- to far-IR, implying both a low dust temperature and high 
column density, the presence of widespread emission from tracers of hot gas suggests that its gas temperature 
may be significantly higher than its measured dust temperature, consistent with previous observations. The 
data show that the gas has sub-structure and complex kinematics. Both the observed broad line-widths and 
the presence of shocks and complex molecular line emission from within \cloud\, is consistent with its location in the CMZ. 
The observed morphology and kinematics of the molecular line emission are tracing gas within 
a single, centrally condensed clump. The absence of gas emission toward the large dust column density peak
at the clump's centre probably results from gas depletion in its cold interior.  The systematic red-shift of
the optically thick transitions compared with the optically thin and hot gas tracers demonstrate
that \cloud\, exhibits radial motions.  If expanding, the outer portions have lower excitation temperatures
than the inner portions.  If contracting, the outer portions have higher excitation temperatures than the
inner portions.  Because the dust temperatures do in fact indicate external heating, the collapse model
is consistent with the observations.  No matter which interpretation is correct, the fact that the optically thick lines all peak at redder velocities than
the optically thin and hot core lines independent of the local systemic velocity clearly demonstrates systematic radial motions
in \cloud.   

With such a high density \cloud\, should be undergoing 
gravitational collapse and fragmentation, however, its rapid star formation may have been delayed 
due to the high turbulence and increased gas temperatures in the CMZ.  Single component
excitation models suggest, from the measured line 
ratios, that the gas appears to be clumped: higher angular resolution data is 
clearly needed to pinpoint these individual regions.

Because \cloud\, is an excellent candidate for the progenitor of a high-mass cluster,  reliably determining 
the distribution, size, mass, and motion of its small-scale fragments will be critical to reveal
important clues about the resulting stellar cluster and how such a cluster is formed. Future ALMA 
observations will reveal these properties: only with this significant increase in sensitivity, angular 
resolution, and dynamic range can we begin to measure the initial conditions within protoclusters 
and test cluster formation models.

%%%%%%%%%%%%%%%%%%%%%%%%%%%%%%%%%%%%%%%%%%%%%%%%%%%%%%%%%%%%%%%%%%%%%%%%%%%%%%%%%%%%
\acknowledgments
We thank the referee for their detailed comments which have improved the clarity of the paper considerably. 
We are grateful to both Peter Schilke and Thomas Moeller for assistance and advice using the RADEX radiative transfer modelling code.
This publication is based on data acquired with the Mopra radio telescope and the Atacama Pathfinder Experiment (APEX).
The Mopra radio telescope is part of the Australia Telescope National Facility which is funded by the 
Commonwealth of Australia for operation as a National Facility managed by CSIRO. 
The University of New South Wales Digital Filter Bank used for the observations with the Mopra 
Telescope was provided with support from the Australian Research Council.
APEX is a collaboration between the Max-Planck-Institut fur Radioastronomie, the European Southern Observatory, and the Onsala Space Observatory

%%%%%%%%%%%%%%%%%%%%%%%%%%%%%%%%%%%%%%%%%%%%%%%%%%%%%%%%%%%%%%%%%%%%%%%%%%%%%%%%%%%%
{\it Facilities:} \facility{Mopra, APEX}.
%%%%%%%%%%%%%%%%%%%%%%%%%%%%%%%%%%%%%%%%%%%%%%%%%%%%%%%%%%%%%%%%%%%%%%%%%%%%%%%%%%%%
%\bibliography{../../apjmnemonic,../../bibtex-references}

\begin{thebibliography}{57}
\expandafter\ifx\csname natexlab\endcsname\relax\def\natexlab#1{#1}\fi

\bibitem[{{Allison} {et~al.}(2009){Allison}, {Goodwin}, {Parker}, {de Grijs},
  {Portegies Zwart}, \& {Kouwenhoven}}]{Allison09}
{Allison}, R.~J., {Goodwin}, S.~P., {Parker}, R.~J., {de Grijs}, R., {Portegies
  Zwart}, S.~F., \& {Kouwenhoven}, M.~B.~N. 2009, \apjl, 700, L99

\bibitem[{{Ao} {et~al.}(2013){Ao}, {Henkel}, {Menten}, {Requena-Torres},
  {Stanke}, {Mauersberger}, {Aalto}, {M{\"u}hle}, \& {Mangum}}]{Ao13}
{Ao}, Y., {Henkel}, C., {Menten}, K.~M., {Requena-Torres}, M.~A., {Stanke}, T.,
  {Mauersberger}, R., {Aalto}, S., {M{\"u}hle}, S., \& {Mangum}, J. 2013, \aap,
  550, A135

\bibitem[{{Bania} {et~al.}(1986){Bania}, {Stark}, \& {Heiligman}}]{Bania86}
{Bania}, T.~M., {Stark}, A.~A., \& {Heiligman}, G.~M. 1986, \apj, 307, 350

\bibitem[{{Bastian}(2008)}]{Bastian08}
{Bastian}, N. 2008, \mnras, 390, 759

\bibitem[{{Breen} \& {Ellingsen}(2011)}]{Breen11}
{Breen}, S.~L. \& {Ellingsen}, S.~P. 2011, \mnras, 416, 178

\bibitem[{{Churchwell}(2002)}]{Churchwell02}
{Churchwell}, E. 2002, \araa, 40, 27

\bibitem[{{Clark} {et~al.}(2005){Clark}, {Negueruela}, {Crowther}, \&
  {Goodwin}}]{Clark05}
{Clark}, J.~S., {Negueruela}, I., {Crowther}, P.~A., \& {Goodwin}, S.~P. 2005,
  \aap, 434, 949

\bibitem[{{Clark} {et~al.}(2013){Clark}, {Glover}, {Ragan}, {Shetty}, \&
  {Klessen}}]{Clark13}
{Clark}, P.~C., {Glover}, S.~C.~O., {Ragan}, S.~E., {Shetty}, R., \& {Klessen},
  R.~S. 2013, \apjl, 768, L34

\bibitem[{{Contreras} {et~al.}(2013){Contreras}, {Schuller}, {Urquhart},
  {Csengeri}, {Wyrowski}, {Beuther}, {Bontemps}, {Bronfman}, {Henning},
  {Menten}, {Schilke}, {Walmsley}, {Wienen}, {Tackenberg}, \&
  {Linz}}]{Contreras13}
{Contreras}, Y., {Schuller}, F., {Urquhart}, J.~S., {Csengeri}, T., {Wyrowski},
  F., {Beuther}, H., {Bontemps}, S., {Bronfman}, L., {Henning}, T., {Menten},
  K.~M., {Schilke}, P., {Walmsley}, C.~M., {Wienen}, M., {Tackenberg}, J., \&
  {Linz}, H. 2013, \aap, 549, A45

\bibitem[{{Davies} {et~al.}(2011){Davies}, {Bastian}, {Gieles}, {Seth},
  {Mengel}, \& {Konstantopoulos}}]{Davies11}
{Davies}, B., {Bastian}, N., {Gieles}, M., {Seth}, A.~C., {Mengel}, S., \&
  {Konstantopoulos}, I.~S. 2011, \mnras, 411, 1386

\bibitem[{{Elmegreen} \& {Efremov}(1997)}]{Elmegreen97}
{Elmegreen}, B.~G. \& {Efremov}, Y.~N. 1997, \apj, 480, 235

\bibitem[{{Figer} {et~al.}(1999){Figer}, {Kim}, {Morris}, {Serabyn}, {Rich}, \&
  {McLean}}]{Figer99}
{Figer}, D.~F., {Kim}, S.~S., {Morris}, M., {Serabyn}, E., {Rich}, R.~M., \&
  {McLean}, I.~S. 1999, \apj, 525, 750

\bibitem[{{Figer} {et~al.}(2006){Figer}, {MacKenty}, {Robberto}, {Smith},
  {Najarro}, {Kudritzki}, \& {Herrero}}]{Figer06}
{Figer}, D.~F., {MacKenty}, J.~W., {Robberto}, M., {Smith}, K., {Najarro}, F.,
  {Kudritzki}, R.~P., \& {Herrero}, A. 2006, \apj, 643, 1166

\bibitem[{{Foster} {et~al.}(2011){Foster}, {Jackson}, {Barnes}, {Barris},
  {Brooks}, {Cunningham}, {Finn}, {Fuller}, {Longmore}, {Mascoop}, {Peretto},
  {Rathborne}, {Sanhueza}, {Schuller}, \& {Wyrowski}}]{Foster-malt90}
{Foster}, J.~B., {Jackson}, J.~M., {Barnes}, P.~J., {Barris}, E., {Brooks}, K.,
  {Cunningham}, M., {Finn}, S.~C., {Fuller}, G.~A., {Longmore}, S.~N.,
  {Mascoop}, J.~L., {Peretto}, N., {Rathborne}, J., {Sanhueza}, P., {Schuller},
  F., \& {Wyrowski}, F. 2011, \apjs, 197, 25

\bibitem[{Garay \& Lizano(1999)}]{Garay99}
Garay, G. \& Lizano, S. 1999, PASP, 111, 1049

\bibitem[{{Gerin} {et~al.}(2011){Gerin}, {Ka{\'z}mierczak}, {Jastrzebska},
  {Falgarone}, {Hily-Blant}, {Godard}, \& {de Luca}}]{Gerin11}
{Gerin}, M., {Ka{\'z}mierczak}, M., {Jastrzebska}, M., {Falgarone}, E.,
  {Hily-Blant}, P., {Godard}, B., \& {de Luca}, M. 2011, \aap, 525, A116

\bibitem[{{Ginsburg} {et~al.}(2012){Ginsburg}, {Bressert}, {Bally}, \&
  {Battersby}}]{Ginsburg12}
{Ginsburg}, A., {Bressert}, E., {Bally}, J., \& {Battersby}, C. 2012, \apjl,
  758, L29

\bibitem[{{Guesten} {et~al.}(1981){Guesten}, {Walmsley}, \& {Pauls}}]{Gusten81}
{Guesten}, R., {Walmsley}, C.~M., \& {Pauls}, T. 1981, \aap, 103, 197

\bibitem[{{G{\"u}sten} {et~al.}(2006){G{\"u}sten}, {Nyman}, {Schilke},
  {Menten}, {Cesarsky}, \& {Booth}}]{Gusten06}
{G{\"u}sten}, R., {Nyman}, L.~{\AA}., {Schilke}, P., {Menten}, K., {Cesarsky},
  C., \& {Booth}, R. 2006, \aap, 454, L13

\bibitem[{{Immer} {et~al.}(2012){Immer}, {Menten}, {Schuller}, \&
  {Lis}}]{Immer12}
{Immer}, K., {Menten}, K.~M., {Schuller}, F., \& {Lis}, D.~C. 2012, \aap, 548,
  A120

\bibitem[{{Jackson} {et~al.}(2013){Jackson}, {Rathborne}, {Foster}, {Whitaker},
  {Sanhueza}, {Claysmith}, {Mascoop}, {Wienen}, {Breen}, {Herpin},
  {Duarte-Cabral}, {Csengeri}, {Longmore}, {Contreras}, {Indermuehle},
  {Barnes}, {Walsh}, {Cunningham}, {Brooks}, {Britton}, {Voronkov}, {Urquhart},
  {Alves}, {Jordan}, {Hill}, {Hoq}, {Finn}, {C.}, {Bains}, {Bontemps},
  {Bronfman}, {Caswell}, {Deharveng}, {Ellingsen}, {Fuller}, {Garay}, {Green},
  {Hindson}, {Jones}, {Lenfestey}, {Lo}, {Lowe}, {Mardones}, {Menten},
  {Minier}, {Morgan}, {Motte}, {Muller}, {Peretto}, {Purcell}, {Schilke},
  {Schneider-Bontemps}, {Schuller}, {Titmarsh}, {Wyrowski}, \&
  {Zavagno}}]{Jackson-malt90}
{Jackson}, J.~M., {Rathborne}, J.~M., {Foster}, J.~B., {Whitaker}, J.~S.,
  {Sanhueza}, P., {Claysmith}, C., {Mascoop}, J.~L., {Wienen}, M., {Breen},
  S.~L., {Herpin}, F., {Duarte-Cabral}, A., {Csengeri}, T., {Longmore}, S.,
  {Contreras}, Y., {Indermuehle}, B., {Barnes}, P.~J., {Walsh}, A.~J.,
  {Cunningham}, M.~R., {Brooks}, K.~J., {Britton}, T.~R., {Voronkov}, M.~A.,
  {Urquhart}, J.~S., {Alves}, J., {Jordan}, C.~H., {Hill}, T., {Hoq}, S.,
  {Finn}, S., {C.}, S., {Bains}, I., {Bontemps}, S., {Bronfman}, L., {Caswell},
  J.~L., {Deharveng}, L., {Ellingsen}, S.~P., {Fuller}, G.~A., {Garay}, G.,
  {Green}, J.~A., {Hindson}, L., {Jones}, P.~A., {Lenfestey}, C., {Lo}, N.,
  {Lowe}, V., {Mardones}, D., {Menten}, K.~M., {Minier}, V., {Morgan}, L.~K.,
  {Motte}, F., {Muller}, E., {Peretto}, N., {Purcell}, C.~R., {Schilke}, P.,
  {Schneider-Bontemps}, N., {Schuller}, F., {Titmarsh}, A., {Wyrowski}, F., \&
  {Zavagno}, A. 2013, ArXiv e-prints

\bibitem[{{Jones} {et~al.}(2012){Jones}, {Burton}, {Cunningham},
  {Requena-Torres}, {Menten}, {Schilke}, {Belloche}, {Leurini},
  {Mart{\'{\i}}n-Pintado}, {Ott}, \& {Walsh}}]{Jones12}
{Jones}, P.~A., {Burton}, M.~G., {Cunningham}, M.~R., {Requena-Torres}, M.~A.,
  {Menten}, K.~M., {Schilke}, P., {Belloche}, A., {Leurini}, S.,
  {Mart{\'{\i}}n-Pintado}, J., {Ott}, J., \& {Walsh}, A.~J. 2012, \mnras, 419,
  2961

\bibitem[{{Kauffmann} {et~al.}(2013){Kauffmann}, {Pillai}, \&
  {Zhang}}]{Kauffmann13}
{Kauffmann}, J., {Pillai}, T., \& {Zhang}, Q. 2013, \apjl, 765, L35

\bibitem[{{Kruijssen}(2012)}]{Kruijssen12b}
{Kruijssen}, J.~M.~D. 2012, \mnras, 426, 3008

\bibitem[{{Kruijssen} \& {Cooper}(2012)}]{Kruijssen12}
{Kruijssen}, J.~M.~D. \& {Cooper}, A.~P. 2012, \mnras, 420, 340

\bibitem[{{Kruijssen} {et~al.}(2013){Kruijssen}, {Longmore}, {Elmegreen},
  {Murray}, {Bally}, {Testi}, \& {Kennicutt}}]{Kruijssen13a}
{Kruijssen}, J.~M.~D., {Longmore}, S.~N., {Elmegreen}, B.~G., {Murray}, N.,
  {Bally}, J., {Testi}, L., \& {Kennicutt}, Jr., R.~C. 2013, MNRAS submitted, arXiv:1303.6286

\bibitem[{{Kudryavtseva} {et~al.}(2012){Kudryavtseva}, {Brandner}, {Gennaro},
  {Rochau}, {Stolte}, {Andersen}, {Da Rio}, {Henning}, {Tognelli}, {Hogg},
  {Clark}, \& {Waters}}]{Kudryavtseva12}
{Kudryavtseva}, N., {Brandner}, W., {Gennaro}, M., {Rochau}, B., {Stolte}, A.,
  {Andersen}, M., {Da Rio}, N., {Henning}, T., {Tognelli}, E., {Hogg}, D.,
  {Clark}, S., \& {Waters}, R. 2012, \apjl, 750, L44

\bibitem[{{Kurtz} {et~al.}(2000){Kurtz}, {Cesaroni}, {Churchwell}, {Hofner}, \&
  {Walmsley}}]{Kurtz00}
{Kurtz}, S., {Cesaroni}, R., {Churchwell}, E., {Hofner}, P., \& {Walmsley},
  C.~M. 2000, Protostars and Planets IV, 299

\bibitem[{{Ladd} {et~al.}(2005){Ladd}, {Purcell}, {Wong}, \&
  {Robertson}}]{Ladd05}
{Ladd}, N., {Purcell}, C., {Wong}, T., \& {Robertson}, S. 2005, \pasa, 22, 62

\bibitem[{{Lis} \& {Carlstrom}(1994)}]{Lis94}
{Lis}, D.~C. \& {Carlstrom}, J.~E. 1994, ApJ, 424, 189

\bibitem[{{Lis} \& {Menten}(1998)}]{Lis98}
{Lis}, D.~C. \& {Menten}, K.~M. 1998, \apj, 507, 794

\bibitem[{{Lis} {et~al.}(2001){Lis}, {Serabyn}, {Zylka}, \& {Li}}]{Lis01}
{Lis}, D.~C., {Serabyn}, E., {Zylka}, R., \& {Li}, Y. 2001, \apj, 550, 761

\bibitem[{{Lo} {et~al.}(2009){Lo}, {Cunningham}, {Jones}, {Bains}, {Burton},
  {Wong}, {Muller}, {Kramer}, {Ossenkopf}, {Henkel}, {Deragopian}, {Donnelly},
  \& {Ladd}}]{Lo09}
{Lo}, N., {Cunningham}, M.~R., {Jones}, P.~A., {Bains}, I., {Burton}, M.~G.,
  {Wong}, T., {Muller}, E., {Kramer}, C., {Ossenkopf}, V., {Henkel}, C.,
  {Deragopian}, G., {Donnelly}, S., \& {Ladd}, E.~F. 2009, \mnras, 395, 1021

\bibitem[{{Longmore} {et~al.}(2013{\natexlab{a}}){Longmore}, {Bally}, {Testi},
  {Purcell}, {Walsh}, {Bressert}, {Pestalozzi}, {Molinari}, {Ott}, {Cortese},
  {Battersby}, {Murray}, {Lee}, {Kruijssen}, {Schisano}, \&
  {Elia}}]{Longmore13}
{Longmore}, S.~N., {Bally}, J., {Testi}, L., {Purcell}, C.~R., {Walsh}, A.~J.,
  {Bressert}, E., {Pestalozzi}, M., {Molinari}, S., {Ott}, J., {Cortese}, L.,
  {Battersby}, C., {Murray}, N., {Lee}, E., {Kruijssen}, J.~M.~D., {Schisano},
  E., \& {Elia}, D. 2013{\natexlab{a}}, \mnras, 429, 987

\bibitem[{{Longmore} {et~al.}(2013{\natexlab{b}}){Longmore}, {Kruijssen},
  {Bally}, {Ott}, {Testi}, {Rathborne}, {Bastian}, {Bressert}, {Molinari},
  {Battersby}, \& {Walsh}}]{Longmore13b}
{Longmore}, S.~N., {Kruijssen}, J.~M.~D., {Bally}, J., {Ott}, J., {Testi}, L.,
  {Rathborne}, J., {Bastian}, N., {Bressert}, E., {Molinari}, S., {Battersby},
  C., \& {Walsh}, A.~J. 2013{\natexlab{b}}, \mnras, 433, L15

\bibitem[{{Longmore} {et~al.}(2012){Longmore}, {Rathborne}, {Bastian}, {Alves},
  {Ascenso}, {Bally}, {Testi}, {Longmore}, {Battersby}, {Bressert}, {Purcell},
  {Walsh}, {Jackson}, {Foster}, {Molinari}, {Meingast}, {Amorim}, {Lima},
  {Marques}, {Moitinho}, {Pinhao}, {Rebordao}, \& {Santos}}]{Longmore12}
{Longmore}, S.~N., {Rathborne}, J., {Bastian}, N., {Alves}, J., {Ascenso}, J.,
  {Bally}, J., {Testi}, L., {Longmore}, A., {Battersby}, C., {Bressert}, E.,
  {Purcell}, C., {Walsh}, A., {Jackson}, J., {Foster}, J., {Molinari}, S.,
  {Meingast}, S., {Amorim}, A., {Lima}, J., {Marques}, R., {Moitinho}, A.,
  {Pinhao}, J., {Rebordao}, J., \& {Santos}, F.~D. 2012, \apj, 746, 117

\bibitem[{{Mart{\'{\i}}n-Pintado} {et~al.}(2001){Mart{\'{\i}}n-Pintado},
  {Rizzo}, {de Vicente}, {Rodr{\'{\i}}guez-Fern{\'a}ndez}, \&
  {Fuente}}]{Martin-Pintado01}
{Mart{\'{\i}}n-Pintado}, J., {Rizzo}, J.~R., {de Vicente}, P.,
  {Rodr{\'{\i}}guez-Fern{\'a}ndez}, N.~J., \& {Fuente}, A. 2001, \apjl, 548,
  L65

\bibitem[{{McMillan} {et~al.}(2007){McMillan}, {Vesperini}, \& {Portegies
  Zwart}}]{McMillan07}
{McMillan}, S.~L.~W., {Vesperini}, E., \& {Portegies Zwart}, S.~F. 2007, \apjl,
  655, L45

\bibitem[{{Molinari} {et~al.}(2011){Molinari}, {Bally}, {Noriega-Crespo},
  {Compi{\`e}gne}, {Bernard}, {Paradis}, {Martin}, {Testi}, {Barlow}, {Moore},
  {Plume}, {Swinyard}, {Zavagno}, {Calzoletti}, {Di Giorgio}, {Elia},
  {Faustini}, {Natoli}, {Pestalozzi}, {Pezzuto}, {Piacentini}, {Polenta},
  {Polychroni}, {Schisano}, {Traficante}, {Veneziani}, {Battersby}, {Burton},
  {Carey}, {Fukui}, {Li}, {Lord}, {Morgan}, {Motte}, {Schuller},
  {Stringfellow}, {Tan}, {Thompson}, {Ward-Thompson}, {White}, \&
  {Umana}}]{Molinari11}
{Molinari}, S., {Bally}, J., {Noriega-Crespo}, A., {Compi{\`e}gne}, M.,
  {Bernard}, J.~P., {Paradis}, D., {Martin}, P., {Testi}, L., {Barlow}, M.,
  {Moore}, T., {Plume}, R., {Swinyard}, B., {Zavagno}, A., {Calzoletti}, L.,
  {Di Giorgio}, A.~M., {Elia}, D., {Faustini}, F., {Natoli}, P., {Pestalozzi},
  M., {Pezzuto}, S., {Piacentini}, F., {Polenta}, G., {Polychroni}, D.,
  {Schisano}, E., {Traficante}, A., {Veneziani}, M., {Battersby}, C., {Burton},
  M., {Carey}, S., {Fukui}, Y., {Li}, J.~Z., {Lord}, S.~D., {Morgan}, L.,
  {Motte}, F., {Schuller}, F., {Stringfellow}, G.~S., {Tan}, J.~C., {Thompson},
  M.~A., {Ward-Thompson}, D., {White}, G., \& {Umana}, G. 2011, \apjl, 735, L33

\bibitem[{{M{\"u}ller} {et~al.}(2005){M{\"u}ller}, {Schl{\"o}der}, {Stutzki},
  \& {Winnewisser}}]{Muller05}
{M{\"u}ller}, H.~S.~P., {Schl{\"o}der}, F., {Stutzki}, J., \& {Winnewisser}, G.
  2005, Journal of Molecular Structure, 742, 215

\bibitem[{{M{\"u}ller} {et~al.}(2001){M{\"u}ller}, {Thorwirth}, {Roth}, \&
  {Winnewisser}}]{Muller01}
{M{\"u}ller}, H.~S.~P., {Thorwirth}, S., {Roth}, D.~A., \& {Winnewisser}, G.
  2001, \aap, 370, L49

\bibitem[{{Negueruela} {et~al.}(2010){Negueruela}, {Clark}, \&
  {Ritchie}}]{Negueruela10}
{Negueruela}, I., {Clark}, J.~S., \& {Ritchie}, B.~W. 2010, \aap, 516, A78

\bibitem[{{Portegies Zwart} {et~al.}(2010){Portegies Zwart}, {McMillan}, \&
  {Gieles}}]{Portegies-Zwart10}
{Portegies Zwart}, S.~F., {McMillan}, S.~L.~W., \& {Gieles}, M. 2010, \araa,
  48, 431

\bibitem[{{Rathborne} {et~al.}(2013){Rathborne}, {Longmore}, {Jackson},
  {Alves}, {Bally}, {Bastian}, {Bressert}, {Contreras}, {Foster}, {Garay},
  {Kruijssen}, {Testi}, \& {Walsh}}]{Rathborne-alma}
{Rathborne}, J.~M., {Longmore}, S.~N., {Jackson}, J.~M., {Alves}, J., {Bally},
  J., {Bastian}, N., {Bressert}, E., {Contreras}, Y., {Foster}, J.~B., {Garay},
  G., {Kruijssen}, J.~M.~D., {Testi}, L., \& {Walsh}, A.~J. 2013, submitted

\bibitem[{{Redman} {et~al.}(2002){Redman}, {Rawlings}, {Nutter},
  {Ward-Thompson}, \& {Williams}}]{Redman02}
{Redman}, M.~P., {Rawlings}, J.~M.~C., {Nutter}, D.~J., {Ward-Thompson}, D., \&
  {Williams}, D.~A. 2002, \mnras, 337, L17

\bibitem[{{Rodr{\'{\i}}guez} \& {Zapata}(2013)}]{Rodriguez13}
{Rodr{\'{\i}}guez}, L.~F. \& {Zapata}, L.~A. 2013, \apjl, 767, L13

\bibitem[{Sault {et~al.}(1995)Sault, Teuben, \& Wright}]{Sault95}
Sault, R.~J., Teuben, P.~J., \& Wright, M. C.~H. 1995, in Astronomical Data
  Analysis Software and Systems IV, ed. R.~Shaw, H.~Payne, \& J.~Hayes, Vol.~77
  (San Francisco: ASP Conference Series), 433

\bibitem[{{Savage} {et~al.}(2002){Savage}, {Apponi}, {Ziurys}, \&
  {Wyckoff}}]{Savage02}
{Savage}, C., {Apponi}, A.~J., {Ziurys}, L.~M., \& {Wyckoff}, S. 2002, \apj,
  578, 211

\bibitem[{{Schilke} {et~al.}(1992){Schilke}, {Walmsley}, {Pineau Des Forets},
  {Roueff}, {Flower}, \& {Guilloteau}}]{Schilke92}
{Schilke}, P., {Walmsley}, C.~M., {Pineau Des Forets}, G., {Roueff}, E.,
  {Flower}, D.~R., \& {Guilloteau}, S. 1992, \aap, 256, 595

\bibitem[{{Sch{\"o}ier} {et~al.}(2005){Sch{\"o}ier}, {van der Tak}, {van
  Dishoeck}, \& {Black}}]{Schoier05}
{Sch{\"o}ier}, F.~L., {van der Tak}, F.~F.~S., {van Dishoeck}, E.~F., \&
  {Black}, J.~H. 2005, \aap, 432, 369

\bibitem[{{Schuller} {et~al.}(2009){Schuller}, {Menten}, {Contreras},
  {Wyrowski}, {Schilke}, {Bronfman}, {Henning}, {Walmsley}, {Beuther},
  {Bontemps}, {Cesaroni}, {Deharveng}, {Garay}, {Herpin}, {Lefloch}, {Linz},
  {Mardones}, {Minier}, {Molinari}, {Motte}, {Nyman}, {Reveret}, {Risacher},
  {Russeil}, {Schneider}, {Testi}, {Troost}, {Vasyunina}, {Wienen}, {Zavagno},
  {Kovacs}, {Kreysa}, {Siringo}, \& {Wei{\ss}}}]{Schuller09}
{Schuller}, F., {Menten}, K.~M., {Contreras}, Y., {Wyrowski}, F., {Schilke},
  P., {Bronfman}, L., {Henning}, T., {Walmsley}, C.~M., {Beuther}, H.,
  {Bontemps}, S., {Cesaroni}, R., {Deharveng}, L., {Garay}, G., {Herpin}, F.,
  {Lefloch}, B., {Linz}, H., {Mardones}, D., {Minier}, V., {Molinari}, S.,
  {Motte}, F., {Nyman}, L., {Reveret}, V., {Risacher}, C., {Russeil}, D.,
  {Schneider}, N., {Testi}, L., {Troost}, T., {Vasyunina}, T., {Wienen}, M.,
  {Zavagno}, A., {Kovacs}, A., {Kreysa}, E., {Siringo}, G., \& {Wei{\ss}}, A.
  2009, \aap, 504, 415

\bibitem[{{van der Tak} {et~al.}(2007){van der Tak}, {Black}, {Sch{\"o}ier},
  {Jansen}, \& {van Dishoeck}}]{vanderTak07}
{van der Tak}, F.~F.~S., {Black}, J.~H., {Sch{\"o}ier}, F.~L., {Jansen}, D.~J.,
  \& {van Dishoeck}, E.~F. 2007, \aap, 468, 627

\bibitem[{{Vasyunina} {et~al.}(2011){Vasyunina}, {Linz}, {Henning},
  {Zinchenko}, {Beuther}, \& {Voronkov}}]{Vasyunina11}
{Vasyunina}, T., {Linz}, H., {Henning}, T., {Zinchenko}, I., {Beuther}, H., \&
  {Voronkov}, M. 2011, \aap, 527, A88

\bibitem[{{Walmsley} {et~al.}(1986){Walmsley}, {Guesten}, {Angerhofer},
  {Churchwell}, \& {Mundy}}]{Walmsley86}
{Walmsley}, C.~M., {Guesten}, R., {Angerhofer}, P., {Churchwell}, E., \&
  {Mundy}, L. 1986, \aap, 155, 129

\bibitem[{{Wilson}(1999)}]{Wilson99}
{Wilson}, T.~L. 1999, Reports on Progress in Physics, 62, 143

\bibitem[{{Wilson} {et~al.}(1982){Wilson}, {Ruf}, {Walmsley}, {Martin},
  {Batrla}, \& {Pauls}}]{Wilson82}
{Wilson}, T.~L., {Ruf}, K., {Walmsley}, C.~M., {Martin}, R.~N., {Batrla}, W.,
  \& {Pauls}, T.~A. 1982, \aap, 115, 185

\bibitem[{{Yusef-Zadeh} {et~al.}(2013){Yusef-Zadeh}, {Wardle}, {Lis}, {Viti},
  {Brogan}, {Chambers}, {Pound}, \& {Rickert}}]{Yusef-zadeh13}
{Yusef-Zadeh}, F., {Wardle}, M., {Lis}, D., {Viti}, S., {Brogan}, C.,
  {Chambers}, E., {Pound}, M., \& {Rickert}, M. 2013, Journal of Physical
  Chemistry A, 117, 9404

\end{thebibliography}
%\bibliographystyle{../../apj}

%%%%%%%%%%%%%%%%%%%%%%%%%%%%%%%%%%%%%%%%%%%%%%%%%%%%%%%%%%%%%%%%%%%%%%%%%%%%%%%%%%%%

%%%%%%%%%%%%%%%%%%%%%%%%%%%%%%%%%%%%%%%%%%%%%%%%%%%%%%%%%%%%%%%%%%%%%%%%%%%%%%%%%%%%
% tables
%%%%%%%%%%%%%%%%%%%%%%%%%%%%%%%%%%%%%%%%%%%%%%%%%%%%%%%%%%%%%%%%%%%%%%%%%%%%%%%%%%%%
\clearpage
\begin{table*}[h]
{\scriptsize
\begin{center}
\caption{Summary of the molecular transitions observed and the detected emission (ordered by increasing excitation energies\tablenotemark{a}. \label{malt90-lines}}
\begin{tabular}{llccccccc}
\tableline
\tableline
& Transition & Frequency  & E$_{u}$/$k$ & \ncrit\,  & \multicolumn{4}{c}{Peak position}  \\
 &                &       &      &  & Long & Lat & Vel &  \tmb \\
 &                 & (GHz) &  (K) &    (\cmc)          & (deg) & (deg)         &  (\kms) & (K)  \\
\tableline
&	\htcop\,     & 86.754330      &    4.16      & 2 $\times$ 10$^{5}$   &        0.248   &        0.004 &        36.68   &         0.39  \\
&	\hntc\,       & 87.090859        &     4.18     &   3 $\times$ 10$^{5}$ &        0.248   &        0.007 &        36.35   &         0.58  \\
&	\cch\,        & 87.316925 	   &     4.19     & 2 $\times$ 10$^{5}$  &        0.243   &        0.007 &        38.69   &         1.00  \\
&	\hcn\,         & 88.631847 	   &     4.25    &  3 $\times$ 10$^{6}$   &        0.248   &        0.007 &        47.51   &         4.67  \\
&	\hcop\,       & 89.188526 	   &    4.28     & 2 $\times$ 10$^{5}$   &        0.246   &        0.007 &        44.48   &         3.31  \\
&	\hnc\,   	  & 90.663572         &     4.35   &  3 $\times$ 10$^{5}$   &        0.248   &        0.009 &        41.21   &         2.91  \\
& 	 \nthp\, 	  &  93.173772  	   &  4.47     & 3 $\times$ 10$^{5}$       &   0.248   &        0.007 &        43.32   &         3.53  \\
&	\sio\,         & 86.847010       &    6.25      &  3 $\times$ 10$^{5}$  &        0.246   &        0.004 &        40.34   &         1.07  \\
&	\tcs\,	        &  92.494303        & 6.66      & 3 $\times$ 10$^{5}$     &        0.226   &       -0.001 &        34.75   &         0.37  \\
&	\tctfs\,         & 90.926036 	   &      7.05   &   4 $\times$ 10$^{5}$   &  -  &  - & -  & -  \\  
&	\hncofz\,   & 87.925238         &   10.55    & 1 $\times$ 10$^{6}$   &        0.246   &        0.007 &        36.86   &         2.52  \\
&	\chtcn\,       & 91.985316 	   &   20.39   & 4 $\times$ 10$^{5}$   &        0.243   &        0.004 &        33.13   &         0.87  \\
&	\hctn\,  	  & 90.978989 	   &   24.01   & 5 $\times$ 10$^{5}$     &        0.246   &        0.007 &        37.52   &         2.98  \\
&	\hctccn\,     & 90.593059         &       24.37  &  1 $\times$ 10$^{6}$    &  - &  - &  - & -  \\  
&	\hcnnt\, (3--2)   & 265.886434  &   25.52 &  7 $\times$ 10$^{7}$      &         0.243   &        0.007 &        43.30   &         3.70  \\
&	\hcopnt\, (3--2) & 267.557619 &    25.68 &  4 $\times$ 10$^{6}$      &         0.243   &        0.007 &        43.00   &         1.74  \\
&	\hncnt\, (3--2)   &  271.981142  &  26.11 &  8 $\times$ 10$^{6}$      &         0.258   &        0.019 &        40.80   &         1.21  \\
&	\nthpnt\, (3--2)  & 279.511701  &   26.83 & 3 $\times$ 10$^{6}$       &        0.243   &        0.007 &        38.40   &         3.16  \\
&	\siont\, (5--4)      &  217.104980   &  31.26 &  5 $\times$ 10$^{6}$   &         0.241   &        0.007 &        39.07   &         0.56  \\
&	\hcnnt\, (4--3)   & 354.505477   &  42.53 & 3 $\times$ 10$^{8}$       &         0.243   &        0.004 &        40.70   &         1.55  \\
&	\hcopnt\, (4--3) & 356.734288 &    42.80 & 9 $\times$ 10$^{6}$       &         0.248   &        0.004 &        35.00   &         0.84  \\
&	\hncnt\, (4--3)   & 362.630303  &  43.51 & 8 $\times$ 10$^{7}$        &        0.251   &        0.009 &        37.60   &         0.84  \\
&	\nthpnt\, (4--3)  & 372.672509  &   44.71 & 8 $\times$ 10$^{6}$      &        0.241   &        0.007 &        40.00   &         0.37  \\
&	\hncofo\,   & 88.239027         &   53.86    &   1 $\times$ 10$^{6}$  &  - &  - &  - &  - \\  
& 	\halpha\,     & 92.034475 	   &               &                                     & -  &  - &  - &  - \\  
\tableline
\end{tabular}
\tablenotetext{a}{ Excitation energies and critical densities (E$_{u}$ and \ncrit) were calculated using values from
the  Leiden Atomic and Molecular Database (LAMDA; \citealp{Schoier05})  and Cologne Database for Molecular Spectroscopy (CDMS; \citealp{Muller01,Muller05}) assuming a gas temperature of 20\,K.}
\end{center}
}
\end{table*} 
%%%%%%%%%%%%%%%%%%%%%%%%%%%%%%%%%%%%%%%%%%%%%%%%%%%%%%%%%%%%%%%%%%%%%%%%%%%%%%%%%%%%
% figures
%%%%%%%%%%%%%%%%%%%%%%%%%%%%%%%%%%%%%%%%%%%%%%%%%%%%%%%%%%%%%%%%%%%%%%%%%%%%%%%%%%%%
\clearpage
%%%%%%%%%%%%%%%%%%%%%%%%%%%%%%%%%%%%%%%%%%%%%%%%%%%%%%%%%%%%%%%%%%%%%%%%%%%%%%%%%%%%
%%%%%%%%%%%%%%%%%%%%%%%%%%%%%%%%%%%%%%%%%%%%%%%%%%%%%%%%%%%%%%%%%%%%%%%%%%%%%%%%%%%%
%%%%%%%%%%%%%%%%%%%%%%%%%%%%%%%%%%%%%%%%%%%%%%%%%%%%%%%%%%%%%%%%%%%%%%%%%%%%%%%%%%%
%%%%%%%%%%%%%%%%%%%%%%%%%%%%%%%%%%%%%%%%%%%%%%%%%%%%%%%%%%%%%%%%%%%%%%%%%%%%%%%%%%%%
%%%%%%%%%%%%%%%%%%%%%%%%%%%%%%%%%%%%%%%%%%%%%%%%%%%%%%%%%%%%%%%%%%%%%%%%%%%%%%%%%%%%
\clearpage
%%%%%%%%%%%%%%%%%%%%%%%%%%%%%%%%%%%%%%%%%%%%%%%%%%%%%%%%%%%%%%%%%%%%%%%%%%%%%%%%%%%%
\clearpage
%%%%%%%%%%%%%%%%%%%%%%%%%%%%%%%%%%%%%%%%%%%%%%%%%%%%%%%%%%%%%%%%%%%%%%%%%%%%%%%%%%%%
\clearpage
%%%%%%%%%%%%%%%%%%%%%%%%%%%%%%%%%%%%%%%%%%%%%%%%%%%%%%%%%%%%%%%%%%%%%%%%%%%%%%%%%%%%
%%%%%%%%%%%%%%%%%%%%%%%%%%%%%%%%%%%%%%%%%%%%%%%%%%%%%%%%%%%%%%%%%%%%%%%%%%%%%%%%%%%%
%%%%%%%%%%%%%%%%%%%%%%%%%%%%%%%%%%%%%%%%%%%%%%%%%%%%%%%%%%%%%%%%%%%%%%%%%%%%%%%%%%%%

%%%%%%%%%%%%%%%%%%%%%%%%%%%%%%%%%%%%%%%%%%%%%%%%%%%%%%%%%%%%%%%%%%%%%%%%%%%%%%%%%%%%
%%%%%%%%%%%%%%%%%%%%%%%%%%%%%%%%%%%%%%%%%%%%%%%%%%%%%%%%%%%%%%%%%%%%%%%%%%%%%%%%%%%%
\clearpage
\section{Appendix : online only material}
\clearpage
%%%%%%%%%%%%%%%%%%%%%%%%%%%%%%%%%%%%%%%%%%%%%%%%%%%%%%%%%%%%%%%%%%%%%%%%%%%%%%%%%%%%
\clearpage
%%%%%%%%%%%%%%%%%%%%%%%%%%%%%%%%%%%%%%%%%%%%%%%%%%%%%%%%%%%%%%%%%%%%%%%%%%%%%%%%%%%%
\clearpage
%%%%%%%%%%%%%%%%%%%%%%%%%%%%%%%%%%%%%%%%%%%%%%%%%%%%%%%%%%%%%%%%%%%%%%%%%%%%%%%%%%%%
\clearpage
%%%%%%%%%%%%%%%%%%%%%%%%%%%%%%%%%%%%%%%%%%%%%%%%%%%%%%%%%%%%%%%%%%%%%%%%%%%%%%%%%%%%
\clearpage
%%%%%%%%%%%%%%%%%%%%%%%%%%%%%%%%%%%%%%%%%%%%%%%%%%%%%%%%%%%%%%%%%%%%%%%%%%%%%%%%%%%%
\clearpage
%%%%%%%%%%%%%%%%%%%%%%%%%%%%%%%%%%%%%%%%%%%%%%%%%%%%%%%%%%%%%%%%%%%%%%%%%%%%%%%%%%%%
\clearpage
%%%%%%%%%%%%%%%%%%%%%%%%%%%%%%%%%%%%%%%%%%%%%%%%%%%%%%%%%%%%%%%%%%%%%%%%%%%%%%%%%%%%
\clearpage
%%%%%%%%%%%%%%%%%%%%%%%%%%%%%%%%%%%%%%%%%%%%%%%%%%%%%%%%%%%%%%%%%%%%%%%%%%%%%%%%%%%%
\clearpage
%%%%%%%%%%%%%%%%%%%%%%%%%%%%%%%%%%%%%%%%%%%%%%%%%%%%%%%%%%%%%%%%%%%%%%%%%%%%%%%%%%%%
\clearpage
%%%%%%%%%%%%%%%%%%%%%%%%%%%%%%%%%%%%%%%%%%%%%%%%%%%%%%%%%%%%%%%%%%%%%%%%%%%%%%%%%%%%
\clearpage
%%%%%%%%%%%%%%%%%%%%%%%%%%%%%%%%%%%%%%%%%%%%%%%%%%%%%%%%%%%%%%%%%%%%%%%%%%%%%%%%%%%%
\clearpage
%%%%%%%%%%%%%%%%%%%%%%%%%%%%%%%%%%%%%%%%%%%%%%%%%%%%%%%%%%%%%%%%%%%%%%%%%%%%%%%%%%%%
\clearpage
%%%%%%%%%%%%%%%%%%%%%%%%%%%%%%%%%%%%%%%%%%%%%%%%%%%%%%%%%%%%%%%%%%%%%%%%%%%%%%%%%%%%
\clearpage
%%%%%%%%%%%%%%%%%%%%%%%%%%%%%%%%%%%%%%%%%%%%%%%%%%%%%%%%%%%%%%%%%%%%%%%%%%%%%%%%%%%%
\clearpage
%%%%%%%%%%%%%%%%%%%%%%%%%%%%%%%%%%%%%%%%%%%%%%%%%%%%%%%%%%%%%%%%%%%%%%%%%%%%%%%%%%%%
\clearpage
%%%%%%%%%%%%%%%%%%%%%%%%%%%%%%%%%%%%%%%%%%%%%%%%%%%%%%%%%%%%%%%%%%%%%%%%%%%%%%%%%%%%
\clearpage
%%%%%%%%%%%%%%%%%%%%%%%%%%%%%%%%%%%%%%%%%%%%%%%%%%%%%%%%%%%%%%%%%%%%%%%%%%%%%%%%%%%%
\clearpage
%%%%%%%%%%%%%%%%%%%%%%%%%%%%%%%%%%%%%%%%%%%%%%%%%%%%%%%%%%%%%%%%%%%%%%%%%%%%%%%%%%%%
\clearpage
%%%%%%%%%%%%%%%%%%%%%%%%%%%%%%%%%%%%%%%%%%%%%%%%%%%%%%%%%%%%%%%%%%%%%%%%%%%%%%%%%%%%\clearpage
%%%%%%%%%%%%%%%%%%%%%%%%%%%%%%%%%%%%%%%%%%%%%%%%%%%%%%%%%%%%%%%%%%%%%%%%%%%%%%%%%%%%
%%%%%%%%%%%%%%%%%%%%%%%%%%%%%%%%%%%%%%%%%%%%%%%%%%%%%%%%%%%%%%%%%%%%%%%%%%%%%%%%%%%%
\end{document}